\documentclass[onecolumn,leqno,10pt,a4paper]{amsart}
\usepackage{amssymb,eucal,mathrsfs,array,setspace,geometry,enumitem,cite}
\usepackage[backref=false]{hyperref}
\usepackage[capitalise,noabbrev]{cleveref} 
\hypersetup{colorlinks=true,citecolor=cyan,linkcolor=black,urlcolor=blue}
\usepackage{color,caption}   
\usepackage{calligra}
\usepackage{lscape}
\usepackage{txfonts}  
\DeclareSymbolFont{largesymbols}{OMX}{zplm}{m}{n} 
\usepackage{mathtools,dsfont} 
\usepackage{stmaryrd} 
\usepackage{latexsym}
\usepackage{amsmath,amsthm,amsopn,amsfonts,stmaryrd,mathtools,dsfont}
\usepackage[english]{babel}
\usepackage[T1]{fontenc}
\usepackage{tabmac}

\usepackage{pb-diagram}
\usepackage{geometry}
\numberwithin{equation}{section}

\newtheoremstyle{Comp}%
  {10pt}{10pt}%
  {}{}%
  {\scshape}{~--~}%
  { }{}

\theoremstyle{Comp}
\newtheorem{theorem}{Theorem}[section]

\newtheorem{proposition}[theorem]{Proposition}
\newtheorem{conjecture}[theorem]{Conjecture}
\newtheorem{open}[theorem]{Open problem}
\newtheorem{corollary}[theorem]{Corollary}

\theoremstyle{remark}

\def\dd{{\mathrm d}}  
\def\QQ{{\mathbb Q}} 
\def\ZZ{{\mathbb Z}}

\def\la{{\lambda}}
\def\La{{\Lambda}}

\def\cal L{{\mathcal L}}

\def\cd{\circledast}

\let\n\noindent

\def\cd{{\circledast}}

\def\lrw{\leftrightarrow}

\let\d\partial

\let\n\noindent

\def\B{{\mathcal B}}

\def\S{\mathcal{S}}

\let\a\alpha

\let\Om\Omega

\let\ta\theta

\def\lrw{\leftrightarrow}

\let\n\noindent

\let\Om\Omega

\let\ta\theta

\let\Ga\Gamma

\newcommand{\llangle}{\ensuremath{\langle\!\langle}}
\newcommand{\rrangle}{\ensuremath{\rangle\!\rangle}}

\def\cd{{\circledast}}

\def\beq{\begin{equation}}
\def\eeq{\end{equation}}

\def\LaA{\La^{\mathsf a}}  
\def\LaS{\La^{\mathsf s}}  

\DeclarePairedDelimiterX{\psc}[2]{\llangle}{\rrangle}{#1 \vert #2}  
\DeclarePairedDelimiterX{\psa}[2]{\langle}{\rangle}{#1 \vert #2}  

\usepackage{tikz}

\usepackage{ytableau}

\makeatletter
\newcommand{\superimpose}[2]{%
  {\ooalign{$#1\@firstoftwo#2$\cr\hfil$#1\@secondoftwo#2$\hfil\cr}}}
  \makeatother

\newcommand{\sFill}[1]{
		\begin{tikzpicture}[scale=.125, remember picture]
	\begin{scope}[overlay]
		\draw[clip] (0,.5) circle [radius=1];
	\end{scope}
	\begin{scope}[overlay]
		\draw[clip] (0,.5) circle [radius=1];
		\node[scale=.8] at (0,.5) {${#1}$};
	\end{scope}
	\end{tikzpicture}
}

\newcommand{\sFillgray}[1]{
		\begin{tikzpicture}[scale=.125, remember picture]
	\begin{scope}[overlay]
		\draw[clip] (0,.5) circle [radius=1];
		\filldraw[fill=gray!50] (0,0.5) circle [radius=1];
	\end{scope}
	\begin{scope}[overlay]
		\draw[clip] (0,.5) circle [radius=1];
		\node[scale=.8] at (0,.5) {${#1}$};
	\end{scope}
	\end{tikzpicture}
}

\newcommand{\yF}[1]{\none[{\sFill{#1}}]}
\newcommand{\yFgray}[1]{\none[{\sFillgray{#1}}]}

\newcommand{\superY}[1]{\begin{array}{c} \scalebox{1}{\begin{ytableau}#1 \end{ytableau} } \end{array}}

\ytableausetup{smalltableaux}

\newcommand{\tcercle}[1]{\ensuremath{\setlength{\unitlength}{1ex}\begin{picture}(2.8,2.8)\put(1.4,1.4){\circle{2.8}\makebox(-5.6,0){#1}}\end{picture}}}

\begin{document}

\title{Symmetric functions in superspace: a compendium  of results and
  open problems (including a SageMath worksheet)}
\author{L~Alarie-V\'ezina}

\address[Ludovic Alarie-V\'ezina]{
D\'epartement de Physique, de G\'enie Physique et d'Optique \\
Universit\'e Laval \\ 
Qu\'ebec, Canada, G1V~0A6.
}

\email{ludovic.alarie-vezina.1@ulaval.ca}

\author{O~Blondeau-Fournier}

\address[Olivier Blondeau-Fournier]{
D\'epartement de Physique, de G\'enie Physique et d'Optique \\
Universit\'e Laval \\ 
Qu\'ebec, Canada, G1V~0A6.
}

\email{olivier.b-fournier.1@ulaval.ca}

\author{P~Desrosiers}

\address[Patrick Desrosiers]{
CERVO Brain Research Center \\
 Québec, Canada, G1J~2G3; 
D\'epartement de Physique, de G\'enie Physique et d'Optique \\
Universit\'e Laval \\ 
Qu\'ebec, Canada, G1V~0A6;
Centre interdisciplinaire en mod\'elisation math\'ematique de l'Universit\'e Laval \\
Québec, Canada, G1V~0A6.}

\email{patrick.desrosiers@phy.ulaval.ca}

\author{L~Lapointe}

\address[Luc Lapointe]{
Instituto de Matem\'atica y F\'{\i}sica \\ 
Universidad de Talca \\
 2 norte 685, Talca, Chile.
}

\email{lapointe@inst-mat.utalca.cl}

\author{P~Mathieu}

\address[Pierre Mathieu]{
D\'epartement de Physique, de G\'enie Physique et d'Optique \\
Universit\'e Laval \\ 
Qu\'ebec, Canada, G1V~0A6.
}

\email{pmathieu@phy.ulaval.ca}

\thanks{\today}
\maketitle


\begin{abstract}
We present a review of the most important results in the theory of symmetric functions in superspace (or symmetric superpolynomials), summarizing all principal contributions since its introduction in 2001 in the context of the supersymmetric Calogero-Moser-Sutherland integrable model.  We also mention some open problems which remain unanswered at this moment, in particular the connection with representation theory.     In addition, we provide a free open access source code, relying on SageMath library, that can be used as a research tool for symmetric superpolynomials.  
The content is directed to an  audience new to this research area, but who is familiar with the classical theory of symmetric functions.   
\end{abstract}



\hypersetup{colorlinks=true,citecolor=cyan,linkcolor=magenta,urlcolor=blue}

\section{Introduction}

Polynomials in the $n$ commuting variables $x_1, \ldots, x_n$ that are invariant under the action of the symmetric group of $n$ elements, $\mathfrak S_n$, are called symmetric polynomials.   
The theory of symmetric polynomials finds important applications in representation theory and in combinatorics,  as is very well highlighted in  Macdonald's seminal book \cite{MacSym95}.   Many relevant quantities in the ring of symmetric functions are for instance described beautifully using fillings of Ferrers' diagram (or tableaux), the best-known example being the product of Schur functions described by the Littlewood-Richardson rule on tableaux, which also gives the multiplicities of tensor product of  irreducible representations of $\mathrm{GL}(n)$. 
Another important point is that the ring of symmetric functions has many interesting bases, some of them being important families of  orthogonal polynomials.      
The most general family is that of the Macdonald polynomials, from which most other bases can be obtained through an appropriate specialization.


Symmetric polynomials have  applications in mathematical-physics where they appear in connection with integrable models.  
  More precisely, the eigenfunctions of the Hamiltonian of the Calogero-Moser-Sutherland model (CMS) are essentially given  by Jack polynomials while the Hamiltonian of the relativistic version of the CMS model, known as the Ruijsenaars-Schneider (RS) model,  has as eigenfunctions the Macdonald polynomials \cite{Ruij99}.\footnote{Here, when referring to the CMS and RS models, we will always understand their trigonometric version. Note that, in the literature, the trigonometric CMS model is also referred to as the Calogero-Sutherland model.}
  Symmetric functions also have important applications in conformal field theory (CFT), mostly due to the fact that a subfamily of Jack polynomials can be used to represent 
  the singular vectors of the Virasoro algebra \cite{Mimachi95,Awata95,SSAFR}.  The Jack polynomials, this time through the Whittaker vector of the conformal algebra,  also play a role in a degenerate case of the AGT conjecture \cite{AGT} relating supersymmetric gauge theory in 4 dimensions with conformal field theory in 2 dimensions \cite{AY,MMS, MMSS,Yan}.

A generalization of symmetric function theory to superspace
has been developed over the last 15 years (see for instance \cite{DLMnpb,DLM_sJack,DLM_class,DLMadv,BF2,ABLM18,GJL}).  
Such a generalization involves the anticommuting variables $\theta_1,\dots,\theta_n$ (that is, such that $\theta_i \theta_j=-\theta_j \theta_i$) 
as well as the usual commuting variables $x_1,\dots,x_n$.
In superspace, the symmetric group $\mathfrak S_n$ now acts diagonally, that is, equally on both sets of variables (the $x_i$'s and the $\ta_i$'s). 
The subring fixed by the symmetric group $\mathfrak S_n$ is called the ring of \emph{symmetric superpolynomials} (or the ring of symmetric functions in superspace).  The anticommuting nature of the indeterminates $\ta_i$ refers to the formalism of supersymmetry.  
In the context of supersymmetric quantum mechanics, each particle now has a bosonic (i.e.~even) and a fermionic (i.e.~odd) degree of freedom.  Supersymmetric versions of quantum integrable many-body systems, such as the CMS model, were first considered by Freedman and Mende, and by Shastry and Sutherland \cite{Freedman90, Shas93}. 
In  \cite{DLMnpb}, three of the authors of the present Compendium introduced a family   of symmetric superpolynomials, called Jack polynomials in superspace (or super-Jacks, for short), which diagonalize the Hamiltonian of the supersymmetric (trigonometric) CMS model.  The combinatorial properties of the super-Jacks, as well as those of the classical superpolynomials such as the power-sums and elementary symmetric functions in superspace, were considered independently of any physical context in \cite{DLM_class, DLM_fauxsJack, DLM_sJack, DLMadv, DLM_eval}.

The relativistic version of the CMS system, namely the RS model mentioned above,
 was studied in \cite{BDM15} from the point of view of supersymmetric quantum integrable models.  
 One of the main results is that the eigenfunctions of the (supersymmetric) Hamiltonian are given by superpolynomials that both generalize the super-Jacks and the Macdonald polynomials, and which are hence called Macdonald superpolynomials.    
 They were introduced earlier in \cite{BF1}, where their combinatorial properties were investigated  (including for instance a generalization to superspace of the original Macdonald positivity conjecture).    It was understood that the underlying symmetry of the supersymmetric RS model is played by the double affine Hecke algebra, which resulted in a connection with the non-symmetric Macdonald polynomials.  
 This link provided the explicit construction of all the conserved commuting quantities (thus proving the complete integrability of the super-RS model), as well as showing the existence of the super-Macdonald polynomials \cite{BF2}.

Families of Schur functions in superspace were then obtained as special cases of the  Macdonald polynomials in superspace \cite{BM1,JL17}.  Even though they have, as in the usual case,  nice tableau expansions and obey simple Pieri rules, much of their combinatorics (such as a Schensted-type insertion) remains to be uncovered. This is yet another example of the fascinating way in which the rich combinatorics of symmetric function theory as well as its deep connections with physics (such as with conformal field theory) extend beautifully to superspace.

Unfortunately, the theory of symmetric functions in superspace has evolved over the years not as systematically as one would have wished, with
the right (canonical) understanding of the concepts often only appearing in the later articles (something as simple as for instance the proper dominance ordering on superpartitions only emerging after many articles had already been published).  Realizing this shortcoming, the main objective of this Compendium is  thus to present a
unified view of the important contributions pertaining to symmetric superpolynomials in the hope that researchers interested in this rich field can easily have an idea of what is known, where to find the proofs, and what are the open problems.   To further facilitate matters,  an interactive free access worksheet, following more or less the presentation of the compendium and 
constructed with the SageMath code library, can be found in \cite{SageComp}.

The structure of the Compendium is as follows. In Section \ref{SsJ}, we introduce the basic properties of the ring of symmetric superpolynomials.  In particular, we first present the two equivalent definitions of superpartitions (Sections \ref{supLano1} and \ref{supLano1}), as well as their diagrammatic representations.    We then turn to the different bases of the ring  of symmetric superpolynomials, and in particular to the multiplicative bases and their generating series (Section  \ref{SecMultbases}).    A natural action of an involutive automorphism, as well as a scalar product, are defined in the next subsection.  

 We then shift our attention to the Jack superpolynomials in Section \ref{SJack}.    The combinatorial definition (Section \ref{sjackcombdef}), the eigenvalue problem definition (Section \ref{charaphys}), and the symmetrization definition (Section \ref{defsjakcsymm}) are first presented,  followed by
their properties in the next subsections.   
  
In Section \ref{SMacs}, we present the most general basis, that of the Macdonald superpolynomials.  We show in particular how the three definitions of Jack superpolynomials can be extended to the Macdonald case.
  In Section \ref{DoubleMM}, we treat a particular subclass of Macdonald superpolynomials that can be written as double symmetric functions.  
  Finally,   we end the compendium in Section \ref{SSchur} with super-Schurs,  where Pieri rules are first presented, followed by a constructive method for building the super-Schurs recursively.

\subsection*{SageMath source code}   
As mentioned earlier, an interactive free access worksheet, following more or less the presentation of the compendium and 
constructed with the SageMath code library, can be found in \cite{SageComp}.

\medskip

\section*{Acknowledgments}

This work is supported by FONDECYT (Fondo Nacional de Desarrollo Científico y Tecnológico de Chile) regular grant \#1170924, and NSERC.

\section{Superpartitions and symmetric superpolynomials}
\label{SsJ}
In this first section, we summarize the basic notions and definitions pertaining to symmetric superpolynomials.

\subsection{Superpartitions}
\label{Spart}  
Recall that  a regular partition $\la=(\la_1, \la_2,\ldots)$ is a non-increasing sequence of positive integers.  We say that $\la$ is partition of $n$, often denoted  $\la \vdash n$, if $\vert \la \vert = \sum_i \la_i = n$.  Alternatively, we may write $\la=(1^{n_\la(1)}, 2^{n_\la(2)},\ldots)$ where the value $n_\la(i)$ denotes the number of parts equal to $i$ in $\la$.   
 
A superpartitions is a generalization of a partition used to label symmetric superpolynomials.  There are two equivalent definitions, which we shall now present.

\subsubsection{Superpartition: the $(\LaA;\LaS)$ description} \label{supLano1}
A superpartition $\La$  is
a pair of partitions written as
\begin{equation}\label{sppa}
\La=(\LaA;\LaS)=(\La_1,\ldots,\La_m;\La_{m+1},\ldots,\La_\ell),
\end{equation}
with the conditions
\begin{equation}\label{sppb}
\La_1>\La_2 > \ldots>\La_m\geq0 \qquad  \text{ and}
\qquad \La_{m+1}\geq \La_{m+2} \geq \ldots \geq
\La_\ell > 0.
\end{equation}
We say that $\ell$ is the length of $\La$.  
We stress that $\LaA$ has distinct parts and that the last part is allowed to be $0$.  
It is often convenient to consider $\La$ with exactly $N>\ell$ parts: in that case we append a sequence of $N-\ell$ parts equal to zero, $\La_{\ell+1}=\ldots=\La_N=0$.
The number $m$, indicating the number of parts in $\LaA$, is called the fermionic degree of $\La$ and $n=|\La|=\sum_i\La_i$ is the bosonic degree.  
Such a superpartition $\La$ is said to be of degree $(n|m)$, which is denoted  $\La\vdash(n|m)$.  Note that the superpartitions of degree $(n|0)$ are the regular partitions.   
The set of  all superpartitions is denoted $\mathrm{SPar}$, and $\text{SPar}(n|m)$ denotes the subset of superpartitions of degree $(n|m)$.
Here is an example of a superpartition of degree $(27|5)$ in this description: $\La=(8,6,3,2,0;5,3)$.

\subsubsection{Superpartition: the $(\La^*,\La^\cd)$ description}\label{supLano2}
A superpartition $\La\vdash(n|m)$ of length $\ell$ (as above) is described by a pair of  partitions, denoted $\La^*$ and $\La^\circledast$, which satisfy the following constraints:
\begin{enumerate}
\item $\La^* \subseteq \La^\circledast$;
\item the degree of $\La^*$ is $n$;
\item the length of $\La^\circledast$ is $\ell$;
\item the skew diagram $\La^\circledast/\La^*$ is both a horizontal and a vertical $m$-strip.
\end{enumerate}
The two definitions are equivalent: the map $(\La^*,\La^\cd) \mapsto (\LaA;\LaS)$ is bijective and given as follows.  The parts of $\LaA$ are the parts $\La^*_i, i=1,2,\ldots$, such that   $\La^*_i\neq \La^\cd_i$ and the parts of $\LaS$, those that satisfy $\La^*_i=\La^\cd_i$.  
The inverse direction is also easy to describe.  Let $(\LaA;\LaS)^+$ be  the partition obtained  by reordering in non-increasing order the concatenation of the entries of $\LaA$ and $\LaS$, and let $\LaA+1^m$ be the partition obtained by adding one to each entry of $\LaA$.  Then $\La^* = (\LaA;\LaS)^+$ and $\La^\cd=(\LaA+1^m;\LaS)^+$.   For example, consider $\La^*=(8,6,5,3,3,2)$ and $\La^\cd=(9,7,5,4,3,3,1)$:
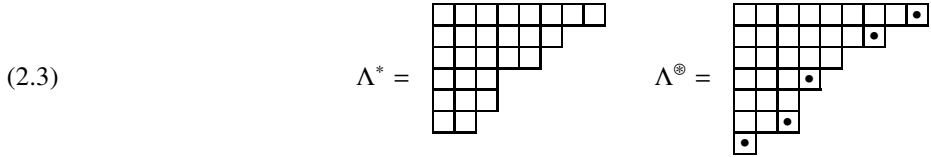
\beq
\La^*=
\scalebox{1.1}{$\superY{
\,&\,&\,&\,&\,&\,&\,& \\
\,&\,&\,&\,&\,& \\
\,&\,&\,&\,&\\
\,&\,&\\
\,&\,&\\
\,&\\
\none \\
}$}
\quad 
\La^\cd=
\scalebox{1.1}{$\superY{
\,&\,&\,&\,&\,&\,&\,&\, & \bullet \\
\,&\,&\,&\,&\,&\, &\bullet \\
\,&\,&\,&\,&\\
\,&\,&\, & \bullet \\
\,&\,&\\
\,&\, & \bullet \\
\bullet \\
}$}
\eeq
where the boxes with a $\bullet$ represent the boxes of the skew diagram $\La^\cd/\La^*$.  This pair of partitions corresponds to the superpartition $\La=(8,6,3,2,0;5,3)$ mentioned earlier.    
As we shall see, the two descriptions of a superpartition are useful when working with superpolynomials.

For two superpartitions $\La$, $\Om$, we say that $\La$ is included in $\Om$, denoted as $\La \subseteq \Om$, if they satisfy $\La^* \subseteq \Om^*$ and $\La^\circledast \subseteq \Om^\cd$.  In other words we have $\La \subseteq \Om$ whenever $\La^*_i\leq \Om^*_i$ and  $\La^\cd_i\leq \Om^\cd_i$ for all $i$.

\subsubsection{Superdiagram} 
The diagrammatic representation of superpartitions is very similar to the usual Young diagrams representing partitions. For a superpartition $\La$, one uses the $(\La^*, \La^\cd)$ description:  we first represent the diagram of $\La^\cd$ and then transform the boxes of the skew diagram $\La^\cd/\La^*$ into circles.  We call the result a superdiagram, or simply a diagram for short.        
 Here is the superdiagram associated to the above example
 \beq \label{ex1Lasuperetoilez}
 \La=(8,6,3,2,0;5,3)  \quad \rightarrow \quad 
\scalebox{1.1}{$\superY{
\,&\,&\,&\,&\,&\,&\,&\,& \yF{} \\
\,&\,&\,&\,&\,&\,&\yF{}\\
\,&\,&\,&\,&\\
\,&\,& \star& \yF{}\\
\,&\,&\\
\,&\,&\yF{}\\
\yF{}
}$}
\eeq
Note that, diagrammatically, $\La^*$ is obtained by removing all circles, and $\La^\circledast$ is obtained by replacing all the circles by boxes. 
Each box or circle in the diagram of $\La$ can be identified by its position $s=(i,j)$, where $i$ denotes the row, numbered from top to bottom, and $j$ denotes the column, numbered from left to right. A row or column ending with a circle is dubbed {\it fermionic}. Other rows and columns are said to be {\it bosonic}.  In the above diagram, the box with a $\star$ has position $s=(4,3)$; it belongs to the third fermionic row and the second fermionic column.

\subsubsection{Conjugate superpartition}  
The conjugate of a superpartition $\La$ is the superpartition that corresponds, 
in the $(\La^*, \La^\cd)$ description, to
\beq
\La' = \bigl{((\La^*)}', {(\La^\cd)}' \bigr).
\eeq  
In the diagrammatic representation,  the conjugate operation amounts to  interchange  the rows and columns of the diagram $\La$.
Using the example above, we have
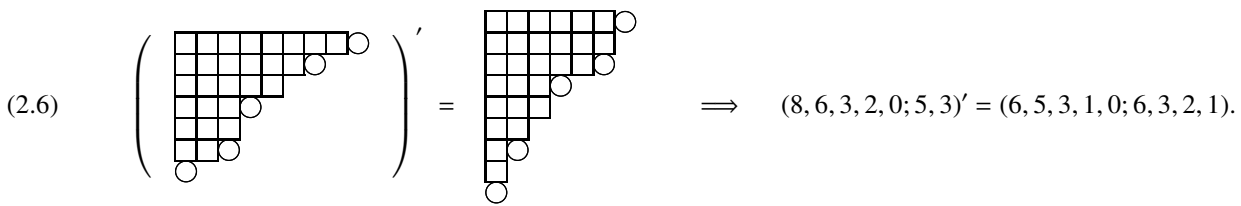
\beq
\left(
\;
\scalebox{1.1}{$\superY{
\,&\,&\,&\,&\,&\,&\,&\,& \yF{} \\
\,&\,&\,&\,&\,&\,&\yF{}\\
\,&\,&\,&\,&\\
\,&\,& \, & \yF{}\\
\,&\,&\\
\,&\,&\yF{}\\
\yF{}
}$}
\right)' \; = \;
\scalebox{1.1}{$\superY{
\,&\,&\,&\,&\,&\,& \yF{} \\
\,&\,&\,&\,&\,&\,\\
\,&\,&\,&\,& \,& \yF{} \\
\,&\,& \,& \yF{}\\
\,&\,&\\
\,&\, \\
\, & \yF{} \\
\, \\
 \yF{}
}$}
\quad \implies \quad (8,6,3,2,0;5,3)'=(6,5,3,1,0 ;6,3,2,1 ).
\eeq

\subsubsection{Dominance order}\label{SecDomOrd111} We now introduce the version of the dominance order that applies to superpartitions. It relies on the $(\La^*,\La^\cd)$ representation.  
Recall the usual dominance order on partitions: 
\begin{equation}
\lambda \geq \mu \quad \iff \quad {|\la|=|\mu|}
\quad \text{and}\quad \lambda_1+ \cdots +\lambda_k \geq
\mu_1+ \cdots +\mu_k\, \quad \text{for }k=1,2,\ldots
\end{equation}
We say that $\Lambda \geq  \Omega$ if and only if 
\begin{equation} \label{eqorder1}
 \quad \La^* \geq \Om^*\quad \text{and}\quad
\La^{\circledast} \geq  \Om^{\circledast} 
\end{equation}
holds true.  For two superpartitions that are comparable, we often write $\La<\Om$ to explicitly indicate that $\La\neq\Om$.  
Pictorially,  $\La<\Om$ if $\Om$ can be obtained from $\La$ successively by moving down a box or a circle.
For example, 
\[
\scalebox{1.1}{$\superY{
\,&\,&\yF{!} \\
\, & ! \\
\none \\
\none
}$}
\geq \quad 
\scalebox{1.1}{$\superY{
\,&\, \\
\, \\
! \\
\yF{!}
}$}
\]
where the boxes with ``!'' are those that were moved down; but
\[
\scalebox{1.1}{$\superY{
\,&\,&\yF{} \\
\, & \, \\
\none
}$}
\not \geq  \quad
\scalebox{1.1}{$\superY{
\,&\, &\, \\
\, \\
\yF{} 
}$}
\qquad \text{and} \qquad  
\scalebox{1.1}{$\superY{
\,&\, &\, \\
\, \\
\yF{} 
}$}
\not \geq  \quad
\scalebox{1.1}{$\superY{
\,&\,&\yF{} \\
\, & \, \\
\none 
}$}.
\]  
The dominance order is of course a partial order (see Figure \ref{Posets41}).
 
 \bigskip

\begin{center}
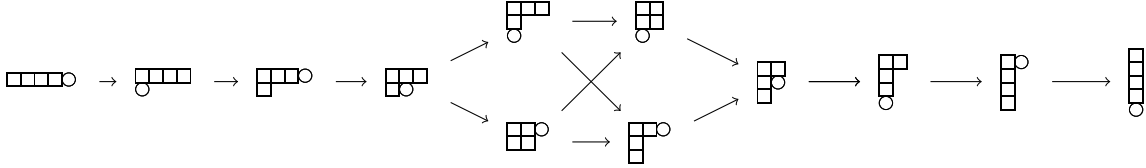
\begin{figure}[ht]
\begin{tikzpicture}[scale=.8]
  \node (4) at (-9,0) {$   \scalebox{0.7}{$\superY{
\,&\,&\,&\,&\yF{} 
}$}
$};
  \node (04) at (-7,0) {$  \scalebox{0.7}{$\superY{
\,&\,&\,& \\
\yF{} 
}$}
  $};
  \node (31) at (-5,0) {$ \scalebox{0.7}{$\superY{
\,&\,&\,&\yF{} \\
 \\ }$}
  $};
  \node (13) at (-3,0) {$  \scalebox{0.7}{$\superY{
\,&\,&\,\\
 \, & \yF{}  }$}
  $};
  \node (22) at (-1,-1) {$      \scalebox{0.7}{$\superY{
\,&\,& \yF{} \\
\, & \\ \none }$}
  $};
  \node  (031) at (-1,1) {$   \scalebox{0.7}{$\superY{
\,&\,&\,\\
  \\  \yF{}  }$}
  $};
  \node (211) at (1,-1) {$    \scalebox{0.7}{$\superY{
\,&\,& \yF{} \\
\,  \\  \\ }$}
  $};
  \node (022) at (1,1) {$    \scalebox{0.7}{$\superY{
\,&\,  \\
\, & \\ \yF{}  \\ }$}
  $};
 \node (121) at (3,0) {$    \scalebox{0.7}{$\superY{
\,&\,  \\
\, & \yF{}  \\  \\}$}
  $};
 \node (0211) at (5,0) {$    \scalebox{0.7}{$\superY{
\,&\,  \\
 \\  \\ \yF{} }$}
  $};
 \node (1111) at (7,0) {$    \scalebox{0.7}{$\superY{
\,&  \yF{} \\
 \\  \\  \\ }$}
  $}; 
   \node (01111) at (9,0) {$    \scalebox{0.7}{$\superY{
 \\
 \\  \\  \\ \yF{} }$}
  $}; 
  \draw [->] (4) -- (04); \draw [->] (04) -- (31);\draw [->] (31) -- (13); \draw [->] (13) -- (22);  \draw [->] (13) -- (031); \draw [->] (22) -- (211);  \draw [->] (031) -- (022); \draw [->] (22) -- (022); \draw [->] (031) -- (211); \draw [->] (211) -- (121); \draw [->] (022) -- (121); \draw [->] (121) -- (0211);   \draw [->] (121) -- (0211);  \draw [->] (0211) -- (1111); \draw [->] (1111) -- (01111);  
\end{tikzpicture}
\caption{{\footnotesize  Hasse diagram representing the dominance ordering of all  superpartitions of degree $(4 | 1)$, which corresponds to the \emph{lowest} degree for which the dominance ordering is no longer total. 
The superpartitions that can be compared are related by arrows pointing
towards the lowest superpartitions.}}
\label{Posets41}
\end{figure}
 \end{center}

\n\textsc{References:}  Superpartitions were first introduced in \cite{DLMnpb}.  However, their diagrammatic representation  only appeared  later in \cite{DLM_class}.  The dominance order presented above  was first introduced in \cite{DLM_eval} (a different order on superpartitions was used before, as for instance in \cite{DLM_fauxsJack}). Note that superpartitions are equivalent to overpartitions \cite{corteel2004overpartitions}.

\bigskip

\subsection{Symmetric superpolynomials}\label{SSP}
Superpolynomials are polynomials in the usual commuting $N$ variables $x=x_1,\ldots ,x_N$  and the $N$ anticommuting variables $\ta=\ta_1,\ldots,\ta_N$.  We denote the ring of superpolynomials in $2N$ variables    over $\mathbb Q$, by $\QQ[x,\ta]$.  
 The diagonal action of the symmetric group $\mathfrak S_N$ acts identically on both sets of variables.  To be more specific, if $\sigma \in \mathfrak S_N$ is a permutation of $\{1,2,\ldots, N\}$, then the action is such that
 \beq
 \sigma \in \mathfrak S_N \; : \; x_i,\ta_i \mapsto x_{\sigma(i)}, \ta_{\sigma(i) }, \qquad i=1,2,\ldots,N. 
 \eeq
 The ring of symmetric superpolynomials $\mathsf A_{N}$ is the subring of  
  $\QQ[x,\ta]$ fixed by the symmetric group, that is,
\beq
\mathsf A_{N} = \QQ[x,\ta]^{\mathfrak S_N}.
\eeq
Let also $\mathsf A_{F,N}= \mathsf A_N \otimes_{\mathbb Z} F$ denote the ring of symmetric superpolynomials over some field $F\neq \QQ$, or over $\ZZ$.  (Note that we will often omit the explicit dependence over the field, and simply write $\mathsf A_{N}$ instead of  $\mathsf A_{F,N}$ even though $F\neq \QQ$).   
The space of symmetric superpolynomials is naturally graded:
\beq
\mathsf A_N = \bigoplus_{n,m \geq 0} \mathsf{A}_N[n|m]
\eeq
where $\mathsf{A}_N[n|m]$ is the space of homogeneous symmetric superpolynomials of degree $n$ in the $x$ variables and degree $m$ in the $\ta$ variables.   
Since $(\ta_i)^k=0$ for any $k\geq2$, the value of $m$ for the degree in the $\ta$ variables implies that each term of the superpolynomial has exactly $m$ distinct  $\ta_i$'s. 
Bases of $\mathsf{A}_N[n|m]$ are labelled by superpartitions of degree $(n|m)$ in at most $N$ parts (i.e.~$\La\in\mathrm{SPar}(n|m)$).

In defining the following bases, we assume that $\La$ is of degree $(n|m)$ with at most $N$ parts.  
The first basis to be introduced is the super-version of the monomial functions:
\begin{equation}
m_\La=m_\La(x,\theta)
=
\frac{1}{\vert\mathrm{Aut}(\LaS) \vert}\sum_{\sigma\in \mathfrak S_N} \sigma.\bigl(\theta_{1}
\cdots\theta_{m}  \, x^\La\bigr)
\end{equation} 
where $x^\La=x_1^{\La_1}\cdots x_N^{\La_N}$,
and 
 the order of the automorphism group that fixes the partition $\LaS$, $\vert\mathrm{Aut}(\LaS) \vert$, is given by
\beq \label{autla1}
\vert\mathrm{Aut}(\la) \vert=
{n_{\la}}(1)!{n_{\la}}(2)!\cdots \qquad \text{if} \qquad
\la = (1^{{n_{\la}}(1)}, 2^{{n_{\la}}(2)}, \ldots).
\eeq

Here is an example, for $N=4$:
\begin{align}
m_{(1,0;1,1)}(x,\theta)=\;&\ta_1\ta_2(x_{1}-x_2)x_3x_4+\ta_1\ta_3(x_{1}-x_3)x_2x_4+\ta_1\ta_4(x_{1}-x_4)x_2x_3\nonumber\\+\,&\ta_2\ta_3(x_{2}-x_3)x_1x_4+\ta_2\ta_4(x_{2}-x_4)x_1x_3+\ta_3\ta_4(x_{3}-x_4)x_1x_2.\nonumber
\end{align}

\bigskip
\n\textsc{References:} The notion of symmetric superpolynomials was first presented in\cite{DLMnpb}, in which the monomial basis was also introduced.
\bigskip

\subsection{Classical multiplicative bases and generating functions}  \label{SecMultbases}
A multiplicative basis of $\mathsf{A}_N[n|m]$ is written,
for a superpartition $\La$, in the form
\beq\label{MultiBaseFF}
f_\La=\tilde{f}_{\La_1} \cdots \tilde{f}_{\La_m} f_{\La_{m+1}} \cdots f_{\La_\ell}
\eeq
where $\tilde{f}_r$ denotes an odd (or fermionic) component and $f_s$ an even (or bosonic) component.  We stress that the ordering of the first $m$ terms is important.  We shall now define three multiplicative bases in superspace: the super-power-sums $p_\La$, the elementary symmetric superfunctions $e_\La$, and the homogeneous superfunctions $h_\La$. 
\begin{enumerate}
\item The super-power-sums:
\begin{equation}\label{Spower}
 \tilde{p}_r=\sum_i\theta_ix_i^r\qquad\text{and}\qquad p_s=
\sum_ix_i^s 
\end{equation}  

\item The elementary superpolynomials:
\beq\label{Selem}
\tilde{e}_r=m_{(0;1^r)}\qquad\text{and}\qquad e_s=
m_{(1^s)}
\eeq

\item The homogeneous  symmetric superpolynomials:
\beq\label{Shomo}
\tilde{h}_r=\sum_{\La\vdash(r|1)}(\La_1+1)m_{\La}\qquad\text{and}\qquad h_s=
\sum_{\la\vdash(s|0)}{m_{\la}}
\eeq
\end{enumerate}
In the three cases, we have $r\geq 0$ and $s\geq 1$.
Here is an example of the 
super-power-sums basis (for $m=2$ and $N=3$) 
\beq \begin{split} \label{p311}
p_{(3,1;1)}&= (\ta_1x_1^3+\ta_2x_2^3+\ta_3x_3^3)(\ta_1x_1+\ta_2x_2+\ta_3x_3)(x_1+x_2+x_3)
\\  &=\bigl(\ta_1\ta_2(x_{1}^3x_2-x_2^3x_1)+
\ta_1\ta_3(x_{1}^3x_3-x_3^3x_1)+\ta_2\ta_3(x^3_{2}x_3-x_3^3x_2) \bigr)(x_1 +x_2+x_3). 
\end{split}
\eeq

In general, for an element $f(x,\ta)\in \mathsf A_N$, the number of variables might be irrelevant since the superpolynomials $f(x,\ta)$  can be expanded in one of the above basis.  For $M>N$, we consider the homomorphism $\mathsf A_M \rightarrow \mathsf A_N$ which acts by setting the variables $x_{N+1}=\ta_{N+1}=\ldots= x_{M}=\ta_M=0$.   This homomorphism maps the monomial $m_\La(x_1, \ldots, x_M, \ta_1, \ldots, \ta_M)$ to the monomial $m_\La(x_1, \ldots, x_N, \ta_1, \ldots, \ta_N)$ if $\ell=\ell(\La) \leq N$, and to zero otherwise.  
Thus, provided the number of variables is large enough, the sequence $\mathsf A_N, \mathsf A_{N+1}, \ldots$ is stable with respect to the increase of variables.  We denote the inverse limit (of infinitely many variables), by
\beq
\mathsf A = \varprojlim \mathsf A_N, \qquad  \mathsf A = \bigoplus_{n,m \geq 0} \mathsf{A}[n|m], \qquad \mathsf{A}[n|m] =  \varprojlim \mathsf A_N[n|m].
\eeq
In the inverse limit, the generators of our multiplicative bases are algebraically independent, and we have
\beq
\mathsf A = F[p_1, p_2,\ldots, \tilde p_0, \tilde p_1, \ldots] = F[e_1, e_2, \ldots, \tilde e_0, \tilde e_1, \ldots] =F[h_1, h_2, \ldots, \tilde h_0,\tilde h_1, \ldots]. 
\eeq

Each bosonic and fermionic generators  of the multiplicative bases corresponds to the expansion coefficient of a certain generating series (cf. \cite[Eqs (3.39), (3.4) and (3.16) resp.]{DLM_class}).  Let $t$ be a formal even indeterminate and $\tau$ an odd indeterminate ($\tau^2=0$).  We then have: 
\begin{align}
P(t,\tau)&=\sum_{n\geq 1}t^{n-1} ( p_n +\tau n \tilde{p}_{n-1}) = \frac{\dd}{\dd t}  \, \log \,  \prod_{i\geq 1} [1-t(x_i+ \tau \ta_i)]^{-1}
\\
E(t,\tau)&=\sum_{n\geq 0} t^n(e_n +\tau\tilde{e}_n)
=\prod_{i\geq 1} (1+t x_i + \tau \theta_i)\,
\\ \label{GenserHH1}
 H(t,\tau)&=\sum_{n\geq0}
t^n(h_n +\tau\tilde{h}_n )=\prod_{i\geq 1} (1-t x_i -
\tau \theta_i)^{-1}
\end{align}
 with $e_0=h_0=1$.  The following three identities can be deduced from the above expressions:
  \begin{align}
 \label{lequationdedualite1}H(t,\tau)E(-t,-\tau)&=1\\
  H(t,\tau)P(t,\tau)&=(t\d_t+\tau\d_\tau)H(t,\tau) \label{idFgen2}\\
    E(t,\tau)P(-t,-\tau)&=-(t\d_t+\tau\d_\tau)E(t,\tau) \label{idFgen3}.
 \end{align}
 Expanding in modes, each of these last identities translate into recurrence relations (Lemma  26 in \cite{DLM_class}).  
 From \eqref{lequationdedualite1}, we obtain
 \beq  \sum_{r=0}^n (-1)^r  e_r h_{n-r}=0, \qquad \quad  \sum_{r=0}^s  (-1)^r(e_r \tilde{h}_{s-r}- \tilde{e}_r h_{s-r})=0, \qquad n=1,2, \ldots, \quad s=0,1,\ldots
\eeq
Then, from \eqref{idFgen2}, \eqref{idFgen3}, we find
\begin{align} 
n h_n=&\sum_{r=1}^n p_r h_{n-r}
, \qquad  n e_n=\sum_{r=1}^n (-1)^{r+1}p_r e_{n-r},  \qquad
n=1,2, \ldots \\
 (n+1)\,
\tilde{h}_n=&\sum_{r=0}^n \bigl[ \, p_r\tilde{h}_{n-r}+ (r+1)\tilde{p}_r
h_{n-r} \, \bigr], \qquad  n=0,1, \ldots  
\\
(n+1) \, \tilde{e}_n=&\sum_{r=0}^n (-1)^{r+1}  \bigl[ \,
p_r\tilde{e}_{n-r}- (r+1)\tilde{p}_r e_{n-r} \, \bigr], 
\qquad n=0,1,\ldots
\end{align}
with the convention that $p_0=0$. Note that, alternatively, using only power-sum superfunctions, one can express both generating series $E(t,\tau), H(t,\tau)$ in a vertex operator form:
\begin{align} \label{Evertex}
E(t,\tau) &= \exp \Bigl[- \sum_{n>0} (-t)^n \Bigl( \frac{1}{n} p_n + \frac{\tau}{t}\tilde{p}_{n-1} \Bigr) \Bigr]  \\
H(t,\tau) &=\exp  \Bigl[\, \sum_{n>0} t^n \Bigl( \frac{1}{n} p_n + \frac{\tau}{t}\tilde{p}_{n-1} \Bigr) \Bigr] . \label{Hvertex}
\end{align}

\subsection{Ring homomorphism I} \label{RHOM1}
We consider the involutive endomorphism $\widehat \omega:  \mathsf A \rightarrow  \mathsf A$, defined by
    \begin{equation}\label{definvolution}
    \widehat{\omega}(e_n)=  h_n \quad\mbox{and}\quad
    \widehat{\omega}(\tilde{e}_n)= \tilde{h}_n; \qquad 
    \qquad 
  \widehat{\omega}(h_n)=  e_n \quad\mbox{and}\quad
 \widehat{\omega}(\tilde{h}_n)= \tilde{e}_n.
 \end{equation}
 Acting with $\widehat{\omega}$ on both sides of equation \eqref{Evertex}, and comparing with equation \eqref{Hvertex}, one deduces that the involution can also be defined as 
\begin{equation} \label{omegasurlesppp1}
  \widehat{\omega}(p_n)= (-1)^{n-1}p_n\quad\mbox{and}\quad
     \widehat{\omega}(\tilde{p}_{n-1}) =(-1)^{n-1}\tilde{p}_{n-1}
\end{equation}
and, for a superpartition $\La$, we have 
\begin{equation}\label{involup}
\widehat{\omega}(p_\Lambda)=\omega_\La \, p_\Lambda, \qquad \omega_\La=(-1)^{|\La|-{\ell}(\La^s)}.
\end{equation}

 \bigskip

\n{\textsc{References}:}  The multiplicative classical bases were presented in \cite{DLM_class}, together with their generating functions. The involution $\widehat{\omega}$ was also introduced there.

\bigskip

\subsection{Scalar product} \label{ScalProd111} As in \eqref{autla1}, 
let $n_\la(i)$ denote the number of parts  equal to $i$ in the
partition $\lambda$.  
We equip the ring $\mathsf A$ with the following scalar product \cite[Proposition 31]{DLM_class}
defined in terms of the power-sums%
\footnote{In fact, this scalar product should be referred to as a bilinear form.  For a given value of the sector $m$, there is a global sign which makes it manifestly not positively defined.  However, we will (often) refer to it as a scalar product in the following.}
,
\begin{equation}\label{scalprod1}
\psc{p_\La}{p_\Om }=(-1)^{\binom{m}{2}}z_\La
\delta_{\La\Om}\, , \qquad {\rm with} \qquad
z_\La =z_{\LaS}=\prod_{k\geq
1}\Bigl[\,k^{n_{\LaS}(k)}\,n_{\LaS}(k)!\,\Bigr] 
\end{equation}
where $\delta_{\La\Om}=1$ if $\La=\Om$, and zero otherwise. 
 For an element $f\in \mathsf A$, we define its adjoint $\perp:f\mapsto f^\perp$ by requiring that
 \beq
 \psc{f *}{*} = \psc{*}{f^\perp*}
 \eeq
 for any elements $*$ of  $\mathsf A$.  From \eqref{scalprod1}, it is clear that
 \beq p_n^\perp = n \frac{\partial}{\partial p_n}, \qquad  \quad \tilde{p}_n^\perp = \frac{\partial}{\partial \tilde{p}_n}.
 \eeq
 We see that the adjoint map transforms an element of the ring $\mathsf A$ to, typically, something that does not belong to the ring but rather to an operator acting on $\mathsf A$.  We thus extend to notion of $\perp$ to any $f:\mathsf A \rightarrow \mathsf A$.  Consider for instance the ring homomorphism I.
  An immediate consequence of \eqref{involup} is that $\widehat\omega^\perp=\widehat\omega$.  Hence, this involution is an isometry \cite[Proposition 32]{DLM_class}: for any superpolynomials $f, g\in \mathsf A$, we have 
\beq
\psc{\widehat{\omega}f}{\widehat{\omega}g}=\psc{\widehat\omega^\perp \widehat\omega f}{g} =  \psc{ \widehat\omega^2 f}{g} = \psc{ f}{g} .
\eeq

\subsection{The kernel \& the Cauchy formula}
\label{sectKerNCaucY}

The orthogonality relation \eqref{scalprod1} is directly related to the reproducing kernel, or Cauchy formula.  Let $y=y_1, y_2, \ldots$ be another set of infinitely many even variables, and let $\phi=\phi_1, \phi_2, \ldots$ another set of infinitely many odd variables (i.e.~$\phi_1^2=\ldots=0$).  The partition function, or kernel, $\Pi=\Pi(x,\theta;y,\phi)$ is given by
\begin{equation}
\label{defK} 
\Pi=    \sum_{\La\in\mathrm{SPar}}  (-1)^{\binom{m}{2}}   z_\La^{-1} \, 
p_\La(x,\theta) \, p_\La(y,\phi)  
=
  \prod_{i,j\geq 1}( {1-x_iy_j-\theta_i\phi_j}  )^{-1}
\end{equation}
which is also referred to as the reproducing kernel since \cite[Corollary 35]{DLM_class}
\begin{equation}
\psc{\Pi(x,\theta;y,\phi)}{f(x,\ta)}=f(y,\phi)\,
,\quad\mbox{for all}\quad f\in\mathsf A.
\end{equation} 
One can show, as a fundamental property of the kernel,  that the following  two conditions are equivalent.  Let $\{u_\La\}_\La$, and $\{v_\La\}_\La$ be independent bases over $\mathsf A[n|m]$ for every $n,m$, then the following are equivalent:
\begin{enumerate}
\item[a.] $\psc{u_\La}{v_\Om} =  \delta_{\La \Om}$
\item[b.] $\Pi = \sum_{\La \in\mathrm{SPar} } u_\La(x,\ta) \, v_\La(y,\phi)$.
\end{enumerate}
For instance, we have  
 \cite[Proposition 38]{DLM_class}
\begin{align} \label{PIasmhzzz}
\Pi=\sum_{\La\in\mathrm{SPar}}(-1)^{\binom{m}{2}}
{m_\La(x,\theta)}{h_\La(y,\phi)}
\end{align}
for which, by virtue of the above property, we can deduce the orthogonality relation 
\beq
 \psc{m_\La}{h_\Om}=(-1)^{\binom{m}{2}}\delta_{\La\Om}. 
\eeq
Observe that by setting $y=(t,0,0,\ldots)$ and $\phi=(-\tau, 0,0,\ldots)$, the kernel $\Pi$ reduces to the generating series $H(t,\tau)$. 
 Hence, expanding in modes, we obtain 
\beq\label{lespetitshenpLa}
h_n = \sum_{\La \vdash(n|0)} z_\La^{-1} p_\La, \qquad \tilde{h}_n= \sum_{\La \vdash (n|1)} z_\La^{-1} p_\La.
\eeq
Applying the homomorphism $\widehat{\omega}$ on the previous relations, we then have
\beq
e_n = \sum_{\La \vdash(n|0)} z_\La^{-1} \omega_\La p_\La, \qquad \tilde{e}_n= \sum_{\La \vdash (n|1)} z_\La^{-1} \omega_\La p_\La
\eeq
where $\omega_\La$ is defined in \eqref{involup}.


\subsection{One parameter deformation of the scalar product \& the homogeneous basis}  \label{SecOnePardef}
Let $\alpha$ be a formal parameter and set $F=\mathbb Q(\alpha)$.  
 We define the following $\alpha$-deformation of the scalar product \eqref{scalprod1} as  
\begin{equation} \label{defscalprodcomb}
\psc{ 
{p_\La}}{  {p_\Om } }_\a =(-1)^{\binom{m}{2}}z_\La
(\a)\delta_{\La\Om}\,,\qquad
z_\La(\a)=\a^{{\ell}(\La)}z_{\La^s}\,. 
\end{equation} 
The Cauchy formula associated to this scalar product is the following kernel  \cite[Theorem 3]{DLMadv}:
\beq
\Pi^\a(x,\ta;y,\phi)=
\sum_{\La\in \mathrm{SPar}}     (-1)^{\binom{m}{2}} z_\La(\a)^{-1}  p_\La(x,\theta) \, p_\La(y,\phi) 
=
\prod_{i,j \geq 1}\frac{1}{(1-x_iy_j-\theta_i\phi_j)^{1/\a}}.  
\eeq
This kernel can now be expanded as in \eqref{PIasmhzzz}, resulting in
\beq\label{ExpandPialphamg}
\Pi^\a(x,\theta;y,\phi)=\sum_{\La} (-1)^{\binom{m}{2}}\, {m_\La(x,\theta)}
{g_\La(y,\phi)}
\eeq
where the (new) basis $\{g_\La\}_\La$ over $\mathsf A_{\mathbb Q(\alpha)}$ is an $\alpha$-deformation of the homogeneous basis
\beq
g_\La= \tilde{g}_{\La_1} \cdots \tilde{g}_{\La_m} g_{\La_{m+1}} \cdots g_{\La_\ell}.
\eeq
Their generating series is given by
\begin{equation}\label{generatriceg} 
G(t,\tau;\a)=\sum_{n\geq0} t^n ( g_n + \tau \tilde{g}_n )=  H(t,\tau)^{1/\alpha}= 
\exp \Bigl[ \frac1\alpha  \sum_{n>0} t^n \Bigl( \frac{1}{n} p_n + \frac{\tau}{t}\tilde{p}_{n-1} \Bigr) \Bigr],
\end{equation}
and they can be given explicitly as
 \cite[Corollary 7]{DLMadv}
\beq\label{ggtexpliciti123}
g_n= \sum_{\La\vdash (n\vert 0)} z_\La(\alpha)^{-1} p_\La 
 , \quad \qquad 
 \tilde g_n = \sum_{\La\vdash (n\vert 1)}    z_\La(\alpha)^{-1}  p_\La  
\eeq
(compare with \eqref{lespetitshenpLa}), with $g_0=1$.    Clearly, $z_\La(1)=z_\La$ for any superpartition $\La$ so that $g_\La$ reduces to $h_\La$ when $\alpha=1$.

The fundamental property of the kernel presented in Section \ref{sectKerNCaucY} also holds true in the case of its $\alpha$-deformation.  Hence, the expression \eqref{ExpandPialphamg} is equivalent to the orthogonality 
  \beq
  \psc{m_\La}{g_\Om}_\alpha = (-1)^{\binom{m}{2}} \delta_{\La \Om}.
  \eeq

\subsection{Ring homomorphism II} 
We shall define an $\alpha$-deformation of the endomorphism $\widehat \omega$ in \eqref{definvolution}, now given by
\beq
\widehat\omega_\alpha (g_n) = e_n, \qquad \widehat\omega_\alpha (\tilde g_n) = \tilde e_n.
\eeq
This maps the basis $g_\La$ to $e_\La$, which are both independent basis over $\mathsf A_{\mathbb Q(\alpha)}$.  Note that this endomorphism is not an involution.  
 Applying $\widehat\omega_\alpha$  on relation \eqref{generatriceg} and comparing with the generating series \eqref{Evertex}, we obtain that its action on the power-sum reads as
\begin{equation}
\label{defendobeta}
\widehat{\omega}_\alpha(p_n)=(-1)^{n-1}\,\alpha\,
p_n, 
\qquad
\widehat{\omega}_\alpha(\tilde{p}_{n-1})=(-1)^{n-1}\,\alpha\,\tilde{p}_{n-1}
 \end{equation} 
 (and we have that $\widehat\omega_1=\widehat\omega$).  
 Hence, the inverse is given by
 \beq
 (\widehat\omega_\alpha)^{-1}= \widehat\omega_{\alpha^{-1}}.
\eeq
From equation \eqref{defendobeta} it is clear that, with respect to the scalar product $\psc{\cdot}{\cdot}_\alpha$, the endomorphism $\widehat\omega_\alpha$ is a self-adjoint map, that is 
\beq
\psc{ \widehat\omega_\alpha f }{  g}_\alpha = \psc{  f }{   \widehat\omega_\alpha g}_\alpha, 
\eeq
for all elements $f,g \in \mathsf A_{\mathbb Q(\alpha)}$. Combining the two previous properties, we see that $ \widehat\omega_\alpha$ is not an isometry as it satisfies instead the relation
\beq
\psc{\widehat{\omega}_{\alpha^{\vphantom{-1}}} f}{\widehat{\omega}_{\alpha^{-1}}g}_\alpha=\psc{f}{g}_\alpha, 
\eeq
for any elements $f,g\in \mathsf A_{\mathbb Q(\alpha)}$.


\section{Jack superpolynomials}
\label{SJack}

The Jack superpolynomials, or Jack polynomials in superspace or
super-Jacks for short, form a basis of the ring of symmetric superpolynomials over the field $\mathbb Q(\alpha)$.  They are different from the $g_\La$'s defined above in that they do not form a multiplicative basis, and there is no known explicit expression such as \eqref{ggtexpliciti123}.  In fact, there are three (all non-explicit) equivalent definitions for the super-Jack functions:
\begin{enumerate}
\item[$\bullet$] a combinatorial definition (i.e.~from triangularity and orthogonality);
\item[$\bullet$] as eigenfunctions of certain commuting operators;
\item[$\bullet$] from the (super)symmetrization of the non-symmetric Jack polynomials.
\end{enumerate}
The first definition holds immediately for infinitely many variables while 
the last two definitions require considering first a finite number $N$  of variables.  The second definition  is related to the supersymmetric extension of the CMS model.  This integrable model describes the dynamics of $N$ particles on a circle, where each particle has bosonic (even) and fermionic (odd) degree of freedom.       
Of course, in the limit $N\rightarrow\infty$, the three cases give rise to the same Jack superpolynomial.  In the following sections, we shall give explicitly these  definitions.

\subsection{Combinatorial characterization}\label{sjackcombdef}
Recall the dominance order, $<$, between superpartitions (Section  \ref{SecDomOrd111}).  For any $n,m$, and two bases $\{ u_\La\}_\La$ and $\{ v_\La\}_\La$ of $\mathsf A_F[n|m]$, we say that the basis  $\{u_\La\}_\La$ is triangular in the basis $\{v_\La\}_\La$ if, for every superpartition $\La\vdash(n|m)$, the superpolynomial $u_\La$ belongs to the subspace
\beq\label{deftriangularbasis}
 \text{span}_F\{ v_\Om \; \vert  \; \Om \leq \La \} \subseteq \mathsf A[n|m].
\eeq

In other words, if we denote by $K(u;v)$ the change of basis matrix between the bases $u$ and $v$, i.e.~$u=K(u;v)v$ where the labelling of the columns and rows is compatible with the dominance ordering, then this matrix is triangular.  Note that this definition also implies that the basis $\{v_\La\}_\La$ is triangular in the basis $\{u_\La\}_\La$.    

\medskip

The Jack superpolynomials $\bigl\{P_\La^{(\alpha)}=P_\La^{(\alpha)}(x,\ta)\bigr\}_\Lambda$ are defined from two properties: (1) triangularity in the monomial basis, and (2) orthogonality.  That is,
\beq\label{defCOmbinaJack}\begin{split}
&(1) \quad P_\La^{(\alpha)}= \sum_{\Om \leq \La} c_{\La \Om}(\alpha)m_\Om, \qquad \text{with} \quad c_{\Lambda \Omega}(\alpha) \in \mathbb Q (\alpha) \quad \text{and} \quad c_{\La\La}(\alpha)=1
\\
&(2) \quad \psc{P_\La^{(\alpha)}}{P_\Om^{(\alpha)}}_\alpha = 0 \quad \text{if} \quad \La\neq \Om.
\end{split}
\eeq
This definition implies of course an overdetermined system (for the coefficients $c_{\La \Om}(\alpha)$).  To show that a solution exists, that is, that the super-Jacks are defined for all superpartitions, one can show that the Jack superpolynomials are the common eigenfunctions of commuting operators which are compatible with the definition \eqref{defCOmbinaJack}.  This will be presented in the sections that follow. 

\medskip

Let $b_\La(\alpha)$ be the proportionality constant in $Q_\La^{(\alpha)} = b_\La(\alpha) P_\La^{(\alpha)}$, and such that
\beq\label{PPPvsQQQsJack}
\psc{P_\La^{(\alpha)}}{Q_\Om^{(\alpha)}}_\alpha = (-1)^{\binom{m}{2}}\delta_{\La \Om} \qquad 
\Rightarrow\quad b_\La(\alpha) = (-1)^{\binom{m}{2}} \psc{P_\La^{(\alpha)} }{P_\La^{(\alpha)} }_\alpha^{-1}
\eeq
The constant $b_\La(\alpha)$ is known as the {reciprocal of the} norm-squared of the super-Jack
$P_\La^{(\alpha)}$.  
 A combinatorial formula for the norm-squared will be presented below.  Note that the Cauchy kernel can then be expanded as
\beq
\Pi^\alpha(x,\ta;y,\phi) = \sum_{\La} (-1)^{\binom{m}{2}} P_\La^{(\alpha)}(x,\ta) Q_\La^{(\alpha)}(y,\phi). 
\eeq

\bigskip
\n{\textsc{References:}} The above definition of the super-Jacks is from \cite[Theorem 1]{DLMadv} while their existence is demonstrated in
\cite[Theorem 21]{DLMadv}.  
\medskip

\subsection{Eigenfunction characterization}
\label{charaphys}
In the next sections, we shall work with a finite number $N$  of variables, and thus consider Jack superpolynomials in $\mathsf A_N$.  Let $D$ and $\Delta$ be two (linear) differential  superoperators acting on  $\mathsf A_N$, and defined by
\begin{align} \label{eqD}
 &D= \frac{\alpha}{2}\sum_{i=1}^N   x_i^2\partial_{x_i}^2
+\sum_{1 \leq i\neq j \leq N}\frac{x_ix_j}{x_i-x_j}\Bigl(\partial_{x_i}-\frac{\theta_i-\theta_j}{x_i-x_j}\partial_{\theta_i}\Bigr), \\ \label{eqDelta}
& \Delta= \alpha \sum_{i=1}^N  x_i\theta_i\partial_{x_i}\partial_{\theta_i}+
\sum_{1 \leq i\neq j \leq N}
\frac{x_i\theta_j+x_j\theta_i}{x_i-x_j}\partial_{\theta_i}.
\end{align}

The Jack superpolynomial $P_\La^{(\alpha)}$ can be characterized by the following
two conditions: (1) triangularity in the monomial basis, as above, and (2) common eigenfunction of $D$ and $\Delta$,
\beq \begin{split} \label{Ptriangular}
& (1)\quad P_\La^{(\a)} =m_\La+\sum_{\Om<\La}
c_{\La\Om}(\alpha)\,m_\Om 
\\
& (2)\quad D\,P_\La^{(\alpha)}=
\varepsilon_{\La}(\alpha)\,P_\La^{(\a)} \qquad \text{and}\qquad \Delta\,
P_\La^{(\alpha)}=\tilde{\varepsilon}_{\La}(\alpha)\,P_\La^{(\alpha)}\, 
\end{split}
\eeq
where the eigenvalues are given by
\begin{equation}
\varepsilon_{\La}(\alpha)= \alpha \, \mathsf n({\La'}^*) - \mathsf n(\La^*) \,  \qquad {\rm and} \qquad
\tilde{\varepsilon}_\La(\alpha)=\alpha\, |\LaA|-|{\La'}^{\mathsf a}| .
\end{equation}
with $\mathsf n(\la)$ given by
  $\mathsf n(\la)= \sum_{i=1}^{\ell(\la)} (i-1) \la_i$ for a regular partition $\la$.  
   Note that in this definition, the eigenvalues can be degenerate, i.e.~one can find distinct superpartitions $\Lambda$ and $\Omega$ such that ${\varepsilon}_\La(\alpha)={\varepsilon}_\Omega(\alpha)$ and $\tilde{\varepsilon}_\La(\alpha)=\tilde{\varepsilon}_\Omega(\alpha)$.  But in this case, it can be shown that $\Lambda$ and $\Omega$  cannot be comparable in the dominance ordering.  As such,  
imposing the dominance ordering fixes uniquely the superpolynomial.  

\medskip

\n{\textsc{References:}} The above definition of the Jack superpolynomials first appeared in \cite[Theorem 1]{DLM_sJack}, with a somewhat different expression for the two operators $D$ and $\Delta$. The origin of this characterization is rooted in physics. Recall that the ordinary Jack polynomials are eigenfunctions of the CMS model (see e.g., \cite{LV}). This statement has a supersymmetric extension: the Jack superpolynomials are eigenfunctions of the supersymmetric version of the CMS model.  The operator $D$ above is related to the model's Hamiltonian.  The operator $\Delta$ is similarly related to the simplest nontrivial operator in an extra  tower of conservation laws that vanishes as the anticommuting variables are set equal to zero.
The above form is presented in \cite[Theorem 2.2]{DLM_eval}.
Warning: in the two articles preceding \cite{DLM_sJack}, the `Jack superpolynomials'  that were constructed are not eigenfunctions of both $D$ and $\Delta$ but only of the former. Consequently, they are not orthogonal.

\medskip

\subsection{The symmetrization construction} The third definition from which the Jack superpolynomials can be characterized is from an appropriate symmetrization of the non-symmetric Jack polynomials, with suitable dressing factors composed of anticommuting variables.  In this section, we introduce the Dunkl operators and we present how their projection onto the space $\mathsf A_N$ connects with the commuting operators defining the super-Jack.    

\subsubsection{The non-symmetric Jack polynomials and the Dunkl operators}  Recall that $\QQ(\alpha)[x_1, \ldots, x_N]$ denotes the ring of polynomials in the variables $x_1, \ldots, x_N$, over $\QQ(\alpha)$.  The operator $K_{ij}$, for $1\leq i,j\leq N$, acts on $\QQ(\alpha)[x_1, \ldots, x_N]$ by exchanging the variables $x_i$ and $x_j$.  Note that we (sometime) write $K_i={K_{i,i+1}}$.   The Dunkl-type operator ${\mathcal D}_i$ is given by\footnote{They are also referred to as Cherednik operators,  \cite{ICher}}   
\beq\label{DDDunk} 
{\mathcal D}_i =\alpha x_i \partial_{x_i} +  \sum_{j< i} \frac{x_i}{x_{i}-x_j}(1- {K_{ij}}) +  \sum_{j> i} \frac{x_j}{x_{i}-x_j}(1- {K_{ij}}) +(1-i), \qquad i=1,2,\ldots, N.
\eeq
The set of operators $\{\mathcal D_i\}_{i}$ forms a family of commuting operators that have common eigenfunctions.   

The non-symmetric Jack polynomial $E_\eta^{(\alpha)}=E_\eta^{(\alpha)}(x)$, indexed by a composition $\eta \in \mathbb Z_{\geq 0}^N$, is an element of $\QQ(\alpha)[x_1, \ldots, x_N]$  defined as follows:  
it is the unique polynomial of the form
\beq 
\label{defE}
E_\eta^{(\alpha)}= x^\eta + \sum_{\nu \prec \eta} a_{\eta \nu}(\alpha) x^\nu, \qquad a_{\eta\nu}(\alpha)\in \QQ(\alpha)
\eeq
where $x^\eta=x_1^{\eta_1} \cdots x_N^{\eta_N}$, and which simultaneously diagonalizes all Dunkl operators
\beq
{\mathcal D}_i E_\eta^{(\alpha)} = (\alpha \, \eta_i - \widehat{\eta}_i)   E_\eta^{(\alpha)},  \qquad \quad i=1,2, \ldots, N
\label{eq.VAP_D_Eeta}
\eeq
with
\beq \label{eigennonsym}
\widehat{\eta}_i = \#\lbrace j=1, \ldots, i-1 | \eta_j \geq \eta_i \rbrace +
\#\lbrace j=i+1, \ldots, N | \eta_j > \eta_i \rbrace.
\eeq
The ordering $*\prec*$ in \eqref{defE} is known as the Bruhat order on (weak) compositions.  For a composition $\eta$, let $\eta^+$ be the unique partition obtained from $\eta$ by ordering the entries, and let $\mathsf w_\eta$ be the unique permutation of minimal length such that $\eta= \mathsf w_\eta \eta^+$ (thus $\mathsf w_\eta$ is an element of the symmetric group $\mathfrak S_N$ which permutes the entries of $\eta$).  Then, we have
\beq \label{bruhatoto}
\nu \prec \eta  \; \iff \;   \nu^+ < \eta^+ \, \, \text{ or } \, \,
\nu^+ = \eta^+ \text{ and } \,  \mathsf w_\nu < \mathsf w_\eta
\eeq   
where comparison on elements of $\mathfrak S_N$ is such that  $\mathsf w_\nu < \mathsf w_\eta$ iff $\mathsf w_\nu \subset \mathsf w_\eta $ (that is, if $\mathsf w_\nu$ can be obtained as a subword $\mathsf w_{\eta}$).

\medskip

\subsubsection{Construction of the Jack superpolynomials in terms of $E_\eta^{(\alpha)}$} \label{defsjakcsymm}Given a superpartition $\La$, one forms the composition $\La^{\mathrm R}=((\LaA)^{\rm R},(\LaS)^{\rm R})$ where  $(*)^{\mathrm R}$ indicates that the entries are written in reversed order.  
Recall that an element $\sigma$ of the symmetric group $\mathfrak S_N$ acts on the space $\QQ(\alpha)[x,\ta]$ by permuting the two sets of variables simultaneously.    Let $m$ be the fermionic degree of the superpartition $\La$.  We have the expansion   \cite[Theorem 41]{DLM_sJack}
\beq \label{sJJJvsEEE}
P^{(\a)}_\La(x,\ta)=\frac{(-1)^{\binom{m}{2}} }{  | \mathrm{Aut}(\LaS) |  } \sum_{\sigma \in \mathfrak S_N}\sigma\bigl( \,\ta_1\cdots \ta_m \, E_{\La^{\mathrm R} }^{(\alpha)}(x)\, \bigr).
\eeq

\subsubsection{The complete set of commuting quantities as projection of the Dunkl operators.}  \label{Dunkkkell}The Dunkl operators $\mathcal D_i$ play the role of the commuting quantities for the non-symmetric Jack polynomials.  Since, as we just saw, the Jack superpolynomials are related to the non-symmetric ones, the commuting quantities for the super-Jacks can be constructed from the $\mathcal D_i$'s.  

We shall consider the action of any given operator $\mathcal O$ on the space of symmetric superpolynomials, which we will denote by $\mathcal O (\mathsf A_N)$.  Let $\kappa_{ij}$ be the operator that interchanges the (odd) variables $\ta_i$ and $\ta_j$.  For any element $f\in \mathsf A_N$ and any $i,j=1,\ldots, N$, we have
\beq
K_{ij}\kappa_{ij} f = f,
\eeq    
which implies that $K_{ij}(\mathsf A_N)= \kappa_{ij}(\mathsf A_N)$ (since both operators are involutions). The action of $\kappa_{ij}$ has the following realization when restricted to $\mathsf A_N$ \cite{DLMnpb} 
\beq\label{kappaij}
\kappa_{ij}= 1-(\ta_i-\ta_j)(\partial_{\ta_i}-\partial_{\ta_j}).
\eeq
Thus, the projection $\mathcal O (\mathsf A_N)$ of an operator $\mathcal O$
involving exchange operators  amounts to moving each operator $K_{ij}$ to the right and then replacing it by the expression for $\kappa_{ij}$ in  \eqref{kappaij}.   Define \cite{DLM_sJack}
\beq
\mathcal H_r' = \sum_{i=1}^N \mathcal D_i^r, \qquad \mathcal I_s' = \sum_{\sigma \in \mathfrak S_N} \sigma \bigl( \ta_1 \partial_{\ta_1} \mathcal D_1^s \bigr) 
\eeq
for $r=1, \ldots, N$,  $s=0,\ldots, N-1$, and  set
\beq
\mathcal H_r = \mathcal H_r' (\mathsf A_N), \qquad \quad
\mathcal I_s = \mathcal I_s' (\mathsf A_N).
\eeq
For instance, we have
\beq
\mathcal H_1 = \alpha \sum_{i=1}^N x_i \partial_{x_i} - N(N-1)/2, \qquad \mathcal I_0= \sum_{i=1}^N \ta_i \partial_{\ta_i}.
\eeq
The following result holds \cite[Proposition 12]{DLMadv}: the $2N$ operators $\mathcal H_r, \mathcal I_s$ are mutually commuting 
\beq
[\mathcal H_r, \mathcal H_s]=[\mathcal H_r, \mathcal I_s]= [\mathcal I_r, \mathcal I_s]=0
\eeq
for all $r,s$, and their common eigenfunctions are the Jack superpolynomials $\{P_\La^{(\alpha)}\}_\La$.  In particular, we have the relations \cite{DLM_eval}
\beq \label{DetDeltafromHI}
D=\frac{1}{2\alpha} \bigl(\mathcal H_2 - N (N-1)^2/2 \bigr) - \frac12 \bigl( \mathcal H_1 + N (N^2+1)/3 \bigr)
, \qquad \quad 
\Delta=\mathcal I_1 + \frac12 ( \mathcal I_0^2-\mathcal I_0^{\phantom2})
\eeq
where $D$ and $\Delta$ are given respectively in \eqref{eqD} and \eqref{eqDelta}.

\subsection{Another scalar product}  In the context of the supersymmetric generalization of the
CMS quantum mechanical $N$-body problem, the Jack superpolynomials in a finite number of variables are orthogonal with respect to an analytic scalar product, to be referred to as the physical scalar product.   This is defined as follows.  Let $f,g\in \QQ(\alpha)[x_1, \ldots, x_N,\ta_1, \ldots, \ta_N]$ and set
\beq
f^\dagger g = f(x^{-1}, \partial_\ta) g(x,\ta) \bigl\vert_{\ta_1=\ldots = \ta_N=0} \; \in \QQ(\alpha) [x_1^{\pm}, \ldots, x_N^{\pm}]
\eeq
where the $f^\dagger$ acts on $g$ by replacing each variable $\ta_i$ in $f$ by $\partial_{\ta_i}$, and then setting all the remaining $\ta_j$'s to zero.  The physical scalar product is given by%
\footnote{Equivalently, the scalar product can be written as
\[
\psa{f}{g}_{\a,N}= \oint \frac{ \dd x_1}{x_1} \ldots  \frac{ \dd x_N}{x_N} \int \dd \ta_1 \ldots \dd \ta_N
\prod_{\substack{1\leq k, l\leq N\\k\neq l}}  
\Bigl(1-\frac{x_k}{x_l}\Bigr)^{1/\a }
\bar{f} g
\]
where the `bar' notation is defined as $\bar{x}=1/x$ and 
\[
\overline{\ta_{i_1} \cdots \ta_{i_k} } \ta_{j_1} \cdots \ta_{j_l}= \ta_1 \cdots \ta_N
\]
if $(i_1, \ldots, i_k)=(j_1, \ldots, j_l)$, and zero otherwise.  
}
\beq
\label{physca}
\psa{f}{g}_{\a,N}= \oint \frac{ \dd x_1}{x_1} \ldots  \frac{ \dd x_N}{x_N}
\prod_{\substack{1\leq k, l\leq N\\k\neq l}}  
\Bigl(1-\frac{x_k}{x_l}\Bigr)^{1/\a }
f^\dagger g.
\eeq
The $2N$ operators $\mathcal H_r, \mathcal I_s$ are self-adjoint with respect to this scalar product \cite[Proposition 5]{DLM_sJack}, i.e.
\beq
\psa{\mathcal H_r *}{*}_{\a,N}=  \psa{*}{\mathcal H_r *}_{\a,N} \quad \text{and} \quad 
\psa{\mathcal I_s *}{*}_{\a,N}=  \psa{*}{\mathcal I_s *}_{\a,N}
\eeq
for all $r=1, \ldots, N$ and $s=0,\ldots, N-1$.  We thus have the following alternative characterization for the super-Jacks: 
 \beq \begin{split}
&(1) \quad P^{(\alpha)}_{\Lambda} = m_{\Lambda} +\sum_{\Om < \La} c_{\La \Om
}(\a) m_{\Om}
\\
&(2) \quad  \psa{P^{(\alpha)}_{\La} }{ P^{(\alpha)}_{\Om} }_{\a,N} =0 \quad {\rm if} \quad \La \neq \Om.
\end{split}
\eeq

Note that the two scalar products (physical and combinatorial) are compatible when the number of variables is infinite. To be more specific,
\beq
\lim_{N\rightarrow \infty} \frac{  \psa{ P^{(\alpha)}_{\La} }{  P^{(\alpha)}_{\Om} }_{\a,N}  }{ \psa{1}{1}_{\a,N}   } = \psc{ P^{(\alpha)}_{\La}  }{  P^{(\alpha)}_{\Om} }_\alpha
\eeq

\medskip

\n{\textsc{References:}} The orthogonality condition $(2)$ is demonstrated
 in \cite[Theorem 2]{DLM_sJack}, and the equivalence between the scalar products is presented in \cite[Remark 22]{DLMadv}.
 
\medskip

\subsection{Further properties of Jack superpolynomials}

We now present some properties of super-Jack functions, and list a few open problems (and conjectures).

\subsubsection{Norm}\label{LaNormdesSJ}

The formula for the (combinatorial) norm of a Jack superpolynomial
involves some  basic diagram data.   Given a cell $s=(i,j)$ in $\lambda$, we let
\begin{equation}
a_{\lambda}(s)=\lambda_i-j\, , \qquad   a'_{\lambda}(s)=j-1 \, , \qquad
l_{\lambda}(s)=\lambda_j'-i \, ,   \quad  {\rm and} \qquad l_{\lambda}'(s)=i-1   .
\end{equation}
The quantities $a_{\lambda}(s),a_{\lambda}'(s),l_{\lambda}(s)$ and $l_{\lambda}'(s)$
are respectively called the arm-length, arm-colength, leg-length and
leg-colength. 
We define two $\a$-deformations of the hook length of a square in a superpartition $\La$,  the upper and  lower-hook lengths respectively given by:
\beq
\begin{split}\label{2hook}
& h^{\rm{up}}_\La(s;\alpha) = \alpha  ( a_{\La^*}(s)+1 ) + {l}_{\Lambda^{\circledast}}(s)   \\
& h^{\rm{lo}}_\La(s;\alpha) = \alpha\,{a}_{\Lambda^{\circledast}}(s) +l_{\La^*}(s)+1.
\end{split}
\eeq 
For example, for the superpartition $\La=(8,6,3,2,0;5,3)$ presented in \eqref{ex1Lasuperetoilez}, the box marked with the $\star$ symbol at position $s=(4,3)$ has data:
\[
h^{\rm{up}}_\La(s;\alpha) = h^{\rm{lo}}_\La(s;\alpha) = \alpha+2.
\]

Let $\B\La$ (the bosonic content of $\La$) be the set of squares in the diagram of  $\La$ that do not lie
at the {intersection of} a row containing a circle {and} a
column containing a circle. The  expression for the norm of a Jack superpolynomial \eqref{PPPvsQQQsJack} reads \cite[Theorem 6.6]{DLM_eval}:
\beq  \label{norm}
{(-1)^{\binom{m}{2}} \psc{P_\La^{(\alpha)} }{P_\La^{(\alpha)} }_\alpha}= \alpha^m \, \prod_{s\in\B\La} \frac{  h^{\rm{up}}_\La(s;\alpha) }{  h^{\rm{lo}}_\La(s;\alpha) }\quad
  \eeq
  where $m$ is the fermionic degree of the superpartition $\La$.   

\medskip

\subsubsection{Evaluation} \label{superEval1}
An evaluation $\epsilon$  of $\mathsf A_{N}$ is a linear map from the algebra $\mathsf A_{N}$ to $\mathbb Q(\alpha)$.  For an element $f\in \mathsf A_{N}$, we denote this map as $f(\epsilon)$.  
The evaluation we consider in this section is to be referred to as the \emph{constant} evaluation, denoted simply as $\epsilon=1$, which maps all indeterminates $x_1, \ldots, x_N$ to 1.  But because of the anticommuting variables, this operation is (most of the time) trivially equal to zero, unless we consider first the following projection onto a given fermionic sector.  
Let $\varrho_{\{1,\ldots, m\}}$ be the projection operator whose action is to pick up the coefficient of the term $\ta_1\cdots \ta_m$, that is
 \beq\label{firstprj}
 \varrho_{(1,\dots, m)} \, \theta_I=\begin{cases}1 &{\rm if~}I=\{1,\dots,m \} \\ 0 &{\rm if~}I\neq (1,\dots,m )
\end{cases}
\eeq
 where $\ta_I=\ta_{i_1}\cdots\ta_{i_k}$ for $I=\{i_1,\dots,i_k \} \subseteq \{1, \ldots, N   \}$ an ordered subset.  We define the normalized projection operator $\hat{\varrho}_m$ by dividing the result of \eqref{firstprj} by the Vandermonde determinant in the first $m$ variables,
 \beq\label{normamamaprojecto}
 \hat{\varrho}_m = \frac{ \varrho_{(1,\dots, m)} }{   \prod_{1\leq i<j \leq m}(x_i-x_j)}.
\eeq
For an element $f\in \mathsf A_N$ of fermionic degree $m$, we set
\beq
f(1) = f(\epsilon=1) = ( \hat{\varrho}_m f) \bigl\vert_{x_1=x_2=\ldots=x_N=1}
\eeq
for the constant evaluation.  
The reason for dividing by the Vandermonde determinant in the evaluation in \eqref{normamamaprojecto} is that it removes the antisymmetric part, which becomes zero when setting all variables to 1 (whenever $m>1$).  
 
 For the Jack superpolynomials, the constant evaluation is given by an explicit combinatorial formula, similar to that of the normalization presented above.    Let $P_\La^{(\alpha)}$ be such that $\La$ is of fermionic degree $m$ and such that $\ell(\La)<N$.  
  The evaluation formula is most simply described in terms of the
skew diagram $\S\La=\La^\cd/(m,m-1,\dots,1)$.  For instance, $\mathcal{S}(4,3,1;4,3)$ is
\smallskip
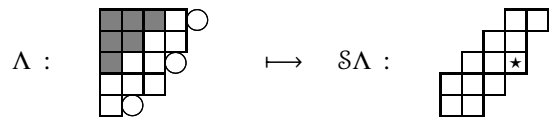
\[
\La\,:\quad 
\scalebox{1.1}{$\superY{
 *(gray) &  *(gray) & *(gray) & \, & \yF{} \\
 *(gray) &  *(gray) &   \,   & \,  \\
*(gray) & \, &  \, & \yF{} \\
     \, & \, & \, \\
\, & \yF{}
  }$} 
\quad \longmapsto \quad 
\S\La \, : \quad 
\ytableausetup{notabloids}
\scalebox{1.1}{$\superY{
 \none &  \none & \none & \, & \, \\
 \none &  \none &   \,   & \,  \\
\none & \, &  \, & \star \\
     \, & \, & \, \\
\, &\, 
  }$}
\]
\smallskip

\noindent
where we note that the position of the $\star$ is $s=(3,4)$.    
 The following formula holds \cite[Theorem 5.7]{DLM_eval} 
\beq 
\label{spe}
P^{(\a)}_{\Lambda}(1)=\frac1{v_\La(\a)}  \prod_{s=(i,j) \in \S\La} \bigl(
N-(i-1)+\alpha(j-1) \big),
\eeq  
where  
\beq
\label{defvla}
v_{\La}(\a) =  \prod_{s\in\B\La} h^{\rm{lo}}_\La(s;\alpha)
\eeq
and with $ h^{\rm{lo}}_\La(s;\alpha)$ being defined in \eqref{2hook}.

\medskip

\subsubsection{Integral form of the Jack superpolynomials}
Consider for example the Jack superpolynomial indexed by $\La=(3,1;\,)$:
\beq \label{unptitsJackzzz} \begin{split}
P_{(3,1;\,)}^{(\alpha)} =& m_{(3,1;\,)}+  \frac{1}{(\alpha+1)} m_{(3, 0; 1)}
+ \frac{-\alpha}{(3 \alpha+2)(\alpha+1)}   m_{(1, 0; 3)}
+
\frac{2}{ (3\alpha+2) (\alpha+1)}    m_{(2, 0; 2)}
+
\frac{(3\alpha+4)}{(3 \alpha+2) (\alpha+1)} m_{(2, 1; 1)}
 \\
&+
\frac{6}{(3 \alpha+2) (\alpha+1) } m_{(2, 0;1, 1)}
+
\frac{2}{(3 \alpha+2) (\alpha+1)^2} m_{(1,0;2,1)}
+
\frac{6}{(3 \alpha+2) (\alpha+1)^2 }m_{(1,0;1,1,1)} .
\end{split}
\eeq
We see that if one multiplies the superpolynomial by $(3\alpha+2)(\alpha+1)^2$, which is the denominator of the coefficient of $m_{(1,0;1^3)}$, then the resulting superpolynomial has coefficients that belong to $\ZZ[\alpha]$.  This normalization is known as the integral form of the Jack superpolynomials.  

\medskip

Among  the superpartitions in $\mathrm{SPar}(n|m)$, there is a minimum one with respect to the dominance ordering.  It is given by 
\beq\label{minide}
\Lambda_{\mathrm{min}} = ( m-1, m-2, \ldots,1,0; \, 1^{\hat n} \,), \qquad \quad \hat n = n-m(m-1)/2.  
\eeq 
 When $P_\La^{(\alpha)}$ with $\La \vdash (n|m)$ is expanded in the monomial basis, the coefficient of $m_{\La_{\mathrm{min}}}$ 
is given by
\beq
\hat{n}! v_\La(\alpha)^{-1}
\eeq
where we recall that $v_\La(\alpha)$ was introduced in \eqref{defvla}  \cite{LLBN, DLM_eval}.

We define the integral form of the  Jack superpolynomials, to be denoted $J_\La^{(\alpha)}$,  as  
\beq
\label{eqnormalJ}
J_\La^{(\alpha)} =  v_\La(\alpha) P_\La^{(\alpha)},
\eeq
When expanded in the monomial basis, i.e.
\beq\label{JJexpandedinvm}
J_\La^{(\alpha)}=\sum_{\Om\leq\La}v_{\La\Om}(\alpha)\,m_\Om,
\eeq
where $ v_{\La\La} (\alpha)=v_\La (\alpha)$, we have the following conjecture.

\begin{conjecture} \cite{DLMadv} The coefficients $v_{\La\Om}(\alpha)$  are polynomials in $\alpha$ with integer coefficients, that is, they belong to $\mathbb Z[\alpha]$.  
\end{conjecture}
The conjecture is illustrated using \eqref{unptitsJackzzz}, in which case $v_{(3,1;\,)}(\alpha)= (3\alpha+2)(\alpha+1)^2$ and $J_{(3,1;\,)}^{(\alpha)}\in \mathsf A_{\mathbb Z[\alpha]}$. 
When $m=0$, the stronger statement that  $v_{\La\Om}(\alpha) \in \mathbb N[\alpha]$ was proven in \cite{Knop}. As can be seen with \eqref{unptitsJackzzz}, this stronger form is not true in general when $m>0$.

 We end this section by mentioning an interesting open problem related to the integral form of the Jack superpolynomials.

\begin{open} Find a combinatorial description, analogous to that of \cite{Knop}, for the expansion coefficients $v_{\La\Om}(\alpha)$ in \eqref{JJexpandedinvm}.  
\end{open}

\medskip

\subsubsection{Duality}   When acting on Jack superpolynomials, the ring homomorphism $\widehat \omega_{\alpha}$ introduced in \eqref{defendobeta} provides the following duality
\beq \label{dual}
\widehat \omega_{\alpha} (P_{\La}^{(\alpha)})
= (-1)^{\binom{m}{2}}
Q_{\La'}^{(1/\alpha)}, \qquad \quad
\widehat\omega_{\alpha} (Q_{\La}^{(\alpha)}) = (-1)^{\binom{m}{2}} P_{\La'}^{(1/\alpha)},
\eeq
which connects the super-Jack indexed by superpartition $\La$ to that indexed by  the conjugate superpartition $\La'$
(up to taking the reciprocal of $\alpha$).

\medskip

\n{\textsc{References:}} The duality relation is proved in \cite[Theorem 27]{DLMadv}.  Note that the relations \eqref{dual} differ slightly from those found in \cite{DLMadv}, since we used in \eqref{dual} the fact that $b_\La(\alpha)= b_{\La'}(1/\alpha)^{-1}$.

\medskip

\subsubsection{Special cases.}
We have \cite[Proposition 28]{DLMadv}
\beq P^{(\a)}_{(;n)}=\frac{\a^n n!}{(1+\a(n-1))_n}g_n\qquad\text{and}\qquad  P^{(\a)}_{(n;)}=\frac{\a^{n+1} n!}{(1+\a n)_{n+1}}\tilde g_n,\eeq
where $(a)_n=a(a-1)\cdots (a-n+1)$.  
When considering special values of $\a$, we connect with the classical bases in the following way \cite[Theorem 30]{DLMadv}
\beq  \lim_{\a\to \infty}P^{(\a)}_{\La}=m_\La\qquad\text{and}\qquad P^{(0)}_\La=(-1)^{\binom{m}{2}}e_{\La'},\eeq
and
\beq P^{(1)}_{(;n)}=h_n\qquad\text{and}\qquad  P^{(1)}_{(n;)}=\frac{1}{n+1}\tilde h_n.\eeq

The Jack superpolynomials at $\a=1$ do not have the nice combinatorial properties that one would expect of Schur superpolynomials (they have for instance negative coefficients when expanded in the monomials, as can be seen in \eqref{unptitsJackzzz} when setting $\alpha=1$).   
Suitable candidates for the Schur superpolynomials will be introduced later.

\medskip

\subsubsection{Pieri rules} \label{PieriJack}
Let the Pieri coefficients for the Jack polynomials in superspace be defined by
\begin{equation} \label{pierifirst}
e_n \, P_\Lambda^{(\alpha)} = \sum_\Omega d_{\Lambda \Omega}(\alpha) \, P_\Omega^{(\alpha)} 
\qquad {\rm and} \qquad \tilde e_n \, P_\Lambda^{(\alpha)} = \sum_\Omega \tilde d_{\Lambda \Omega}(\alpha) \, P_\Omega^{(\alpha)} 
\end{equation}
The Pieri coefficients
$d_{\Lambda \Omega}(\alpha)$  and $\tilde d_{\Lambda \Omega}(\alpha)$ happen to be much more complicated than in the usual Jack polynomials case due to the presence in some cases of a non-linear factor.  For instance, given the superpartitions
\begin{equation} \label{spartsex} 
{\Lambda = \tiny{{{\tableau*[scY]{  & & &&&&\bl  \bigcirc    \cr & &  &&
 \cr  &&&&\bl  \bigcirc  \cr  &&&\bl\bigcirc\cr&\cr \cr }}}}
\qquad {\rm and }\qquad
\Omega = \tiny{{{\tableau*[scY]{  & & &&&&   \cr & &  && & \bl  \bigcirc 
 \cr  &&&& \cr  &&&  \cr&&\bl  \bigcirc \cr \cr\bl  \bigcirc  }}}}  }
\end{equation}
the coefficient
of $P_\Omega^{(\alpha)}$ in the product
$e_3 P_\Lambda^{(\alpha)}$ is given by 
\begin{equation} \label{ugly}
  {\frac{1}{1152}\frac{\alpha^4(2\alpha+3)(3\alpha+4){(416\alpha^6+2000\alpha^5+3484\alpha^4+2608\alpha^3+559\alpha^2-256\alpha-108)}}
    {(4\alpha+3)(5\alpha+4)(7\alpha+6)(2\alpha+1)(\alpha+1)^{10}}}
\end{equation}
Explicit formulas for the  coefficients,
$d_{\Lambda \Omega}(\alpha)$  and $\tilde d_{\Lambda \Omega}(\alpha)$ are provided in \cite{GJL}. Remarkably, the non-linear factor described previously turns out
to be a determinant related to the partition function of the 6-vertex model in statistical mechanics while the remaining linear factors can be expressed,   
as in the usual Jack polynomial case, in terms of certain hook-lengths in
a Ferrers' diagram.
 
\subsection{Admissible Jack superpolynomials} \label{LesZadmissibleJ} For generic values of $\alpha$, the coefficients in the monomial expansion of a Jack superpolynomial are well-defined, i.e.~they are not singular.  But when $\alpha$ is set to a negative rational number, these coefficients can have poles.  This can be observed directly from the super-Jack shown in \eqref{unptitsJackzzz} when letting $\alpha$ be equal to $\alpha_{1,4}=-2/3$ for instance.  Throughout this section, we set
\beq
\alpha_{k,r} = -\frac{k+1}{r-1}
\eeq
with $k,r\in \mathbb N$, $k\geq 1, r\geq 2$, and such that $(k+1)$ and $(r-1)$ are coprime. Jack superpolynomials that are not singular when setting $\alpha=\alpha_{k,r}$ are known as admissible super-Jacks.    We first introduce the notion of admissibility for superpartitions, and we shall then see that the Jack superpolynomials indexed by admissible superpartitions are indeed admissible.  

\medskip

A superpartition $\La$ with exactly $N$ parts is $(k,r,N)$-admissible if it satifies  \cite{DLM_clust}
\beq \label{eqadmisup}
 \Lambda_i^{\cd}- \La_{i+k}^*\geq r \quad {\text{for all}} \quad 1 \leq i \leq N-k 
\eeq
with $k\geq 1$ and $r\geq 2$ (as above).  For example, the superpartition $(7,4,2,1;7,5,0)$ is $(2,3,7)$-admissible, and the superpartition $(4,1,0;8,3)$ is $(1,2,5)$-admissible.  Note that for superpartitions of fermionic degree $m=0$, the condition \eqref{eqadmisup} reduces to the standard admissibility condition on partition considered in  \cite{Feigin}.   Let us denote by $\pi_{(k,r,N)}$ the set of superpartitions that are $(k,r,N)$-admissible.  

\medskip 

Let $\mathcal F_N^{(k)}$ denote the ideal of symmetric superpolynomials (over $\mathbb C$) in $N$ (even) variables $x=x_1, \ldots, x_N$ and in $N$ (odd) variables $\ta=\ta_1, \ldots, \ta_N$ that satisfy
\beq
f(x,\ta)=0 \qquad \text{whenever} \qquad x_1=x_2= \ldots = x_{k+1}.
\eeq
The interest of considering the space $\mathcal F_N^{(k)}$ is that it can be described in terms of admissible Jack superpolynomials.  Super-Jack polynomials that vanish when a certain number of variables coincide provide examples of what is known as the clustering property.      
We have the following results \cite{DLM_clust}.
\begin{itemize}
\item Let $\La\in \pi_{(k,r,N)}$ be admissible.  The Jack superpolynomial $P_\La^{(\alpha)}(x_1, \ldots, x_N; \ta_1, \ldots, \ta_N)$ is well-defined at $\alpha=\alpha_{k,r}$, that is, its expansion into the monomial basis has no poles at $\alpha=\alpha_{k,r}$.
\end{itemize}

From the regularity of the super-Jack just mentioned above, we define the space $I_N^{(k,r)}$ to be the space of complex Jack superpolynomials associated to admissible superpartitions and parameter $\alpha=\alpha_{k,r}$, in other words, let
\beq
 I_N^{(k,r)} = {\rm span}_{\mathbb C} \bigl\{ 
P^{(\alpha_{k,r})}_{\La} \, \big| \, 
\La~{\rm is}~(k,r,N){\text-}{\rm admissible}
\bigl\}.
\eeq

\smallskip

\begin{itemize}
\item  The space $I_N^{(k,r)}$ is an ideal of $\mathsf A_N$ (over $\mathbb C$).  

\medskip
\item The ideal $I_N^{(k,r)}$ is closed under the action of the differential superoperators
\beq
\begin{split}
L_n &= \sum_{i=1}^N x_i^{-n} \bigl( x_i \partial_{x_i} + \frac{1-n}{2} \ta_i \partial_{\ta_i} \bigr), \qquad n\leq 1, \, n \in \mathbb Z, \\
G_r & = \sum_{i=1}^N x_i^{-r+1/2} \bigl( \partial_{\ta_i} + \ta_i \partial_{x_i} \bigr), \qquad r \leq 1/2, \, r \in \mathbb Z+\tfrac12.
\end{split}
\eeq
Therefore, the ideal $I_N^{(k,r)}$ defines a module over a subalgebra of the centerless Neveu-Schwarz algebra (i.e.~the super-Virasoro algebra) given by the (anti)commutation relations:
\[
[L_n,L_m]=(n-m)L_{n+m}, \qquad [L_n, G_r]=(n/2-r)G_{n+r}, \qquad \{G_r, G_s\} = 2L_{r+s}.
\]

\medskip

\item  We have $I_N^{(k,r)} \subseteq \mathcal F_N^{(k)}$ for any $r$.  

\end{itemize}

\medskip

In addition, we conjecture that the inclusion in the last result is in fact an equality 
when $r=2$.

\begin{conjecture} The Jack superpolynomials $P_\La^{(\alpha_{k,2})}(x_1, \ldots, x_N, \ta_1, \ldots, \ta_N)$, with $\La$ $(k,2,N)$-admissible, form a basis of the space $\mathcal F_N^{(k)}$.  

\end{conjecture}

We present a simple example illustrating this last conjecture.  Consider the following symmetric superpolynomial:
\[
\prod_{1\leq i<j\leq 3} (x_i-x_j-\ta_i\ta_j)^2.
\]
Clearly, this superpolynomial vanishes whenever two variables $x_i$ coincide, that is, it belongs to $\mathcal F_3^{(1)}$.  With $\alpha_{1,2}=-2$, one can compute its expansion in super-Jacks, and one finds
\[
\prod_{1\leq i<j\leq 3} (x_i-x_j-\ta_i\ta_j)^2 = P_{(4,2)}^{(-2)} - 2 \bigl( P_{(3,2;\,)}^{(-2)} + \tfrac53 P_{(3,0;2)}^{(-2)}+ P_{(1,0;4)}^{(-2)} \bigr)
\]
where all superpartitions (and the partition) that appear in the expansion are $(1,2,3)$-admissible.

\bigskip

\begin{open}
Connection with superconformal field theory.  The basis of states for the irreducible Neveu-Schwarz modules in minimal superconformal model $\mathcal M(2,4p)$ can be characterized by admissible superpartitions \cite{melzer1994}.
\end{open}

\subsection{Relation with infinite dimensional algebras}
The Hamiltonian defining the supersymmetric CMS model can be formulated into the language of infinite-dimensional free-field algebras.  It can thus be written as a (super)operator acting on the Fock space.  
In particular, the operators $\mathcal H_2$ and $\mathcal I_1$ presented in Section \ref{Dunkkkell} can be realized in the algebra formed by the tensor product of the oscillator algebra (or Heisenberg) with the Clifford algebra (or free fermion).

Let $\mathfrak H$ denote the Heisenberg algebra, that is, the associative algebra over $\mathbb C$ with generators $\mathfrak H= \oplus_{n\in \mathbb Z} \mathbb C a_n \oplus \mathbb C {\bf 1}$, and with defining commutation relations
\beq
[a_n, a_m] = n \delta_{n,-m}  {\bf 1}.
\eeq
In the following, we shall also need the free-field algebra associated to the fermionic variables.  Let $\mathfrak f$ denote the free-fermion algebra with generators  
$\mathfrak f= \oplus_{r\in \mathbb Z + \frac12} \mathbb C b_r \oplus \mathbb C {\bf 1}$, and with defining (anti)commutation relations
\beq
\{b_r,b_s\} = \delta_{r,-s} {\bf 1}.  
\eeq
Consider the tensor product algebra $\mathsf H=\mathfrak H \otimes \mathfrak f$, and define the generators $\widehat{\mathcal H}_2, \widehat{\mathcal I}_1 \in \mathsf H$ given by  
\beq
\begin{split}
\widehat{\mathcal H}_2 &=\alpha (\alpha -1) \sum_{n\geq 1}  n a_{-n} a_n + \alpha  \widehat{\mathcal H}_1
+ \alpha (\alpha-1) \sum_{n\geq 0 }n^2 b_{-n-\frac12} b_{n+\frac12} + N \alpha \sum_{n\geq 0}n b_{-n-\frac12} b_{n+\frac12}
\\
& \quad + \alpha  \sum_{n,m\geq 1} \bigl[ a_{-n} a_{-m} a_{n+m} + \alpha a_{-n-m} a_n a_m \bigr]  
+ 2 \alpha \sum_{n,m\geq 1}  m \bigl[ a_{-n} b_{-m-\frac12} b_{n+m+\frac12} + \alpha b_{-n-m-\frac12}  b_{m+\frac12} a_n \bigr]
\end{split}
\eeq
and
\beq \begin{split}
\widehat{\mathcal I}_1 &= (\alpha-1) \sum_{n\geq 0}n b_{-n-\frac12} b_{n+\frac12} - \frac12  \bigl( \widehat{\mathcal I}_0\bigr)^2  
+
\sum_{m\geq 0, n\geq 1} \bigl[ \alpha b_{-m-n-\frac12} b_{m+\frac12} a_n  + a_{-n} b_{-m-\frac12} b_{m+n+\frac12} \bigr]
\end{split}
\eeq
with
\beq
\widehat{\mathcal H}_1 = \sum_{n\geq 1} a_{-n} a_n, \qquad \quad
 \widehat{\mathcal I}_0 = \sum_{n\geq 0}b_{-n-\frac12} b_{n+\frac12}.
\eeq
Note that we drop the tensor product notation, i.e.~writing $a_n$ for $a_n \otimes {\bf 1}$ and $b_n$ for ${\bf 1}\otimes b_n$.  

These two operators can be viewed as the analogues of $\mathcal H_2$ and $\mathcal I_1$, when formulated in the limit when the number of variables is infinite.  To make this connection more precise, we consider the algebra  homomorphism
\beq
\Xi_{ \la} \; : \; a_n \mapsto  \frac{n}{\la}   \partial_{p_n}, \qquad 
a_{-n} \mapsto  \la p_n, \qquad 
b_{m+\frac12} \mapsto \partial_{\tilde p_{m}}, \qquad b_{-m-\frac12} \mapsto  \tilde p_{m}
\eeq
where $\la\in \mathbb C$ and for $n=1,2, \ldots$, $m=0,1,\ldots$  
 Observe that  this homomorphism preserves both the algebras $\mathfrak H(\alpha)$ and $\mathfrak f$.  Up to a certain rescaling constant (which we ignore here in this section), the relation with the conserved current is
\beq
\Xi_{1} ( \widehat{\mathcal H}_2 ) =\mathcal H_2, \qquad 
\Xi_{1} ( \widehat{\mathcal I}_1 ) =\mathcal I_1.
\eeq

There is yet another interesting connection between the CMS superoperators and the representation theory of superconformal field theory.

\medskip

The complete Neveu-Schwarz algebra (i.e.~the super-Virasoro algebra) is spanned by the following generators, which are realized in the free-field algebra as
\beq \begin{split}
L_n &=-\frac12 \alpha_0 (n+1)a_n+\frac{1}{2}\sum_{m\in\mathbb{Z}}a_m a_{n-m}+\frac{1}{2}\sum_{k\in\mathbb{Z}+{\frac12}}\bigl(k+\frac{1}{2}\bigr)b_{n-k}b_{k} \quad (n\neq 0),
\\
L_0&=\frac12(a_0^2-\alpha_0 a_0)+
\sum_{\substack{m\in\mathbb{Z}\\ m>0}}a_{-m}a_{m}+\sum_{\substack{k\in\mathbb{Z}+{\frac12}\\k>0}}
k\, b_{-k}b_{k},
\\
G_r&=-\alpha_0 \bigl(r+\frac{1}{2}\bigr)b_r+\sum_{m\in\mathbb{Z}}a_mb_{r-m},
\end{split}
\eeq
with $n\in \mathbb Z, r\in \mathbb Z+\tfrac12$.  The constant $\alpha_0\in \mathbb C$ parametrizes the so-called central charge $c$ that appears in the commutation relations:
\beq
\begin{split}
&[L_m, L_n]=(m-n) L_{m+n} + \frac1{12} c (m^3-m)\delta_{m,-n} 
 \\
&[L_m, G_r]=( \frac12 m-r) G_{m+r}\\
&\{ G_r, G_s \} = 2 L_{r+s} + \frac13 c (r^2-\frac14) \delta_{r,-s}
\end{split}
\eeq
given by
\beq
c= \frac32-3 \alpha_0^2.
\eeq
(Compare with the expressions presented in Section \ref{LesZadmissibleJ}).  We will refer to this algebra simply as the (universal) NS algebra.  
Then, we consider the generators formed by the expressions:
\beq
\mathcal L = \sum_{n\geq 1} a_{-n} L_n , \qquad  \qquad \mathcal G = \sum_{m\geq 0} b_{-m-\frac12} G_{m+\frac12}.
\eeq 
After some simple manipulations, we can obtain for $\mathcal G$,
\beq
\mathcal G = -\alpha_0 \sum_{k\geq 0}k b_{-k-\frac12} b_{k+\frac12} + (a_0-\alpha_0) \widehat{\mathcal I}_0 + \sum_{k\geq 0, m\geq 1} \bigl[ \,  a_{-m} b_{-k-\frac12} b_{k+m+\frac12}  + b_{-k-m-\frac12} b_{k+\frac12} a_m \, \bigr].
\eeq
When acting on a module, or on a vector in Fock space, the mode $a_0$ will act as a constant on a given vector.  From the algebra homomorphism $\Xi_\la$, one can obtain the relation:
\beq\label{XialgHomoGG}
\sqrt{\alpha} \, \Xi_{1/\sqrt{\alpha}} \bigl( \mathcal G\bigr) = \mathcal I_1 + \frac{1}{2} \mathcal I_0^2  - \frac12 \mathcal I_0 + \text{cst.} \mathcal I_0 = \Delta + \text{cst.} \mathcal I_0
\eeq 
where `$\text{cst.}$' stands for some constant and using the definition of the operator $\Delta$ in \eqref{DetDeltafromHI}.  In order to hold, the above relation requires to identify the constant $\alpha_0$ in $\mathcal G$ with
\beq
\alpha_0 = \frac{1}{\sqrt\alpha} - \sqrt\alpha.
\eeq

\medskip

Consider the families of superconformal field theory where the central charge of the NS algebra is parametrized by 
\beq
c= \frac32 - 3 \frac{(\alpha-1)^2}{\alpha}.
\eeq
A given highest-weigh vector $|\chi\rangle$, or singular vector, which is annihilated by all positive modes, of the associated NS algebra, must satisfy $\mathcal G |\chi\rangle=0$.  Using the algebra homomorphism $\Xi_\la$, where $\Xi_\la(|\chi\rangle) = \chi\in \mathsf A_{\mathbb C(\alpha)}$, we have 
\beq
\Xi_{1/\sqrt\alpha}\bigl( \mathcal G  |\chi\rangle) = \bigl( \Delta +  \text{cst.} \mathcal I_0) \chi=0,
\eeq
which implies that $\chi$ must be an eigenfunction of the superoperator $\Delta$.  Since Jack superpolynomials are eigenfunction of this operator, this provides a remarkable connection between the representation of singular vectors in the NS algebra and the theory of symmetric superpolynomials.  However, as we have seen above, the super-Jacks are not solely characterized by $\Delta$ since we may find degenerate eigenvalues.

We now look at the generator $\mathcal L$.  We may  write it as
\beq \begin{split}
\mathcal L =& -\frac12\alpha_0 \sum_{n\geq 1} n a_{-n} a_n  + (a_0-\alpha_0/2) \widehat{\mathcal H}_1 + \sum_{n,m\geq 1} \bigl[ \, a_{-n}a_{-m} a_{n+m} + \tfrac12 a_{-n-m}a_n a_m \, \bigr]
\\
& \; + \frac12 \sum_{m,n\geq 0} (n+1) a_{-m-n-1} b_{m+\frac12} b_{n+\frac12}
+ \frac12 \sum_{m\geq 0, n\geq1} (2m+n+1) a_{-n} b_{-m-\frac12} b_{m+n +\frac12}
\end{split}
\eeq 
and with the algebra homomorphism as in \eqref{XialgHomoGG}, we find 
\beq\label{XiAlpjaSuperLL}
\sqrt{\alpha} \, \Xi_{1/\sqrt{\alpha}} \bigl( \mathcal L\bigr) = D + \mathcal M
\eeq
where $D$ is the super-Jack operator in \eqref{DetDeltafromHI} and $\mathcal M$ is the following operator (which contains higher terms):
\beq\begin{split}
\mathcal M =& \text{cst.} \mathcal H_1 - 2 \alpha \sum_{n,m\geq 1} nm \tilde{p}_{n+m} \partial_{\tilde{p}_m} \partial_{p_n} - (\alpha-1) \sum_{n\geq 0} n^2 \tilde p_n \partial_{\tilde p_n} - N \sum_{n\geq 0} n \tilde p_n \partial_{\tilde p_n} + \sum_{n,m\geq 1}(n+m) p_n p_m \partial_{p_{n+m}}
\\
&\; + \sum_{n,m\geq 0} (n+1) p_{n+m+1} \partial_{\tilde p_m} \partial_{\tilde p_n}
+ \sum_{n\geq 1, m\geq 0} (n+1) p_n \tilde p_m \partial_{\tilde p_{n+m}}.
\end{split}
\eeq
As a result, a \emph{single} super-Jack is typically not a representation of a singular vector in NS algebra.  Rather, because of the expression \eqref{XiAlpjaSuperLL}, it is a linear combination of super-Jacks which describes a singular vector.  In particular, this is precisely what was found in the work \cite{DLM_SCFT}.

\bigskip

\n
\textsc{References:}  The Hamiltonian of the supersymmetric CMS model expressed in infinitely many variables can be found in \cite{DLMadv} (see also \cite{LM}).  
The connection between singular vectors of superconformal field theory and Jack superpolynomials was initiated in \cite{DLM_SCFT}.  A similar situation was discovered for the second sector of the superconformal algebra, namely the Ramond sector (which corresponds to consider integer modes expansion of the fermionic generators), in \cite{ADM}.  The relation was revisited recently \cite{BMRW} where the operator which generates the linear combination of super-Jacks was obtained.  There is another scheme of representation for the singular vector of the NS algebra based on a different bosonization, and homomorphism, related to a certain $q$-deformation of the Virasoro algebra \cite{Yana2015,Bel2013}.  In the latter work, a single Uglov polynomial represents a singular vector.

\bigskip

\section{Macdonald superpolynomials}
\label{SMacs}
The next level of generalization amounts to construct a one parameter deformation of the Jack superpolynomials.  By analogy with standard symmetric function theory, they are the Macdonald superpolynomials, or super-Macdonald for short.  They form a basis of the ring of superpolynomials $\mathsf A$ with underlying field $F= \mathbb Q(q,t)$, the field of rational functions in $q$ and $t$.  
   These two parameters are viewed as formal (real) free parameters.  For specific values of these parameters, the super-Macdonalds reduce to other families of superpolynomials previously introduced.   

Recall that in the case of  Jack superpolynomials we have three different characterizations, which are described at the beginning of Section \ref{SJack}.  We shall see in this section that this also holds for the super-Macdonalds.

\subsection{The combinatorial scalar product}
The first proposed definition uses the triangularity in the monomial basis
and the orthogonality.  The latter requires a deformation of the scalar product \eqref{defscalprodcomb}, which now reads 
\beq\label{ppspqt}
\psc{ p_\La}{ p_\Om }_{q,t} = (-1)^{\binom{m}{2}}     z_\La(q,t) \, \delta_{\La\Om}
\eeq
where
\beq\label{zLaqt}
z_\La(q,t) =  q^{|\La^{\mathsf a}|} \, z_{\LaS}   \prod_{i} \frac{1-q^{\La^{\mathsf s}_i}}{1-t^{\La^{\mathsf s}_i}}.
\eeq   
with the value of $z_{\LaS}= ({\prod_i \La_i^{\mathsf s}})
{|\mathrm{Aut}(\La^{\mathsf s})|}$, defined in \eqref{scalprod1}.  
The Macdonald superpolynomials  $\bigl\{ P_\La(q,t)=P_\La(x,\ta;q,t) \bigr \}_\Lambda$ are the unique family of superpolynomials that satisfy the two following conditions:
\beq\label{defPPPqqqtttt}
\begin{split}
&(1) \quad P_\La(q,t) =  \sum_{\Om\leq\La}c_{\La\Om}(q,t) m_\Om,
\quad c_{\La\La}(q,t)=1, \quad c_{\La\Om}(q,t) \in \mathbb Q(q,t); \\
&(2) \quad  \psc{P_\La(q,t)}{P_\Om(q,t)}_{q,t}=0 \quad {\rm if} \quad \La \neq \Om.
\end{split}
\eeq
As in the super-Jack's case,  the two conditions define an overdetermined system.  The existence of the $P_\La(q,t)$'s thus requires a proof (whose idea is given below).  
For  superpartitions that are ordinary partitions, i.e.~when $m=0$,
the scalar product is the usual Macdonald scalar product, so that 
the Macdonald superpolynomials are then simply the Macdonald polynomials.

Note that we have lost the usual invariance under the map $q \mapsto q^{-1}$ and $t \mapsto t^{-1} $ in the scalar product (up to a global constant equal to the total degree). Indeed, we have
\beq\label{invqtps1}
(q/t)^{|\La|} \psc{ p_\La}{ p_\Om }_{q^{-1},t^{-1}} = (qt)^{-|\La^{\mathsf a}|} \psc{ p_\La}{ p_\Om}_{q,t},
\eeq
and as a consequence, we have that
\beq \label{nosymenqt}
P_\La(x,\ta;q,t)\ne P_\La(x,\ta;q^{-1},t^{-1}).
\eeq
when $m>0$. 
However, this invariance of the usual Macdonald polynomials has an intriguing extension to superspace.  As we will see, it requires a transformation that only affects the even variables $x_i$'s which are paired with the odd variables $\ta_i$'s.  For a Macdonald superpolynomial of fermionic degree $m$, the corrected version of \eqref{nosymenqt} reads
\beq\label{inv_prop_123_stos}
\rho_{(1, \ldots, m)}\bigl( P_\La\bigr) (q x_1, \ldots, qx_m, x_{m+1}, \ldots, x_N;q,t)= q^{|\LaA|}  \rho_{(1, \ldots, m)}\bigl( P_\La\bigr)(x;q^{-1}, t^{-1}).
\eeq
where we recall that $\rho_{(1, \ldots, m)}$ is defined in \eqref{firstprj}.  
Let $\mathsf T_q$ be the ring endomorphism determined by the following action:  for each fermionic sector $I$ of $f\in \mathsf A$, the action $\mathsf T_qf$ transforms the $x_i$ of the corresponding sector $I$ as
\beq\label{TqTq_inversion}
x_i \mapsto q x_i \quad (i\in I), \qquad x_i \mapsto x_i \quad (i\not\in I), \qquad i=1,2 \ldots
\eeq
 For instance, its action on the monomial basis (of fermionic degree $m$) is given by
\beq\label{TqTqsurmLa}\begin{split}
\mathsf T_q m_\La(x,\ta) &= { \sum_{\sigma \in \mathfrak S_n}}' \sigma \bigl[   \,  \ta_1 \cdots \ta_m  (q^{\La_1} x_1^{\La_1} )  \cdots (q^{\La_m} x_m^{\La_m} ) (  x_{m+1}^{\La_{m+1}}  )  \cdots (x_{N}^{\La_{N}} ) \,  \bigr]
\\
& = q^{|\LaA|} m_\La(x,\ta)
\end{split}
\eeq
where the prime over the sum indicates that the sum is restricted to distinct terms.  
Using the operator $\mathsf T_q$,  relation \eqref{inv_prop_123_stos} can be reformulated as
\beq\label{inv_super123}
\mathsf T_q P_\La(x,\ta;q,t) = q^{|\LaA|} P_\La(x,\ta;q^{-1},t^{-1}).
\eeq
We will present a proof of this inversion formula in Section~\ref{sMacdoEigen1}.

\bigskip

\n
\textsc{References:}  The idea of defining super-Macdonald polynomials with conditions of triangularity and orthogonality with respect to a scalar product  was first presented as a conjecture in \cite[Conjecture 34]{DLMadv}. In \cite{BF1}, it was shown that this conjecture was false and
  the present definition was introduced (albeit only conjecturally). The proof that such a definition indeed defines a family of Macdonald superpolynomials 
  was given in   \cite{BF2}.

\bigskip

\subsection{Special cases}   Before discussing the properties of the super-Macdonald polynomials, we summarize their various limits.

\begin{itemize}
\item Set $q=t^\alpha$ with $\alpha\in \mathbb R, \alpha>0$, and consider the limit $t\rightarrow 1$ (so that $q\rightarrow 1$ also).  We see that
\beq
\frac{1-q^{\La^{\mathsf s}_i}}{1-t^{\La^{\mathsf s}_i}} = \frac{1-t^{\alpha \La^{\mathsf s}_i}}{1-t^{\La^{\mathsf s}_i}} \rightarrow \alpha, \qquad \text{as $t\rightarrow 1$}
\eeq
for all $\La_i^{\mathsf s}$.  Therefore, the scalar product \eqref{ppspqt} reduces to that characterizing the Jack superpolynomials (up to a global constant $\alpha^m$) and thus \beq \lim_{t\rightarrow 1} P_\La(t^\alpha, t)=P_\La^{(\alpha)}.\eeq
 
\medskip

\item When $q=0$, we obtain the analogue of the Hall-Littlewood polynomials in superspace: $P_\La(0,t)=P_\La(t)$.  The relation \eqref{nosymenqt} implies the existence of another distinct family of super Hall-Littlewood polynomials, given by  $P_\La(\infty,t^{-1}) = \bar P_\La(t)$.    

\medskip

\item When $t=1$ (and $q$ is arbitrary) we have \beq P_\La(q,1)= m_\La.\eeq

\medskip

\item When $q=1$ (and $t$ is arbitrary) we have \beq P_\La(1,t) = (-1)^{\binom{m}{2}} e_{\La'}.\eeq

\medskip

\item When we set $q=t$, there remains a residual dependence over $t$ in the scalar product \eqref{ppspqt}. This defines a one-parameter family of pseudo-Schur superpolynomials, denoted by $s_\La(t)$.
Three particular values are of special interest.  First, when $q=t=1$ this defines the Schur-Jack symmetric superpolynomials 
\beq 
s_\La^{\mathrm{Jack}}= P_\La(1,1).
\eeq
Note that these superpolynomials correspond to super-Jacks at $\alpha=1$.  
Two much more interesting cases  are those defined from the specializations 
\beq \label{limitSMactoSS}
s_\La=P_\La(0,0)\qquad\text{and}\qquad
 \bar s_\La= P_\La(\infty, \infty)
 \eeq
which are both called Schur superpolynomials.  
For instance, we have
\[\begin{split}
\label{beau}s_{(1,0;3)}&=m_{(1,0;3)}+m_{(2,0;2)}+m_{(1,0;2,1)} +m_{(2,0;1,1)}
+m_{(1,0;1,1,1)},\\
\bar s_{(1,0;3)}&=m_{(1,0;3)}+m_{(1,0;2,1)} 
+m_{(1,0;1,1,1)}.\end{split}
\]
In contrast with the Schur-Jacks (for instance set $\alpha=1$ in \eqref{unptitsJackzzz}), those two versions of the Schurs have positive integral coefficients expansion in the monomials basis, and have combinatorial descriptions \cite{BM1, JL17} (see also below in Section \ref{SSchur}). 
 Observe that when $q=t=0$ or $q=t=\infty$, the scalar product becomes singular.     
Therefore, each family of Schur superpolynomials cannot be defined solely from triangularity and orthogonality.  Instead, as shown below,  the super-Schur functions $s_\La$ are essentially dual to the $\bar s_\La$'s.

\end{itemize}

\medskip

\n
The different limiting cases are represented graphically in the following Figure \ref{FigLimSMac}.

\begin{figure}[ht]
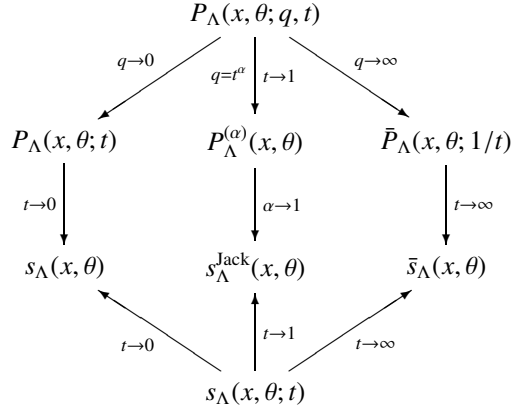

\caption{Limiting cases of the super-Macdonalds}
\label{FigLimSMac}
$$
\begin{diagram}
\node{}\node{P_\La(x,\theta;q,t)} \arrow{sw,t}{q\to 0}\arrow{s,lr}{q=t^\alpha}{t\to 1}\arrow{se,t}{q\to \infty} \node{}\\
\node{P_\La(x,\theta;t)} \arrow{s,l}{t\to 0} \node{P^{(\alpha)}_\La(x,\theta)} \arrow{s,r}{\alpha\to 1} \node{\bar P_\La(x,\theta;1/t)} \arrow{s,r}{t\to \infty} \\
\node{s_\La(x,\theta)}  \node{s^\mathrm{Jack}_\La(x,\theta)}   \node{\bar s_\La(x,\theta)}  \\
\node{} \node{s_\La(x,\theta;t)}\arrow{nw,b}{t\to 0}\arrow{n,r}{t\to 1}\arrow{ne,b}{t\to \infty} \node{}
\end{diagram}
$$
\end{figure}

\bigskip

\subsection{The Cauchy kernel} The scalar product introduced in \eqref{ppspqt} can  be obtained from a generating series (the Cauchy kernel), which naturally expands in the super power-sum basis.  
Let $x=(x_1, x_2, \ldots)$ and $y=(y_1, y_2, \ldots)$ be two independent sets of infinitely many even variables and let $\ta=(\ta_1, \ta_2, \ldots)$ and $\phi=(\phi_1, \phi_2, \ldots)$ be two independent sets of infinitely many odd variables.  
We define the following reproducing kernel  $\Pi=\Pi(x,\ta;y,\phi;q,t)$, 
\beq
\Pi = \prod_{i,j\geq1} \frac{(t x_iy_j;q)_\infty}{(x_iy_j;q)_\infty} \Bigl( 1 + \frac{\ta_i \phi_j}{1-q^{-1}x_iy_j} \Bigr) 
\eeq
where $(*;q)_\infty$ stands for the $q$-Pochamer symbol $(z;q)_r= (1-z)(1-zq)\cdots(1-zq^{r-1})$ with $r \to \infty$.  With standard algebraic manipulations, one obtains the expansion
\beq
\Pi =  \sum_{\La \in \text{SPar}} (-1)^{\binom{m}{2}} z_\La(q,t)^{-1} p_\La(x,\ta) p_\La(y,\phi) 
\eeq  
  where the weight $z_\La(q,t)$ is precisely \eqref{zLaqt}.

\medskip

The expansion of the Cauchy kernel and the orthogonality with respect to the scalar product \eqref{ppspqt} are related. The following property is a direct generalization of the result of Section~\ref{sectKerNCaucY}.  
Let $\{u_\La\}_\La$ and $\{v_\La\}_\La$ be bases of $\mathsf A$ (now over $\mathbb Q(q,t)$). We have that the two statements 
\begin{equation} \label{lemfond1qt}
\begin{split}
\mathrm{(a)}& \quad  \psc{ u_\La}{ v_\Om}_{q,t}= \delta_{\La \Om} \\
\mathrm{(b)}& \quad \sum_\La u_\La(x,\ta) v_\Om(y,\phi) = \Pi(x,\ta;y,\phi;q,t)
\end{split} \end{equation}
are equivalent \cite{BF2}.  For instance, the basis orthonormal  to the monomial basis, 
\beq
\psc{g_\La(q,t)}{m_\Om}_{q,t}= (-1)^{\binom{m}{2}}\delta_{\La\Om}
\eeq
is explicitly obtained by using  (b) of \eqref{lemfond1qt}.  This  results in a multiplicative basis:
\beq\label{basisgqtone}
g_\La(q,t) =  \tilde{g}_{\La_1}(q,t) \cdots \tilde{g}_{\La_m}(q,t) \, g_{\La_{m+1}}(q,t) \cdots   g_{\La_\ell}(q,t)
\eeq
where  $g_n(q,t)$ and $\tilde{g}_n(q,t)$ can be obtained from the generating series 
\beq\label{GenSGqt}
G(y,\phi;q,t)= \sum_{n\geq 0} y^n ( g_n(q,t) + \phi \tilde{g}_n(q,t)) = \exp \sum_{n>0}y^n \Bigl( \frac1n \Bigl( \frac{1-t^n}{1-q^n} \Bigr) p_n + \frac{\phi}{y q^{n-1} } \tilde{p}_{n-1} \Bigr)
\eeq
with $g_0(q,t)=1$.  The explicit expressions of the modes are 
\beq
g_n(q,t) = \sum_{\la \vdash n} \frac{p_\la}{z_\la(q,t)} ; \qquad \quad 
\tilde g_n(q,t) = \sum_{\La \vdash (n|1)} \frac{p_\La}{z_\La(q,t)}.
\eeq
The basis $\{g_\La(q,t)\}_\La$ is a $(q,t)$ deformation of the homogenous basis in superspace.  With $q=t^\a$ and $t\to 1$, we have that $g_\La(q,t)\to g_\La$ defined in Section \ref{SecOnePardef}.

\medskip

\subsection{Ring homomorphism III} \label{tiringhomotrois}  Let  $\widehat \omega_{q,t}$ denote the $\QQ(q,t)$-algebra automorphism of the superpolynomial ring $\mathsf A$ given by the following action on the power-sums:
\beq
\widehat{\omega}_{q,t}( p_r) = (-1)^{r-1} \frac{1-q^r}{1-t^r} p_r, \qquad \quad
\widehat{\omega}_{q,t}( \tilde{p}_{r-1}) = (-1)^{r-1}q^{r-1} \tilde{p}_{r-1} 
\eeq
for each $r\geq 1$.  Hence, for any superpartition $\La$ we have
\beq
\widehat{\omega}_{q,t}(p_\La) = \omega_\La q^{|\LaA|} \prod_i \frac{1-q^{\LaS_i}}{1-t^{\LaS_i}} p_\La
\eeq
where $\omega_\La$ was introduced in \eqref{involup}.  We denote the inverse as $\widehat{\omega}_{q,t}^{-1}$, and remark that when acting on the subring $\mathsf A[n|m]$ we can write
\beq
\widehat{\omega}_{q,t}^{-1} = (t/q)^n \widehat{\omega}_{t^{-1},q^{-1}}
\eeq
(it is enough to verify the relation for $p_\La$ with $\La\vdash (n|m)$).    

From the expression of the generating series \eqref{GenSGqt}, acting with the homomorphism $\widehat{\omega}_{q,t} $, and comparing with equation \eqref{Evertex}, it follows that
\beq
\widehat{\omega}_{q,t}( g_n(q,t)) = e_n, \qquad \quad \widehat{\omega}_{q,t}( \tilde{g}_n(q,t)) = \tilde{e}_n.
\eeq

The automorphism $\widehat{\omega}_{q,t}$ is self-adjoint w.r.t~the $q,t$-scalar product, i.e.~
\beq
\psc{\widehat{\omega}_{q,t}f}{g}_{q,t}= \psc{f}{\widehat{\omega}_{q,t}g}_{q,t}
\eeq
for all superpolynomials $f,g\in \mathsf A$.  We have also that
\beq\label{rhIIIqtto11}
\psc{\widehat{\omega}_{q,t}^{-1}f}{g}_{q,t}  = \psc{\widehat{\omega} f}{g}
\eeq
where $\widehat{\omega}$ is the ring homomorphism I (Section \ref{RHOM1}) and $\psc{\cdot}{\cdot}$ is the scalar product at $q=t=1$ of Section \ref{ScalProd111}.   Again, by linearity, it is enough to verify these properties for $f=p_\La$ and $g=p_\Om$, which is straightforward.

\medskip

\subsection{Macdonald superpolynomials as solutions of an eigenvalue problem}\label{sMacdoEigen1}
As mentioned above, the two conditions (triangularity and orthogonality) define the Macdonald superpolynomials by a Gram-Schmidt process, and for generic $n,m$, this leads to an overdetermined system. To prove the existence of the super-Macdonald basis, one needs to show that they can alternatively be obtained through an eigenvalue problem.  We present this approach in this section.

\medskip

Let $E : \mathsf A \rightarrow \mathsf A$ be a $\QQ(q,t)$-linear (super)operator.   Throughout this section, we denote the adjoint of $E$ with respect to the scalar product $\psc{\cdot}{\cdot}_{q,t}$ by
 $E^\perp$.   
As a consequence of \eqref{lemfond1qt},  the two following properties are equivalent:
\beq
\begin{split}
\mathrm{(a)}& \quad E = E^\perp\\
\mathrm{(b)}& \quad  E^{(x,\ta)} \Pi(x,\ta;y,\phi;q,t) = E^{(y,\phi)} \Pi(x,\ta;y,\phi;q,t),
\end{split}
\eeq
where the notation $E^{(x,\ta)}$ (resp.~$E^{(x,\ta)}$) indicates that the operator $E$ is acting on the sets of variables $x$ and $\ta$ (resp.~$y$ and $\phi$).  

We will construct Macdonald superpolynomials as eigenfunctions of such a superoperator $E$, which we require to have the following three properties: 
\begin{itemize}
\item[i)] its action on the monomial basis is triangular:
$Em_\La = \sum_{\Om \leq \La} v_{\La \Om}m_\Om$ with $ v_{\La\Om}\in F$;

\item[ii)] it is a self-adjoint operator, $E^\perp=E$; 

\item[iii)]   $v_{\La\La}\neq v_{\Om\Om}$ whenever $\La\neq \Om$.  

\end{itemize}

The argument is standard and can be found for instance in \cite{MacSym95}.  Its extension to superspace is straightforward.  
We now present the explicit expression for $E$.  The definition \eqref{defPPPqqqtttt} holds in infinitely many variables, i.e.~for the ring $\mathsf A$.  
We may first work in finitely many variables $(x_1, \ldots, x_N, \ta_1, \ldots, \ta_N)$, namely in the ring $\mathsf A_{N}$. Elements of this ring are indexed by superpartitions with at most $N$ parts.    Since superpartitions are uniquely characterized by two ordinary partitions, say $\La^*$ and  $\La^\circledast$, we expect that the superoperator $E$ to break into two independent parts $E_1$ and $E_2$.  

Let $\tau_i$ be the $q$-shift operator such that $x_i\mapsto qx_i$ and $x_j\mapsto x_j$ if $j\neq i$.  For an ordered subset $I\subseteq \{1, \ldots, N\}$, recall that $\varrho_I$ picks up the coefficient of $\ta_I$ in a superpolynomial, i.e., $\varrho_I\ta_J=\delta_{IJ}$, and set
\beq
\xi_i = \sum_{I \subseteq \{1, \ldots, N\} } \prod_{j\in I, j\neq i} \frac{(qtx_i-x_j)(x_i-x_j)}{(qx_i-x_j) (tx_i-x_j)} \ta_I \rho_I
\eeq
for $i=1,\ldots, N$.  Let also
\beq \label{AAAiiitttt} A_i(t) = \prod_{j\neq i} \frac{(tx_i-x_j)}{(x_i-x_j)},\eeq
and define the following four operators, to be referred to  as supercharges
\beq \label{supercharges}
{Q}_1 = \sum_i \ta_i \tau_i^{-1} , \qquad  {Q}_2 =\sum_i A_i(t^{-1}) \partial_{\ta_i}, \qquad    {Q}_3 = \sum_{i} A_i(t)    \xi_i  \tau_i  \partial_{\ta_i}, \qquad  {Q}_4 = \sum_i \ta_i.
\eeq
From these supercharges, we construct the operators
\beq\label{op_DN1}
D_{N,1}=t^{N-1} \{Q_1,Q_2\}, \qquad \bar{D}_{N,1} = t^{1-N} \{ Q_3,Q_4\} .
\eeq
We can show \cite{BF2, OBFphdth} that both operators have a triangular action on the monomial basis and
 that both have the Macdonald superpolynomials as eigenfunctions:
\beq
D_{N,1}P_\La = d_{N, \La^*}(q,t) P_\La, \qquad \quad \bar{D}_{N,1}P_\La = d_{N,\La^*}(q^{-1},t^{-1}) P_\La
\eeq
 where $d_{N,\la}(q,t) = \sum_{i=1}^N q^{-\la_i}t^{i-1}$ for a regular partition $\la$.    Their eigenvalues thus only depend on $\La^*$.  (Remember that $P_\La(q,t)\neq P_\La(q^{-1},t^{-1})$ so that $D_{N,1}$ and $\bar D_{N,1}$ are not simply related to each other by inverting  $q$ and $t$).  Thus, two superpolynomials $P_\La$ and $P_\Om$ for which $\La^* = \Om^*$ will have the same eigenvalues.  To lift the degeneracy, we introduce the auxiliary operators: 
\beq
D_{N,2}= \frac{(q-1)(t-1)}{qt^{1-N}} \sum_{i,j} \tau_i^{-1} \frac{x_j A_j(t^{-1})}{x_j-tx_i} \ta_i \partial_{\ta_j} + t^{N-1} \{Q_1, Q_2\} 
\eeq
and
\beq\label{eq_DN2_sMacdo}
\bar D_{N,2} = \frac{(q-1)(t-1)}{t^{N-1}} \sum_{i,j}  \frac{x_i A_i(t)}{tx_i-x_j}  \ta_j \partial_{\ta_i} \xi_i \tau_i +t^{1-N} \{Q_3,Q_4\}
\eeq
which also have the superpolynomials $P_\La$ as eigenfunctions.  
The eigenvalues of these two new operators now depend on $\La^\circledast$, 
\beq
D_{N,2} P_\La = d_{N, \La^\circledast}(q,t) P_\La, \qquad \quad \bar{D}_{N,2}P_\La = d_{N,\La^\circledast}(q^{-1},t^{-1}) P_\La.
\eeq
Thus, in principle, the set of superoperators $D_{N,1}, D_{N,2}, \bar D_{N,1}, \bar D_{N,2}$ uniquely fix $P_\La$.   However,  for different values of $N$ they are not compatible with the restriction homomorphism $\mathsf A_{N+1} \rightarrow \mathsf A_{N }$ (and the projective limit of infinitely many variables).  Instead, we consider a modified version of the operators given by
\beq\label{E1E2}
E_{N,1} = \frac{q}{q-1} (D_{N,1} - D_{N,2}), \qquad E_{N,2}= \frac{1}{q-1}(q \bar D_{N,1} - \bar D_{N,2}) - \sum_{i=1}^N t^{1-i}
\eeq
and with a new set of eigenvalues
\beq
E_{N,1} P_\La = e_{N,1,\La} P_\La, \qquad E_{N,2} P_\La = e_{N,2,\La} P_\La.
\eeq
which, from \eqref{E1E2}, are such that $e_{N+1,1,\La} = e_{N,1,\La}=e_{1,\La}$  and $e_{N+1,2,\La} = e_{N,2,\La}=e_{2,\La}$.  We can thus introduce 
\beq E_1=\varprojlim E_{N,1},\qquad\text{and}\qquad  E_2=\varprojlim E_{N,2}\eeq
that are now well-defined since their eigenvalues  do not depend upon $N$.    The Macdonald superpolynomial $P_\La$ is uniquely defined from the eigenvalue problem
\beq E_1 P_\La = e_{1,\La}P_\La, \qquad E_2 P_\La = e_{2,\La}P_\La
\eeq
with
\beq
e_{1,\La} = \sum_{i \, : \, \La_i^\circledast \neq \La_i^*} q^{-\La^*_i} t^{i-1}, \qquad 
e_{2,\La} = \sum_{i \, : \, \La_i^\circledast = \La_i^*} (q^{\La_i^*}-1)t^{1-i}.
\eeq
Since the operators $E_1, E_2$ act triangularly on the monomials basis, are self-adjoint, and have $P_\La$ as eigenfunctions with distinct eigenvalues, that is  $(e_{1,\La}, e_{2,\La})\neq(e_{1,\Om}, e_{2,\Om})$ for $\La\neq \Om$ , this establishes the validity of \eqref{defPPPqqqtttt}.  In other words, this solves the existence issue.

We end this section with a proof of the inversion formula presented in \eqref{inv_super123}, given that this result is new.  Using the notation of this section, we define the action of the operator $\mathsf T_q$  as
\beq
\mathsf T_q = \sum_{I\subseteq \{1, \ldots, N\}} \tau_I \ta_I \rho_I, \qquad  \quad \tau_{\{i_1, i_2, \ldots\}} = \tau_{i_1} \tau_{i_2} \cdots 
\eeq
which does the transformation \eqref{TqTq_inversion} for any given fermionic sector.   
Then, we consider the action of $\mathsf T_q$ on the supercharges defined above.  
Acting on a generic term of fermionic sector $\ta_J$, we have
\beq
\mathsf T_q Q_4 \ta_J = \sum_{j \not\in J} \tau_J \tau_j \ta_j \ta_{J}  = \Bigl( \sum_j \ta_j \tau_j \Bigr) \Bigl( \sum_I \tau_I \ta_I \rho_I \Bigr) \ta_J
\eeq
so that
\beq
\mathsf T_q Q_4 = Q_1^{(1/q,1/t)} \mathsf T_q 
\eeq
where the notation $Q_1^{(1/q,1/t)}$ refers to the supercharge $Q_1$ but with the parameters $q$ and $t$ replaced by $1/q$ and $1/t$ respectively (although in this case $Q_1$ does not depend on $t$).   Likewise, we have 
\beq\begin{split}
\mathsf T_q Q_3 {\ta_J}  &
=  \Bigl( \sum_I \tau_I \ta_I \rho_I \Bigr) \, \sum_{i\in J} \Bigl( \prod_{j\in J/i} \frac{qtx_i-x_j}{qx_i-x_j}\Bigr)
\Bigl( \prod_{{k \not\in J}} \frac{tx_i-x_k}{x_i-x_k}\Bigr) \tau_i \partial_{\ta_i}
\ta_{J}
 \\
 & = \sum_{i\in J} \Bigl( \prod_{j\in J/i} \frac{tx_i-x_j}{x_i-x_j}\Bigr)
 \Bigl( \prod_{{k \not\in J}} \frac{tx_i-x_k}{x_i-x_k}\Bigr) \tau_{J/i} \tau_i \partial_{\ta_i} \ta_{J}
 \\
 & = \Bigl( \sum_i A_i(t) \partial_{\ta_i} \Bigr) \, \Bigl( \sum_I \tau_I \ta_I \rho_I \Bigr) \, \ta_J,
\end{split}
\eeq
so that
\beq
\mathsf T_q Q_3 = Q_2^{(1/q,1/t)} \mathsf T_q.
\eeq
Hence, we can deduce that $\mathsf T_q P_\La(q,t)$ is an eigenfunction of the superoperator $D_{N,1}^{(1/q,1/t)}$: 
\beq
D_{N,1}^{(1/q,1/t)} \bigl( \mathsf T_q P_\La(q,t) \bigr) =t^{1-N} \{Q_1^{(1/q,1/t)}, Q_2^{(1/q,1/t)} \}  \mathsf T_q P_\La(q,t) =  \mathsf T_q \bar{D}_{N,1} P_\La(q,t) = d_{N,\La^*}(q^{-1}, t^{-1}) \bigl(  \mathsf T_q P_\La(q,t) \bigr),
\eeq
 To complete the proof, one needs to show that $\mathsf T_q P_\La(q,t)$ is also an eigenfunction of the superoperator $D_{N,2}^{(1/q,1/t)}$, since each Macdonald superpolynomial is completely characterized by these two operators (up to a constant).   From expression \eqref{eq_DN2_sMacdo}, it is almost immediate to obtain that
\beq
\mathsf T_q \bar{D}_{N,2}= D_{N,2}^{(1/q,1/t)}\mathsf T_q
\eeq
using, as an intermediate step,
\beq\begin{split}
\mathsf T_q   \sum_{i,j}  \frac{x_i A_i(t)}{tx_i-x_j}  \ta_j \partial_{\ta_i} \xi_i \tau_i  {\ta_J}
&=
\Bigl( \sum_I \tau_I \ta_I \rho_I \Bigr) \sum_{i\in J} \sum_{j=i {\rm ~or~} j\not \in J} \frac{x_i}{tx_i-x_j}\Bigl[ \prod_{l\in J, l\neq i} \frac{tx_i-x_l}{x_i-x_l} \prod_{l \not \in J}  \frac{tx_i-x_l}{x_i-x_l} \Bigr]
\\
& \quad \times 
\Bigl( \prod_{k\in J, k\neq i} \frac{(qtx_i-x_k)(x_i-x_k)}{(qx_i-x_k)(tx_i-x_k)} \Bigr) \tau_i \ta_j \partial_{\ta_i} \ta_J
\\
& = 
\sum_{i\in J} \sum_{j=i {\rm ~or~} j\not \in J} \tau_j \tau_i^{-1} \tau_J  \frac{x_i}{tx_i-x_j}
\Bigl[ \prod_{l\in J, l\neq i} \frac{qtx_i-x_l}{qx_i-x_l} \prod_{l \not \in J}  \frac{tx_i-x_l}{x_i-x_l} \Bigr] \tau_i \ta_j \partial_{\ta_i} \ta_J
\\
&=
\sum_{i,j} \tau_j \frac{x_i A_i(t)}{tx_i-x_j} \ta_j \partial_{\ta_i} \mathsf T_q {\ta_J}.
\end{split}
\eeq
Therefore, we have
\beq
D_{N,2}^{(1/q,1/t)} \bigl(\mathsf T_q P_\La(q,t) \bigr) = d_{N,\La^\cd}(q^{-1}, t^{-1}) \bigl(\mathsf T_q P_\La(q,t) \bigr),
\eeq
as wanted. Finally, to complete the proof, we observe that the proportionally constant is fixed by acting with $\mathsf T_q$ on the dominant monomial $m_\La$ in $P_\La(q,t)$, which was obtained in \eqref{TqTqsurmLa}.

\bigskip

\noindent
\textsc{References:}  The material of this subsection is extracted from \cite{BF2} with small improvements. In particular, the charges introduced in \eqref{supercharges} are taken from
\cite{BDM15, OBFphdth} where
the Macdonald superpolynomials are shown to be eigenfunctions of the commuting conservation laws of the integrable supersymmetric version of the  trigonometric Ruijsenaars-Schneider model.

\bigskip

\subsection{Further properties of the $P_\La(q,t)$'s}  \label{propofsmacdo} In this section, we collect some properties satisfied by the superpolynomials $P_\La(q,t)$.

\subsubsection{One-part superpartitions}  
The $P_\La(q,t)$'s are upper triangular when expanded in  the $\{g_\La(q,t)\}_\La$  basis defined earlier (see \eqref{basisgqtone}), i.e.~$P_\La(q,t) = \sum_{\Om \geq \La}u_{\La\Om}\, g_\Om(q,t)$ with $u_{\La\La}=u_{\La\La}(q,t) \neq 1$.  For one-part superpartitions, that is, for superpartitions of the form $\La=(n;\,)$ or $\La=(\, ; n)$, there is thus exactly one term in the decomposition.  It thus only remains to fix the proportionality constants.  Let $\chi_n = (t;q)_n/(q;q)_n$.  We have \cite{OBFphdth}
\beq
P_{(n;\,)}= \bigl( \sum_{i=0}^n q^{i-n} \chi_i\bigr)^{-1} \tilde{g}_n(q,t), \qquad \quad  P_{(\, ; n)}= \chi_n^{-1} g_n(q,t).
\eeq

\subsubsection{The two dualities} As we have seen, the Macdonald superpolynomials form an orthogonal basis over $\mathsf A$ using the scalar product \eqref{ppspqt}.  Let $Q_\La=Q_\La(q,t)$ denote the superfunction determined by 
\beq
\psc{P_\La}{Q_\Om}_{q,t} = (-1)^{\binom{m}{2}}\delta_{\La \Om} \qquad \Leftrightarrow \qquad \sum_\La (-1)^{\binom{m}{2}} P_\La(x,\ta;q,t) Q_\La(y,\phi;q,t) = \Pi(x,\ta;y,\phi;q,t).
\eeq
Clearly, each $Q_\La$ is proportional to $P_\La$. Set $Q_\La = b_\La(q,t) P_\La$ where
\beq
b_\La(q,t) = (-1)^{\binom{m}{2}} \psc{ P_\La}{ P_\La}_{q,t}^{-1}.
\eeq
The  coefficient $b_\La(q,t)$ is the reciprocal of the norm-squared. An explicit expression for the norm-squared is presented in the next section.

From the automorphism $\widehat{\omega}_{q,t}$ on $\mathsf A$, given at Section \ref{tiringhomotrois}, we have the duality: 
\beq\label{dualiqtMacs}
\widehat{\omega}_{q,t} ( P_\La(q,t)) = (-1)^{\binom{m}{2}} (q/t)^{|\La|} Q_{\La'}(t^{-1},q^{-1}), \qquad
\quad
\widehat{\omega}_{q,t} ( Q_\La(q,t)) = (-1)^{\binom{m}{2}} P_{\La'}(t^{-1},q^{-1}).  
\eeq

There is a second duality relation which is defined as follow.  Let $\widehat{\rho}_{q,t}$ be the automorphism determined by
\beq
\widehat{\rho}_{q,t}( p_r) = (-1)^{r-1} \frac{1-q^r}{1-t^r} p_r, \qquad
\quad
\widehat{\rho}_{q,t}( \tilde p_r) = \sum_{\La \vdash (r|1) } z_\La^{-1} \omega_\La  \prod_i (1-q^{\LaS_i}) \, p_\La
\eeq
on the power-sums.  Observe that $\widehat{\rho}_{q,t}$ acts exactly as $\widehat{\omega}_{q,t}$ on the even part of the power-sums, i.e.~on the $p_r$'s, which means that the two automorphisms are equal in the absence of anticommuting variables.  The non-trivial part of $\widehat{\rho}_{q,t}$ is its action on the odd parts of the power-sums, i.e.~on the $\tilde{p}_r$'s, which is expressed as a linear combination of power-sums.  

\begin{conjecture}  From the action of $\widehat{\rho}_{q,t}$, we have the (second) duality relation
\beq
\widehat{\rho}_{q,t} ( P_\La(q,t))  =  t^{|{\La'}^{\mathsf a}|} Q_{\La'}(t,q) .
\eeq
\end{conjecture}

\bigskip

\noindent
\textsc{References:}  The (first) duality relation \eqref{dualiqtMacs} is presented, and proved, in \cite[Theorem 18]{BF2}, while the second duality is conjectured in \cite{OBFphdth}.  

\medskip

\subsubsection{Norm} We now give a combinatorial formula for the norm-squared.   Introduce the $(q,t)$-deformation of the hook length of a box $s$ in a superpartition $\La$,
\beq
h^{\mathrm{up}}_\La(s; q,t) =   1-q^{a_{\La^*}(s)+1} \, t^{ l_{\La^\circledast}(s)}, \qquad
\quad
h^{\mathrm{lo}}_\La(s; q,t)= 1-q^{ a_{\La^\circledast}(s)} \, t^{l(s)+1},
\eeq
(using the notation introduced in Section \ref{LaNormdesSJ}), and set
\beq
h^{\mathrm{up}}_\La(q,t) = \prod_{s\in \mathcal B \La}h^{\mathrm{up}}_\La(s; q,t), \qquad \quad
h^{\mathrm{lo}}_\La(q,t) = \prod_{s\in \mathcal B \La}h^{\mathrm{lo}}_\La(s; q,t) = h^{\mathrm{up}}_{\La'}(t,q).  
\eeq
We have \cite{GL2018}
\beq
\label{conjNSM1}
 (-1)^{\binom{m}{2}} \psc{ P_\La}{ P_\La}_{q,t} =  q^{ |\La^{\mathsf a}|} \, \frac{h^{\mathrm{up}}_\La(q,t)}{h^{\mathrm{lo}}_\La(q,t)}
= 
q^{ |\La^{\mathsf a}|} \prod_{s\in \mathcal B \La} \frac{1-q^{a_{\La^*}(s)+1} \, t^{ l_{\La^\circledast}(s)}  }{1-q^{ a_{\La^\circledast}(s)} \, t^{l(s)+1}}.
\eeq

\medskip

\subsubsection{Specialization} \label{specializationSM} Here we will work in $\mathsf A_{N}$.    We present a generalization of the evaluation map from Section \ref{superEval1}.    Let $\epsilon_{q,t}$ denote the specialization, that is the linear map from the algebra $\mathsf A_N$ to $\mathbb Q(q,t)$, which consists in letting the $N$ variables $x_1, \ldots, x_N$ take the following values:
 \beq\label{specialepsqt1}
\epsilon_{q,t} \, : \,  x_i \mapsto q^{-\delta_i^k}\, t^{i-1}, \qquad i=1, \ldots, N
\eeq
where, for any positive integer $k$ we define the staircase partition
\beq
\delta^k = (k-1, k-2, \ldots, 1, 0, 0,\ldots)
\eeq
with exactly $N$ parts.  For a superpolynomial $f\in \mathsf A_N$ of fermionic degree $m$, we denote its specialization according to the map $\epsilon_{q,t}$ as $f(\epsilon_{q,t})$, and given by
\beq
f(\epsilon_{q,t}) = (\hat\varrho_m f) \bigl\vert_{x_i = q^{-\delta^m_i}  t^{i-1}  }.
\eeq
In other words, the specialization first projects $f$ on the fermionic sector $(1,\ldots, m)$ using the normalized projector $\hat\varrho_m$ (see \eqref{normamamaprojecto}), and then specializes the $x$ variables in the remaining expression according to \eqref{specialepsqt1} with $k=m$.

When acting on a Macdonald superpolynomial, this specialization has a beautiful  combinatorial formula.   We shall need one more combinatorial element.  We have already seen the set of bosonic boxes, $\mathcal B \La$.  Boxes that are not bosonic are called fermionic, and denoted $\mathcal F \La$.  
For each fermionic box $s$ in the diagram of $\La$, let $\zeta_\La(s)$ denote the number of bosonic boxes above $s$.   The sum of all such terms is $\zeta_\La$, i.e.~$\zeta_\La= \sum_{s\in \mathcal F \La} \zeta_\La(s)$.  For example, we have for the superpartition $(4,3,1,0; 6,3)$ 
\[
\scalebox{1.1}{$\superY{
 \, &  \, & \, & \, & \, & \, \\
 1 &  1 &   \,   & 1 & \yF{}   \\
1 & 1 &  \, & \yF{} \\
     \, & \, & \, \\
2 & \yF{} \\
\yF{} 
  }$} 
  \qquad
 \zeta_{(4,3,1,0; 6,3)}= 7.
\]

Let $\La$ be a superpartition of fermionic degree $m$ and such that $\ell(\La)\leq  N$.   We have \cite{GL2018}
\beq\label{ConjSpecSM}
P_\La(\epsilon_{q,t}) =   
\frac{  t^{\zeta_\La+ \mathsf n( \La^\circledast/\delta^{m+1})}   }   
{q^{(m-1) |\Lambda^{\mathsf a}/\delta^{m}| - \mathsf{n}(\Lambda^{\mathsf a}/\delta^{m})}} \,   
 h^{\mathrm{lo}}_\La(q,t)^{-1} 
  \prod_{s=(i,j) \in \mathcal S\La}  (1-q^{j-1}  t^{N-(i-1)} )   ,
\eeq
with  $\mathsf n(\la) = \sum_i (i-1) \la_i$.

\medskip

\subsubsection{Symmetry} We can view the specialization of the previous section more generally, with the linear map $\epsilon_{q,t}$ replaced by a map depending on a given superpartition.  This will result in an important symmetry property for the Macdonald superpolynomials.    
Let $\La$ be an arbitrary superpartition and  fix $\mathsf w_\La \in \mathfrak S_N$ such that $ \La = \mathsf w_\La \La^*$ ($\mathsf w_\La$ acts by permuting the entries of $\La$).  Denote by  $\mathsf w_\La(i)$ the image of the $i$th entry under the permutation $\mathsf w_\La$.   The specialization $\mathsf u_\La$ gives the variables $x_1, \ldots, x_N$ the values
\beq
\mathsf u_\La  \; : \;   x_{\mathsf w_\La(i)} = q^{-\La^*_i} \, t^{i-1}, \qquad i=1,2, \ldots, N,
\eeq
and for an element $f\in \mathsf A_{N}$ of fermionic degree $m$, we set
\beq
f(\mathsf u_\La) =   (\hat{\varrho}_m  f) \bigl\vert_{ x_{\mathsf w_\La(i)} = q^{-\La^*_i} \, t^{i-1}   }.
\eeq
To illustrate the specialization of the variables, we let $\La=(2;4,1)$ so that
\[
\mathsf u_{(2;4,1)} \; : \; (x_1, x_2, x_3, x_4, x_5, \ldots) = (  q^{-2} t, q^{-4},q^{-1} t^2, t^3, t^4, \ldots).
\]
Observe that for $(\delta^m; \,) = (m-1, \ldots, 0; \,)$ we have
\beq
f(\mathsf u_{(\delta^m; \,)})= f(\epsilon_{q,t}),
\eeq 
that is, we recover the specialization of  Section \ref{specializationSM}.

We now define the following normalized version of $P_\La$ as 
\beq
\tilde P_\La = P_\La / P_\La ( \mathsf u_{(\delta^m ; \,)} ), 
\eeq
which is such that $ \tilde{P}_\La(\epsilon_{q,t}) =1$.  With this normalization, the Macdonald polynomials in superspace
  appear to also satisfy a remarkable symmetry property held by the usual Macdonald polynomials.  

\begin{conjecture} (Symmetry).  \cite{OBFphdth}  For any superpartitions $\La,\Om$ of  the same degree,  we have 
\beq
 \tilde{P}_\La({\mathsf u}_\Om ) =  \tilde{P}_\Om({\mathsf u}_\La ).  
\eeq
\end{conjecture}

\bigskip

\subsubsection{Integral form and positivity} We present a generalization of the original Macdonald positivity conjecture.  The integral form associated to  the super-Macdonald $P_\La(q,t)$  
is given by 
\beq
J_\La(q,t) = h_\La^{\mathrm{lo}}(q,t) P_\La(q,t).
\eeq
When expanded in the monomial basis, the coefficients of $J_\La(q,t)$  seem to be polynomials in $\mathbb Z[q,t]$. In fact, they appear to obey a much stronger property that will be formalized in the next conjecture.

Let $\varphi_t$ be the automorphism of $\mathsf A$ determined by 
\beq
\varphi_t (p_r) = (1-t^r)^{-1} p_r, \quad (r\geq 1); \qquad \varphi_t(\tilde{p}_s) = \tilde{p}_s, \quad (s\geq 0),
\eeq
and then define the modified Macdonald superpolynomial $H_\La(q,t)$ as 
\beq\label{ModifSMH}
H_\La(q,t) = \varphi_t(J_\La(q,t))=  h^{\mathrm{lo}}_\La(q,t)   \,  \varphi_t(P_\La(q,t)).
\eeq
The expansion of $H_\La(q,t)$ in the Schur superpolynomial basis $s_\Om$ (see \eqref{limitSMactoSS}) defines the $(q,t)$-Kostka coefficients,
\beq
H_\La(q,t)=\sum_\Omega K_{\Om \La} (q,t)\, s_\Omega
\eeq
denoted by $K_{\Om \La} (q,t)$.  

\begin{conjecture} (Positivity). \cite{BF1} The coefficients $K_{\Om \La}(q,t)$ are, for all superpartitions $\La$ and  $\Om$, polynomials in $q$ and $t$ with nonnegative integer coefficients,
  that is, they belong to $\mathbb N[q,t]$.
\end{conjecture}

\medskip

Here are tables for the coefficients  $K_{\Om \La}(q,t)$ of small degrees.

\begin{table}[ht]
\caption{  $K(q,t)'$ for degree $(1|1)$.  } 
\label{tab11}
\begin{center}
\begin{tabular}{c| c |c | } 
 & $(1;\,) $ & $(0;1)$ \\ \hline
$(1;)$ & $1$ & $q$ \\ \hline
$(0;1\,)$ & $t$ & $1$ \\ \hline
\end{tabular}
\end{center}
\end{table}
\begin{table}[ht]
\caption{ $K(q,t)'$ for  degree $(2|1)$.} 
\label{tab21}
\begin{center}
\begin{tabular}{c| c |c | c | c | } 
 & $(2;\,) $ & $(0;2 )$ & $(1;1 )$ & $(0;1,1 )$ \\ \hline
$(2;\,)$ & $1$ & $q^2$ & $q $ & $q^3 $ \\ \hline
$(0;2)$ & $t$ & $ 1$ & $qt$ & $q$\\ \hline
$(1;1)$ & $t$ & $ qt$ & $1$ & $q$\\ \hline
$(0;1,1)$ & $t^3$ & $ t$ & $t^2$ & $1$\\ \hline
\end{tabular}
\end{center}
\end{table}
\begin{table}[ht]
\caption{$K(q,t)'$ for   degree $(2|2)$.} 
\label{tab22}
\begin{center}
\begin{tabular}{c| c |c | } 
 & $(2,0;\,) $ & $(1,0;1)$ \\ \hline
$(2,0;\,)$ & $1$ & $q$ \\ \hline
$(1,0;1)$ & $t$ & $1$ \\ \hline
\end{tabular}
\end{center}
\end{table}
\begin{table}[ht]
\caption{ $K(q,t)'$ for  degree $(3|1)$.} 
\label{tab31}
\begin{center}
\begin{tabular}{c| c |c | c | c | c| c| c|} 
                  & $(3;\,) $ & $(0;3 )$ & $(2;1 )$ & $(1;2 )$ & $(0;2,1 )$ & $(1;1,1 )$ & $(0;1,1,1 )$ \\ \hline
$(3;\,)$ & $1$ & $q^3$ & $q+q^2 $ & $q^2+q^4 $  & $q^4+q^5$ & $q^3$ & $q^6$  \\ \hline
$(0;3)$ & $t$ & $ 1$ & $qt + q^2t$ & $q+q^2t$  & $q+q^2$  & $q^3t$ & $q^3$ \\ \hline
$(2;1)$ & $t$ & $ q^2t$ & $1+qt$ & $q+q^2t$  & $q^2+q^3t$  & $q$ & $q^3$\\ \hline
$(1;2)$ & $t^2$ & $ qt$ & $t+ qt^2$ & $1+q^2 t^2$ & $q+q^2t$  & $qt$  & $q^2$\\ \hline
$(0;2,1)$ & $t^3$ & $ t$ & $t^2+qt^3$ & $t+qt^2$ & $1+qt$  & $qt^2$  & $q$\\ \hline
$(1;1,1)$ & $t^3$ & $ qt^3$ & $t+t^2$ & $t+qt^2$ & $qt+qt^2$  & $1$  & $q$\\ \hline
$(0;1,1,1)$ & $t^6$ & $ t^3$ & $t^4+t^5$ & $t^2+t^4$ & $t+t^2$  & $t^3$  & $1$\\ \hline
\end{tabular}
\end{center}
\end{table}
\begin{table}[ht]
\caption{ $K (q,t)'$ for  degree $(3|2)$.} 
\label{tab32}
\begin{center}
\begin{tabular}{c| c |c | c | c |c| } 
 & $(3,0;\,) $ & $(2,1;\, )$ & $(2,0;1 )$ & $(1,0;2 )$ & $(1,0;1,1 )$ \\ \hline
$(3,0;\,)$ & $1$ & $q$ & $q+q^2 $ & $q^2 $  & $q^3$ \\ \hline
$(2,1;\,)$ & $qt$ & $ 1$ & $q+q^2t$ & $q^3t$ & $q^2$\\ \hline
$(2,0;1)$ & $t$ & $ t$ & $1+qt$ & $q$ & $q$\\ \hline
$(1,0;2)$ & $t^2$ & $ qt^3$ & $t+qt^2$ & $1$ & $qt$\\ \hline
$(1,0;1,1 )$ & $t^3$ & $ t^2$ & $t+t^2$ & $t$ & $1$\\ \hline
\end{tabular}
\end{center}
\end{table}

\begin{table}[ht]
\caption{ $K (q,t)'$ for  degree $(4|1)$.}
\label{tab41}
\begin{center}
\begin{tabular}{l  | c | c | c | c |c | c |} 
 &$(4;)$  & $(0;4)$ & $(3;1)$ & $(1;3)$  &  $(0;3, 1)$ & $(2;2)$   \\ \hline
$(4;)$  &  $1$ & $q^4 $& $q+q^2+q^3 $& $q^3+q^5+q^6 $& $q^5+q^6+q^7$ & $q^2+q^4$  \\ \hline
$(0;4)$ & $t$ & $1$ & $qt+q^2t+q^3t$ & $q+q^2+q^3t$ & $q+q^2+q^3$ & $q^2t+q^4t$ \\ \hline
$(3;1)$ & $t$ & $q^3t$ & $1+qt+q^2t$ & $q^2+q^3t+q^4t$ & $q^3+q^4t+q^5t$ & $q+q^2t$ \\ \hline
$(1;3)$  & $t^2$ & $qt$ & $t+qt^2+q^2t^2$ & $1+q^2t+q^3t^2$ & $q+q^2t+q^3t$ & $qt+q^2t^2$  \\ \hline
$(0;3, 1)$ & $t^3$ & $t$ & $t^2+qt^3+q^2t^3$ & $t+qt+q^2t^2$ & $1+qt+q^2t$ & $qt^2+q^2t^3$ \\ \hline
$(2;2)$  & $t^2$ & $q^2t^2$ & $t+qt+qt^2$ & $qt+q^2t+q^3t^2$ & $q^2t+q^3t+q^3t^2$ & $1+q^2t^2$ \\ \hline
$(0;2, 2)$ & $t^4$ & $t^2$ & $t^3+qt^3+qt^4$ & $t+qt^2+qt^3$ & $t+qt+qt^2$ & $t^2+q^2t^4$ \\ \hline
$(2;1, 1)$  &$ t^3$ & $q^2t^3$ & $t+t^2+qt^3$ & $qt+q^2t^2+q^2t^3$ & $q^2t+q^2t^2+q^3t^3$ & $t+qt^2$\\ \hline
$(1;2, 1)$  & $t^4$ & $qt^3$ & $t^2+t^3+qt^4$ & $t+qt^2+q^2t^4$ & $qt+qt^2+q^2t^3$ & $t^2+qt^3$ \\ \hline
$(0;2, 1, 1)$ & $t^6$ & $t^3$ & $t^4+t^5+qt^6$ & $t^2+t^3+qt^4$ & $t+t^2+qt^3$ & $t^4+qt^5$ \\ \hline
$(1;1^3)$  & $ t^6$ & $qt^6$ & $t^3+t^4+t^5$ & $t^3+qt^4+qt^5$ & $qt^3+qt^4+qt^5$ & $t^2+t^4$ \\ \hline
$(0;1^4)$ & $t^{10}$ & $t^6$ & $t^7+t^8+t^9$ & $t^4+t^5+t^7$ & $t^3+t^4+t^5$ & $t^6+t^8$ \\ \hline
\end{tabular}

\vspace{1cm}

\begin{tabular}{l  | c | c | c | c |c | c | } 
 & $(0;2, 2)$ &  $(2;1, 1)$  & $(1;2, 1)$  & $(0;2, 1, 1)$  & $(1;1^3)$  &  $(0;1^4)$ \\ \hline
$(4;)$  &  $q^6+q^8$ & $q^3+q^4+q^5$& $q^4+q^5+q^7$ & $q^7+q^8+q^9$ &$ q^6$& $q^{10}$ \\ \hline
$(0;4)$ & $q^2+q^4$ & $q^3t+q^4t+q^5t$ & $q^3+q^4t+q^5t$ & $q^3+q^4+q^5$ & $ q^6t$ &$ q^6$ \\ \hline
$(3;1)$ & $q^4+q^5t$ & $q+q^2+q^3t$ & $q^2+q^3+q^4t$ & $q^4+q^5+q^6t$ & $q^3 $&$ q^6$\\ \hline
$(1;3)$  & $q^2+q^3t$ & $qt+q^2t+q^3t^2$ & $q+q^2t+q^4t^2$ & $q^2+q^3+q^4t$ & $q^3t$ & $q^4$\\ \hline
$(0;3, 1)$ & $q+q^2t$ & $qt^2+q^2t^2+q^3t^3$ & $qt+q^2t^2+q^3t^2$ & $q+q^2+q^3t$ & $q^3t^2$ & $q^3$ \\ \hline
$(2;2)$  & $q^2+q^4t^2$ & $q+qt+q^2t$ & $q+q^2t+q^3t$ & $q^3+q^3t+q^4t$ & $q^2$ & $q^4$ \\ \hline
$(0;2, 2)$ & $1+q^2t^2$ & $qt^2+qt^3+q^2t^3$ & $qt+qt^2+q^2t^3$ & $q+qt+q^2t$ & $q^2t^2$ & $q^2$\\ \hline
$(2;1, 1)$  & $q^2t+q^3t^2$ & $1+qt+qt^2$ & $q+qt+q^2t^2$ & $q^2+q^3t+q^3t^2$ & $q$ & $q^3$ \\ \hline
$(1;2, 1)$  & $qt+q^2t^2$ & $t+qt^2+qt^3$ & $1+qt^2+q^2t^3$ & $q+q^2t+q^2t^2$  & $qt$ & $q^2$ \\ \hline
$(0;2, 1, 1)$ & $t+qt^2$ & $t^3+qt^4+qt^5$ & $t^2+qt^3+qt^4$ & $1+qt+qt^2$ & $qt^3$ & $q$ \\ \hline
$(1;1^3)$  & $qt^2+qt^4$ & $t+t^2+t^3$ & $t+t^2+qt^3$ & $qt+qt^2+qt^3$ & $1$ & $q$ \\ \hline
$(0;1^4)$ & $t^2+t^4$ & $t^5+t^6+t^7$ & $t^3+t^5+t^6$ & $t+t^2+t^3$ & $t^4$ & $1$\\ \hline
\end{tabular}
\vspace{0.5cm}

\end{center}
\end{table}

\vspace{1cm}
       {{
          \begin{table}[ht]
\caption{ $K (q,t)'$ for  degree $(4|2)$.} 
\label{tab42}
\footnotesize
\begin{tabular}{l | c | c | c | c | c | c | c | c | c |} 
  & $(4, 0;)$ & $(3, 1;)$ & $(3, 0; 1)$ & $(1, 0;3)$ & $(2, 0;2)$ & $(2, 1;1)$ & $(2, 0;1, 1)$ & $(1, 0;2, 1)$ & $(1, 0;1^3)$  \\ \hline
$(4, 0;)$ & $1$ &$ q+q^2$ & $q+q^2+q^3$ & $q^3$ & $q^2+q^4$ & $q^3$ & $q^3+q^4+q^5$ & $q^4+q^5$ & $q^6$ \\ \hline
$(3, 1;)$ & $qt$ & $1+q^2t$ & $q+q^2t+q^3t$ & $q^4t$ & $q^2+q^3t$ & $q$ & $q^2+q^3+q^4t$ & $q^3+q^5t$ & $q^4$  \\ \hline
$(3, 0; 1)$ &  $t$ & $t+qt$ & $1+qt+q^2t$ & $q^2$ & $q+q^2t$ & $qt$ & $q+q^2+q^3t$ & $q^2+q^3$ & $q^3$ \\ \hline
$(1, 0;3)$ & $t^2$ & $qt^2+q^2t^3$ & $t+qt^2+q^2t^2$ & $1$ & $qt+q^2t^2$ & $q^3t^3$ & $qt+q^2t+q^3t^2$ & $q+q^2t$ & $q^3t$  \\ \hline
$(2, 0;2)$ &   $t^2$ & $t+qt^2$ & $t+qt+qt^2$ & $qt$ & $1+q^2t^2$ & $qt$ & $q+qt+q^2t$ & $q+q^2t$ & $q^2$ \\ \hline
$(2, 1;1)$ &   $qt^3$ & $t+qt^2$ & $qt+qt^2+q^2t^3$ & $q^3t^3$ & $qt+q^2t^2$ & $1$ & $q+q^2t+q^2t^2$ & $q^2t+q^3t^2$ & $q^2$\\ \hline
$(2, 0;1, 1)$ &   $t^3$ & $t^2+t^3$ & $t+t^2+qt^3$ & $qt$ & $t+qt^2$ & $t^2$ & $1+qt+qt^2$ & $q+qt$ & $q$\\ \hline
$(1, 0;2, 1)$ & $t^4$ & $t^3+qt^5$ & $t^2+t^3+qt^4$ & $t$ & $t^2+qt^3$ & $qt^4$ & $t+qt^2+qt^3$ & $1+qt^2$ & $qt$  \\ \hline
$(1, 0;1^3)$ &  $t^6$ & $t^4+t^5$ & $t^3+t^4+t^5$ & $t^3$ & $t^2+t^4$ & $t^3$ & $t+t^2+t^3$ & $t+t^2$ & $1$\\ \hline
\end{tabular}
\end{table}
}}

\subsubsection{Pieri rules}  Pieri rules for the Macdonald polynomials in superspace that generalize those of the Jack polynomials in superspace (see Section~\ref{PieriJack}) were conjectured in \cite{GJL}.

\subsubsection{Another scalar product}
Macdonald superpolynomials in finitely many variables satisfy another orthogonality relation, arising from a physical scalar product.  This physical, or analytical, scalar product has its origin in the context of the supersymmetric quantum Ruijsenaars-Schneider (RS) integrable model \cite{BDM15}.  

Let $f,g \in \QQ(q,t)[x_1, \ldots, x_N, \ta_1, \ldots, \ta_N]$, and set
\beq
f^\dagger g = f(x_i^{-1}, A_i(t^{-1}) \partial_{\ta_i};q^{-1}, t^{-1})g(x, \ta;q,t) \bigl\vert_{\ta_1 = \ldots = \ta_N=0} \; \in \QQ(q,t)[x_1^\pm, \ldots, x_N^\pm].
\eeq
In this last expression, $f^\dagger$ is defined by first sending each variable $x_i$ to $x_i^{-1}$ and the parameters $q,t$ to $1/q,1/t$ (respectively), then by acting on $g$ by replacing each occurrence of $\ta_i$ in $f$ by $A_i(t^{-1})\partial_{\ta_i}$, where (see \eqref{AAAiiitttt})
\[
A_i(t) = \prod_{j\neq i} \frac{tx_i-x_j}{x_i-x_j},
\]
and finally by setting each remaining variable $\ta_i$ to zero.    Let $\Delta_N(x;q,t)$ denote the  weight function
\beq
\Delta_{N}(x;q,t) =  \prod_{i\neq j} \frac{(x_i/x_j;q)_\infty}{(tx_i/x_j; q)_\infty}.
\eeq
The analytic scalar product is defined by
\beq
\psa{f}{g}_{N,q,t} = \oint \frac{\dd x_1}{x_1} \ldots \frac{\dd x_N}{x_N} \Delta_N(x;q,t) f^\dagger g.
\eeq

Recall from Section \ref{sMacdoEigen1} the construction of the superoperators $D_{N,1}$ and $\bar{D}_{N,1}$, of which the super-Macdonalds are common eigenfunctions.  With respect to the physical scalar product, we have \cite{BDM15}
\beq
\psa{D_{N,1} f }{g}_{N,q,t} = \psa{ f }{ \bar{D}_{N,1} g}_{N,q,t}
\eeq
for any $f,g\in \mathsf A_N$.   In particular, the superoperator 
\beq
\mathfrak D=t^{-(N-1)/2} D_{N,1} + t^{(N-1)/2} \bar{D}_{N,1}
\eeq
is self-adjoint w.r.t.~to the physical scalar product and is related to the Hamiltonian of the super-RS model.

Finally, it is shown in \cite{BF2} that for any superpartitions $\La, \Om$ such that $\La \neq \Om$, we have
\beq
\psa{ P_\La}{P_\Om}_{N,q,t} =0,
\eeq
which gives the second orthogonality relation for the Macdonald superpolynomials.  
\medskip

\subsection{Constructing the Macdonald superpolynomials from the non-symmetric Macdonald polynomials}

We now present the third characterization of the super-Macdonald functions.  In analogy with the super-Jack case, this characterization is obtained using symmetrizers acting this time on the non-symmetric Macdonald polynomials.  Since the non-symmetric Macdonald polynomials are the common eigenfunctions
of the Cherednik operators, we can obtain all the conserved currents associated to the super-Macdonalds.  

\subsubsection{The Cherednik operators}

 The elementary transposition $K_i=K_{i,i+1}$, for $i=1,2, \ldots, N-1$, acts on the space $\mathbb Q(q,t)[x_1, \ldots, x_N]$ by interchanging the variables $x_j$ and $x_{j+1}$ if $j=i$  or as the identity otherwise. 
 The Demazure-Lusztig operator is defined in terms of the $K_i$ as
 \beq
 T_i = t + \frac{tx_i-x_{i+1}}{x_i-x_{i+1}}(K_i-1).
 \eeq
The set of operators $(T_i)_{i=1,\ldots, N-1}$ forms a representation of the Hecke algebra on the space of polynomials. They satisfy the relations
\beq   
\mathrm{(i)} \quad (T_i-t)(T_i+1)=0, \qquad 
\mathrm{(ii)} \quad T_{i}T_{i+1}T_{i} = T_{i+1}T_i T_{i+1} \qquad
\mathrm{(iii)} \quad T_i T_j = T_j T_i \quad |i-j|>1.
\eeq
The first relation defines the inverse $T_i^{-1}= t^{-1}-1+ t^{-1} T_i$.  Recall the action of $\tau_1: x_1 \mapsto qx_1$, from the previous section, and set $\omega = K_{N-1}\cdots K_1 \tau_1$.  From the Hecke algebra (and $\tau_1)$, we shall build commuting operators called the Cherednik operators \cite{ICher_NS}:
\beq
\mathsf Y_i = t^{-N+i} T_i \cdots T_{N-1} \omega T_1^{-1}\cdots T_{i-1}^{-1}
\eeq
for $i=1,2, \ldots, N$.  We reiterate that, quite remarkably, these $N$ operators commute among themselves 
: $[\mathsf Y_i, \mathsf Y_j]=0$.

\subsubsection{The non-symmetric Macdonald polynomials}

The common eigenfunctions of the $\mathsf Y_i$ operators are the non-symmetric Macdonald polynomials,
\beq
E_\eta(q,t) = E_\eta(x;q,t) = x^\eta + \sum_{\nu \prec \eta} b_{\eta \nu}  x^\nu, \qquad b_{\eta \nu}\in \mathbb Q(q,t)
\eeq
defined for each composition $\eta\in \mathbb Z_{\geq 0}^N$, where $*\prec*$ is the Bruhat order on compositions \eqref{bruhatoto}, and which are such that
\beq
\mathsf Y_i E_\eta(q,t) = q^{\eta_i} t^{- \widehat{\eta}_i} E_\eta(q,t).
\eeq
with $\widehat{\eta}_i$ defined in \eqref{eigennonsym}.   

\subsubsection{The symmetrization process}\label{SymmmProsMacdo}
From the non-symmetric polynomials $E_\eta(q,t)$, we can obtain the superpolynomials $P_\La(q,t)$ by a suitable symmetrization  process using the $t$-symmetrizer and $t$-antisymmetrizer  which we now introduce.
Suppose that $\sigma\in \mathfrak S_N$ has a reduced decomposition  into elementary transpositions that reads $\sigma=s_{i_1}\cdots  s_{i_k}$.   
Given a such an element $\sigma$, set $T_\sigma=T_{i_1}\ldots T_{i_k}$.  For any subset $I \subseteq (1,\ldots, N)$, we can define a Hecke symmetrization $U^+$ and a Hecke antisymmetrization $U^-$ in the following way:
\beq
U^+_I =  \sum_{\sigma \in \mathfrak S_{|I|}} T_\sigma, \qquad \qquad U_I^- = \sum_{\sigma \in \mathfrak S_{|I|}} (-t)^{-\ell(\sigma)} T_\sigma,
\eeq
(where it is understood that the operator $T_\sigma= T_\sigma^{(I)}$ acts on the set of variables $I$, and $|I|$ denotes the cardinality of $I$.)   
Let $\La\vdash(n|m)$ with $N$ parts.  Then, up to a global constant (to be found in \cite[Eq. (31)]{BF2}), we have
\beq
P_\La(q,t) \propto \sum_{\sigma \in \mathfrak S_N} \sigma. \ta_1\cdots \ta_m \, \prod_{1\leq i<j\leq m} \frac{x_i-x_j}{tx_i-x_j} \, U^-_{(1,\ldots, m)}  U^+_{(m+1, \ldots, N)} \, E_{\La^{\mathrm R}}(q,t)
\eeq
where the permutation $\sigma$ acts as before, i.e.~by mapping $x_i,\ta_i \mapsto x_{\sigma(i)}, \ta_{\sigma(i)}$.   
Having such a construction is particularly useful when constructing the commuting conserved quantities of the superpolynomials $P_\La(q,t)$ using the Cherednik operators.  Define
\beq\begin{split}
\bar D_{N,1,r}&=  \sum_{\sigma \in \mathfrak S_N} \sigma. \ta_1\cdots \ta_m  \,   \prod_{1\leq i<j\leq m} \frac{x_i-x_j}{tx_i-x_j} \, e_r(\mathsf Y_1, \ldots, \mathsf Y_N)  \,  \prod_{1\leq i<j\leq m} \frac{tx_i-x_j}{x_i-x_j} \varrho_{(1\ldots m)}
\\
\bar{D}_{N,2,r}& = \sum_{\sigma \in \mathfrak S_N} \sigma. \ta_1\cdots \ta_m  \,   \prod_{1\leq i<j\leq m} \frac{x_i-x_j}{tx_i-x_j} \, e_r(q \mathsf Y_1, \ldots, q \mathsf Y_m, \mathsf Y_{m+1}, \ldots \mathsf Y_N)  \,  \prod_{1\leq i<j\leq m} \frac{tx_i-x_j}{x_i-x_j} \varrho_{(1\ldots m)}
\end{split}
\eeq
for $r=1,\ldots, N$, 
where $e_r(*)$ is  the $r$th elementary symmetric function.  
At fixed fermionic degree $m$, these operators generalize the ones defined above, i.e.~$\bar D_{N,1,1}=\bar D_{N,1}$ and $\bar D_{N,2,1}=\bar D_{N,2}$  (see \eqref{op_DN1}).  Acting on superpolynomials, we have
\beq
\bar{D}_{N,1,r} P_\La(q,t) = d_{N,\La^*,r}(q^{-1},t^{-1}) P_\La(q,t); \qquad \quad
\bar{D}_{N,2,r} P_\La(q,t) =  d_{N,\La^\cd,r}(q^{-1},t^{-1}) P_\La(q,t)
\eeq
where the expression for the eigenvalues are given by the specialization
\beq
d_{N,\la,r}(q,t) = e_r({\mathsf u}_\la).
\eeq

\bigskip

\n {\textsc{References:}} The relation between the non-symmetric Macdonald polynomials and the Macdonald superpolynomials is presented in \cite{BF2}.

\medskip

\section{The double Macdonald polynomials}
\label{DoubleMM}
Consider the superpolynomial ring $\mathsf{A}_{N}[n|m]$ and its basis elements. Instead of taking the (usual) projective limit $N\rightarrow \infty$, one can associate to each superpolynomial 
a bisymmetric polynomials in the  variables $(x_1, \ldots, x_m)$ and $(x_{m+1}, \ldots x_N)$ and then take both limits $m, N\rightarrow\infty$.  In a certain regime (to be described below) this limit is well-defined and is called the bisymmetric function limit of the superfunctions.  When considered for the Macdonald superpolynomials, this limit leads to a connection with the hyperoctohedral  group $\mathrm B_n$.   

For a given subset of superpartitions, that is, the superpartitions for which the associated super-Macdonald functions have a well-defined double projective limit (as we just mentioned), the super-Macdonald functions factorize into regular Macdonald polynomials.  This allows in particular to prove many of the conjectures of Section \ref{propofsmacdo}.  In this section, we  summarize the main results of \cite{BLM15}.

\subsection{From superpolynomials to bisymmetric polynomials}

To any superpolynomial, we can associate a bisymmetric 
polynomial.
For a superpolynomial of fermionic degree $m$, this is done in the following two steps: (1)-extract the coefficient of $\ta_1\cdots \ta_m$ of the superpolynomial,  and (2)-divide the resulting coefficient by the Vandermonde determinant in the variables $x_1,\ldots,x_m$. (Note that this simply amounts to applying the  operator $\hat{\varrho}_m$ defined in \eqref{normamamaprojecto}).
By construction, the resulting polynomial is symmetric in 
both the variables $x_1,\ldots, x_m$ and the variables $x_{m+1},\ldots, x_N$, which in the following we will be denoting as
\beq
x=(x_1,\ldots, x_m), \qquad y=(y_1, \ldots, y_{N-m})=(x_{m+1},\ldots, x_N).
\eeq
The bisymmetric polynomial associated to the power-sum $p_{(3,1;1)}$ for $N=3$ given in \eqref{p311}  is $(x_1^2x_2+x_1x_2^2)(x_1+x_2+y_1)$.

A bisymmetric polynomial is naturally indexed by a pair of partitions, which
is easily extracted from the superpartition.  To be more precise,
the correspondence  between a superpartition $\La$ of fermionic degree $m$ and a pair of partitions $\lambda,\mu$ is
 \beq
 \label{cora}
\La =(\LaA ; \LaS) \lrw (\la, \mu) = (\LaA - \delta^{m}, \LaS),
\eeq
where $\delta^{m}=(m-1,\ldots, 0)$ stands as usual
for the staircase partition.   
  It is clear that $\La$ is fully characterized by $m$ and the pair
$ (\lambda,\mu)$.  For $\La\vdash (n|m)$, let $\hat{n}$ denote the total degree of the corresponding partitions $(\la,\mu)$, i.e.~$\hat{n}=|\la|+|\mu|= n - m(m-1)/2$.  

With this correspondence, the monomial and power-sum symmetric 
superpolynomials are
then associated to the following bisymmetric polynomials
respectively (cf. \cite[Eqs (1.6)--(1.7)]{BLM15})
\beq
\label{modea123}
m_{\La} \; \leftrightarrow \; m_{\la,\mu}(x,y)=s_\la(x) \,m_\mu(y);
\quad
\quad
p_{\La} \; \leftrightarrow \; 
p_{\la,\mu}(x,y) = s_\la(x) \,  p_\mu (x,y),
\eeq
where $s_\la$, $m_\la$, and $p_\la$ are respectively the Schur, monomial and power-sum symmetric functions ($p_{\mu}(x,y)$
is the usual power-sum symmetric functions in the union of the variables $x$ and $y$).

The bisymmetric polynomials associated to the Macdonald superpolynomials 
$P_\La(x_1,\ldots,x_N,\ta_1,\ldots,\ta_N;q,t)$  will simply be denoted $P_\La(x,y;q,t)$.
(The two forms -- the superpolynomial or its bisymmetric version -- are distinguished by their explicit variable-dependence: either $(x,\ta)$ or $(x,y)$.)

The following result is a mere translation of \eqref{defPPPqqqtttt} (see \cite[Theorem 1]{BLM15}). Let $\Lambda \leftrightarrow (\lambda,\mu)$
be a superpartition of fermionic degree $m$, and set
$x=(x_1,\dots,x_m)$ and $y=(x_{m+1},\dots,x_N)$.  Then, for 
$N -m\geq |\lambda|+|\mu|$, there exists
a unique bisymmetric polynomial  $P_{\Lambda}=P_{\Lambda}(x,y;q,t)$ such that: 
\beq\label{mac1}
\begin{split}
&1) \quad  
P_{\Lambda} = m_{\lambda,\mu} + \text{lower terms},\\
&2) \quad
\psc{P_{\Lambda}}{ P_{\Omega}}_{q,t}' =0\quad\text{if}\quad 
\Lambda \ne \Omega.
\end{split}\eeq
The dominance  ordering on pairs of partitions is such
that $(\lambda,\mu) \geq (\omega,\nu)$ whenever the corresponding superpartitions
$\Lambda$ and $\Omega \leftrightarrow (\omega,\nu)$ are such that
$\Lambda \geq \Omega$.  
The scalar product $\psc{*}{*}'_{q,t}$ on the basis of power-sums \eqref{modea123} reads
\beq \label{newsp}
\psc{p_{\lambda,\mu} }{  p_{\omega,\nu} }_{q,t}'= \delta_{\lambda \omega} 
\delta_{\mu \nu} q^{|\lambda|}
\,z_\mu(q,t),
 \eeq  
where $z_{\mu}(q,t)$ is given in \eqref{zLaqt}  (when taking the fermionic degree to be zero there).    Note that this scalar product is basically the same as that defined in the super-Macdonald case, except that we have dropped a 
factor $(-q)^{-\binom{m}{2}}$ which does not affect the orthogonality.

We stress that the monomial expansion of $P_{\Lambda}$ is independent of $N-m$ 
(granted that $N-m$ is large enough) and thus $N-m$ can be considered infinite.
Unexpectedly, a similar independence upon $m$ holds.

\subsection{The stable sector}
It turns out that  the monomial expansion \eqref{mac1} 
of the bisymmetric polynomial $P_\La$ 
does not depend on $m$  
whenever $m \geq \hat{n}$.  This will be referred to as the  {\it stable sector}
of
the  bisymmetric Macdonald polynomials.   Let us illustrate the stability property of
$P_\La$, where $\Lambda \leftrightarrow (\emptyset,(2))$, by displaying the monomial
 decompositions when $\hat{n}=2$ for three different values of $m\geq 2$: 
\begin{align*}
P_{(1,0;2)}&=m_{\emptyset,(2)}+{\frac { \left( 1-t \right)  \left(1+ qt \right) }{1-q{t}^{2}}}
\,m_{\emptyset,(1,1)}
\nonumber\\
P_{(2,1,0;2)}&=m_{\emptyset,(2)}+{\frac { \left( 1-t \right)  \left(1+ qt \right) }{1-q{t}^{2}}}
\,m_{\emptyset,(1,1)}
\nonumber\\
P_{(3,2,1,0;2)}&=m_{\emptyset,(2)}+{\frac { \left( 1-t \right)  \left(1+ qt \right) }{1-q{t}^{2}}}
\,m_{\emptyset,(1,1)}.
\end{align*}
In the three cases, 
one recovers three identical expressions (even though the corresponding
monomials depend on different sets of variables).   
By contrast, for $m=1$, we have
\[
 P_{(0;2)}=m_{\emptyset,(2)}
+{\frac { \left( 1-t \right)  \left(1+ q\right) }{1-qt}}\,m_{\emptyset,(1,1)}+{\frac { \left( 1-t \right) }{1-qt}}\,m_{{(1),(1)}}.
\]

In the stable sector, it will thus be more natural to index 
the bisymmetric polynomial $P_\La= P_{\lambda,\mu}$  
by the pair of partitions $(\lambda,\mu)$, even more so that in this sector
the ordering on superpartitions can be replaced by the
following dominance ordering on pairs of partitions: 
for $(\la,\mu)$ and $(\omega, \nu)$ both of total degree $\hat{n}$, 
\begin{equation}\label{domibi}
(\la,\mu) \geq (\omega,\nu)  \quad {\rm iff}  \quad
 \la_1+ \cdots + \la_i \geq \omega_1 +\cdots +\omega_i \quad \text{and} \quad  |\la|+ \mu_1 + \cdots + \mu_j \geq    |\omega |+ \nu_1 + \cdots + \nu_j \,  \, \, \forall i,j  ,
\end{equation}
where it is understood that $\lambda_k=0$ if $k>\ell(\lambda)$ (and similarly
for $\mu, \omega$ and $\nu$).

\subsection{The double Macdonald polynomials}
Because they are labelled by two partitions and they are naturally viewed as functions  of two sets of (commuting) variables, the bisymmetric
Macdonald polynomials in the stable sector
are called double Macdonald polynomials.
Therefore, in the stable sector
we have the following (which is \cite[Theorem 2]{BLM15}).
Let $(\lambda,\mu)$ be of total degree $\hat{n}$.
Then the double Macdonald polynomials $P_{\lambda,\mu}(x,y;q,t)$,
where  $x=(x_1,\ldots,x_m)$  and $y=(x_{m+1},\ldots,x_N)$ (with  $m \geq \hat{n}$ and $N-m \geq \hat{n}$) are the unique bisymmetric polynomials such that
\beq \begin{split}
\label{Macsta}
1) & \quad P_{\la,\mu}(x,y;q,t) = m_{\la,\mu}(x,y) + \sum_{\omega,\nu < \la,\mu} c_{\la,\mu;\omega,\nu}(q,t) \, m_{\omega,\nu}(x,y), \\
2)  & \quad \psc{ P_{\la,\mu} }{ P_{\omega,\nu}}_{q,t}' = 0 \quad\text{if}\quad
(\la,\mu) \neq (\omega,\nu),
\end{split}
\eeq
where the ordering on pairs of partitions 
and the scalar product are respectively defined in
\eqref{domibi} and \eqref{newsp}.
 It should also be stressed that there is no solution to the two conditions \eqref{Macsta} in the non-stable sector if the ordering \eqref{domibi} is used. Therefore, in order to interpolate between the usual and double
Macdonald polynomials (corresponding respectively to the cases 
$m=0$ and $m\geq \hat{n}$), the construction relying on the super-dominance ordering is necessary.

\subsection{ The double Macdonald polynomials as products of two Macdonald polynomials}
When $m$ is large enough, that is, when $m \geq \hat n$,
the correspondence $\Lambda \leftrightarrow (\lambda,\mu)$ provides a natural factorization of $\Lambda$ into the two partitions $\lambda$ and $\mu$ (in the sense that the two partitions are far enough so that they do not interact anymore). For instance,
 consider for $m=8$, the superpartition $\La=(9,7,5,4,3,2,1,0;3,1)$, the identification is given by 
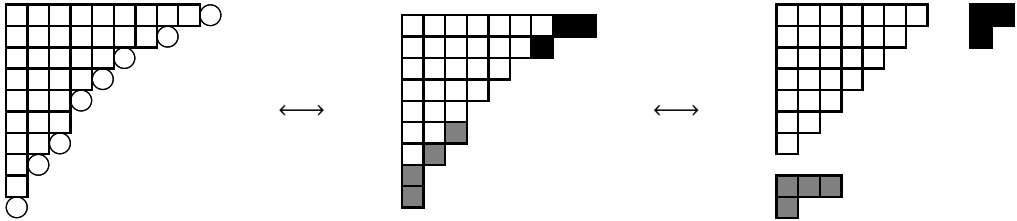
\[
\scalebox{1.1}{$\superY{
 \, & \, & \, & \, &  \, & \, & \, &  \, & \, & \yF{} \\
 \, & \, & \, & \, &  \, & \, & \, & \yF{}\\ 
 \, &  \, &   \,   & \, & \, & \yF{}   \\
 \, &  \, &   \,   & \, &  \yF{}   \\
 \, &  \, &   \,    &  \yF{}   \\
\, & \, &  \, \\
     \, & \, & \yF{}   \\
\, & \yF{}
\, \\
\\
\yF{}
  }$} 
  \quad  \longleftrightarrow \qquad
  \scalebox{1.1}{$\superY{
 \, & \, & \, & \, &  \, & \, & \, &  *(black) & *(black)  \\
 \, & \, & \, & \, &  \, & \, & *(black)  \\ 
 \, &  \, &   \,   & \, & \,    \\
 \, &  \, &   \,   & \,    \\
 \, &  \, &   \,       \\
\, & \, &  *(gray) \\
     \, & *(gray)   \\
*(gray) \\
*(gray) 
  }$} 
  \quad  \longleftrightarrow \qquad
  \scalebox{1.1}{$\superY{
 \, & \, & \, & \, &  \, & \, & \, & \none & \none &  *(black) & *(black)    \\
 \, & \, & \, & \, &  \, & \, & \none & \none & \none & *(black)  \\ 
 \, &  \, &   \,   & \, & \,    \\
 \, &  \, &   \,   & \,    \\
 \, &  \, &   \,       \\
\, & \,  \\
 \,    \\
 \none \\
*(gray) & *(gray) & *(gray) \\
*(gray) 
  }$} 
  \]
so that $\la,\mu = (2,1), (3,1)$. Note that in the first correspondence, the circles are removed and the white cells (in the second diagram) build the partition $\delta^m$.  The last correspondence simply says that we read the upper extra cells (resp. lower extra cells)  row-wise (resp. column-wise) to obtain the partition $\lambda$ (resp. $\mu'$). As we will now see, the double Macdonald polynomial $P_\Lambda (x,y;q,t)$ also factors as a product of Macdonald polynomials indexed by $\lambda$ and $\mu$, albeit in a non-obvious way.

Through this section, we shall use the so-called plethystic notation \cite{Lasc03, Hag08, Berg09}, where the power-sum functions $p_1, p_2, \ldots$ act on the ring of rational functions in $x_1, \ldots, x_N,q,t$ over $\mathbb Q$.  The action is  written as
\beq
p_k[*] \; : \; x_i,q,t \mapsto x_i^k, q^k,t^k.
\eeq
For instance, we have
\[
p_k[x_1 + x_2 + \ldots + x_N] = x_1^k + x_2^k + \ldots + x_N^k = p_k(x_1, \ldots, x_N),
\]
so that a symmetric function $f(x_1, \ldots, x_N)$ is equal to $f[x_1+ \ldots + x_N]$ in that notation.  For any positive numbers $m,N$, with $m<N$, we set:
\beq
X=x_1+ \ldots + x_m, \qquad \quad
Y= y_1 + \ldots + y_{N-m}= x_{m+1} + \ldots  + x_N.
\eeq

In the stable sector, that is, given any two partitions $\la,\mu$, such that $m\geq |\la|+|\mu|$, $N-m\geq   |\la|+|\mu|$, we have the following properties of bisymmetric polynomials:
\begin{itemize}
\item The scalar product $\psc{*}{*}_{q,t}'$ is equivalent to \cite[Lemma 4]{BLM15}
\beq\label{psdmacdoppp}
\psc{ p_\la\bigl[ X+\frac{q(1-t)}{1-qt} Y \bigr] p_\mu \bigl[Y\bigr] }{  p_\omega\bigl[ X+\frac{q(1-t)}{1-qt} Y \bigr] p_\nu \bigl[Y\bigr]   }_{q,t}''=  \delta_{\la \omega} \delta_{\mu \nu}q^{|\la|} z_\la(q,qt) z_{\mu}(qt,t);
\eeq

\medskip

\item The double Macdonald polynomials \eqref{Macsta} factorize \cite[Theorem 5]{BLM15} as
\beq 
\label{factor}
P_{\la,\mu}(x,y;q,t) = P_{\la}^{(q,qt)} \Bigl[ X + \frac{q(1-t)}{1-qt} Y\Bigr] \, P_{\mu}^{(qt,t)} \bigl[ Y\bigr],  
\eeq
where $P_{\la}^{(q,t)}(x)$ denotes the usual Macdonald polynomial  $P_\la(x;q,t)$.  
\end{itemize}
It is immediate to see that the double Macdonald polynomials are orthogonal with respect to the scalar product \eqref{psdmacdoppp}, which reads
\beq
\psc{  P_{\la,\mu}(q,t)  }{  P_{\omega,\nu}(q,t)  }_{q,t}'' = q^{|\la|} \,  \psc{ P_{\la}(q,qt)   }{    P_{\omega}(q,qt)  }_{q,qt} \; \cdot \;  \psc{ P_{\mu}(qt,t)   }{    P_{\nu}(qt,t)  }_{qt,t}.
\eeq
In particular, setting $\omega=\la$ and $\nu=\mu$ in the previous equation, we obtain the norm-squared of the double Macdonalds,
\beq\label{normdoublemacdo}
 \psc{  P_{\la,\mu}  }{  P_{\la,\mu}  }_{q,t}'' = q^{|\la|} b_\la(q,qt)^{-1} b_\mu(qt,t)^{-1}
\eeq
with 
\[
b_\la(q,t) = \prod_{s\in \la} \frac{1-q^{ a_\la(s) } t^{l_\la(s) +1} }{1-q^{a_\la(s) +1} t^{l_\la(s)} }.
\]
 the standard normalization constant of Macdonald polynomias \cite{MacSym95}. Note that, upon the correspondence $\La\leftrightarrow (\la,\mu)$, the norm-squared \eqref{normdoublemacdo} is consistent with the result \eqref{conjNSM1}  whenever the superpartition lies in the stable sector.    The precise connection is shown in \cite[App. C]{BLM15}.

For any superpartition $\La$ of fermionic degree $m$ in the stable sector, the factorization \eqref{factor} can be reformulated as
\beq\label{projsMacasDouble}
\hat{\varrho}_m( P_\La(x,\ta;q,t)   ) =   P_{\la}^{(q,qt)} \Bigl[ X + \frac{q(1-t)}{1-qt} Y\Bigr] \, P_{\mu}^{(qt,t)} \bigl[ Y\bigr].
\eeq
Using the connection with the non-symmetric Macdonald polynomials described in Section~\ref{SymmmProsMacdo}, we also have that in the stable sector  
\beq
P_\la^{(q,qt)}\Bigl[ X + \frac{q(1-t)}{1-qt} Y\Bigr] \, P_\mu^{(qt,t)}\bigl[Y\bigr] \propto 
\left(\prod_{1\leq i<j\leq m} \frac{1}{tx_i-x_j} \right) \, U^-_{(1,\ldots, m)} \, U^+_{(m+1, \ldots, N)} \, E_{(\la+\delta^m; \mu)^{\mathrm R}}(x;q,t),
\eeq
which says that when $t$-antisymmetrizing and $t$-symmetrizing in  a proper way a non-symmetric Macdonald polynomial, we get a product of Macdonald polynomials.

Relation \eqref{projsMacasDouble} allows us to compute in the double Macdonald case the specialization of Section~\ref{specializationSM} for the super-Macdonalds.  First, the map $\epsilon_{q,t}$ has the following action on functions $X$ and $Y$:
\beq
\begin{split}
&\epsilon_{q,t} \; : \; X \mapsto  q^{-m+1}+ q^{-m+2} t+ \ldots + t^{m-1}= q^{-m+1}\frac{1-(qt)^m}{1-qt} \\
&\epsilon_{q,t} \; : \; Y \mapsto   t^{m-1}+ t^{m-2}+\ldots + t^{N-1} = t^m \frac{1-t^{N-m}}{1-t}.
\end{split}
\eeq
We then have,
\beq
P_\La(\epsilon_{q,t}) = \frac{t^{m |\mu |} }{q^{(m-1)|\la|}} P_\la^{(q,qt)} \Bigl[ \frac{1-q^mt^N}{1-qt} \Bigr] \,    P_\mu^{(qt,t)} \Bigl[ \frac{1-t^{N-m} }{1-t} \Bigr],
\eeq
corresponding to the stable sector $\La\leftrightarrow (\la,\mu)$ of fermionic degree $m$.  
Note that this expression is explicit since the specialization of a power-sum $p_n$ as
\[
p_n\Bigl[ \frac{1-u}{1-t} \Bigr] \; : \; p_n \mapsto \frac{1-u^n}{1-t^n }
\]
is known for Macdonald polynomials \cite{MacSym95}, and given by the expression
\[
P_\la^{(q,t)}\Bigl[ \frac{1-u}{1-t} \Bigr]  = \prod_{s\in \la} \frac{t^{l'_\la(s) } -q^{a'_\la(s)} u }{1-q^{a_\la(s) }  t^{l_\la(s) +1}}.
\]
Again, this is consistent  with the result at equation \eqref{ConjSpecSM} in the stable sector, the precise connection being shown in  \cite[App. C]{BLM15}.

We end this section by mentioning a few special cases (in the stable sector):
\begin{itemize}
\item The double Jack functions are given by
\beq
P_{\la,\mu}^{(\alpha)}(x,y) = P_\la^{(\alpha/(\alpha+1))} \Bigl[ X + \frac{1}{\alpha+1} Y \Bigr] \,P_\mu^{(\alpha+1)} \bigl[  Y \bigr],
\eeq
where $P_\la^{(\alpha)}$ stands for the standard Jack symmetric function, and note that the parameter $\alpha$  is not affected by the plethysm (since it belongs to the base field $\mathbb Q(\alpha)$), hence
\[
p_n\bigl[ X + \frac{1}{\alpha+1} Y \bigr] = p_n\bigl[ X\bigr]+ \frac{1}{\alpha+1} p_n \bigl[Y\bigr].
\]

\medskip

\item The double Schur functions are given by

\beq \label{doubleSSfun}
s_{\la,\mu}(x,y) = s_\la(x) s_\mu(y), \qquad \quad
\bar{s}_{\la,\mu}(x,y) = s_\la(x,y) s_\mu(y). 
\eeq

\end{itemize}

\medskip

\subsubsection{Connection with the hyperoctahedral group }  We now sketch the connection between the double Macdonald  functions and the hyperoctahedral group $\mathrm B_n$.       

In the stable sector, the modified Macdonald superpolynomial $H_\La(q,t)$ of \eqref{ModifSMH} becomes 
\beq
H_{\la,\mu}(x,y;q,t)= J_\la^{(q,qt)} \Bigl[ \frac{X+qY}{1-qt} \Bigr]\, J_\mu^{(qt,t)} \Bigl[ \frac{tX+Y}{1-t} \Bigr] = H_\la^{(q,qt)}\bigl[X+qY\bigr] \, H_\mu^{(qt,t)}\bigl[tX+Y\bigr] 
\eeq
 where $J_\la^{(q,t)}$ (resp.~$H_\la^{(q,t)}$) stands for the integral form of the Macdonald polynomial (resp.~for the modified Macdonald polynomials).  The expansion of  $H_{\la,\mu}(x,y;q,t)$ into the basis of double Schur functions \eqref{doubleSSfun}, written as
 \beq
 H_{\la,\mu}(x,y;q,t) = \sum_{\omega, \nu} K_{\omega,\nu; \la,\mu}(q,t) s_{\omega, \nu}(x,y)
\eeq
defines the coefficients $K_{\omega, \nu; \la,\mu}(q,t)$ which we refer to as the $(q,t)$-Kostka coefficients of type $\mathrm B$.  
From the usual Macdonald positivity, we get immediately that the $(q,t)$-Kostka coefficients of type $\mathrm B$
  belong to $\mathbb N[q,t]$.   Moreover, given the proposition that follows,
the positivity suggests that the double Macdonald polynomials are in correspondence with natural bigraded regular modules of the 
hyperoctahedral group $\mathrm B_n$, where $n=|\la|+|\mu|= |\omega|+|\nu|$.

 \begin{proposition}  \cite[Proposition 11]{BLM15}.  
Let $\la$ and $\mu$ be two partitions such that $n=|\la|+|\mu|$. The coefficient $K_{\omega, \nu; \la,\mu}(1,1)$ is the dimension of the irreducible representation of $\mathrm B_n$ indexed by the pairs of partitions $\omega, \nu$ (it does not depend on $\la$ and $\mu$).  Note that 
$K_{\omega, \nu; \la,\mu}(1,1)$ is also the number of pairs of standard Young tableaux of respective shapes $\omega$ and $\nu$ filled (without repetitions) with the numbers $(1,2,3,\ldots,n)$.  
\end{proposition}

\bigskip

\n {\textsc{References:}} The double Macdonald polynomials were studied in \cite{BLM15}.  The specialization to the Jack case and its connection with physics are worked out in \cite{LM}.

\medskip

\section{The super-Schurs}
\label{SSchur}

As was saw in \eqref{limitSMactoSS}, the Schur superpolynomials are obtained from the Macdonald superpolynomials either by setting the $(q,t)$ parameters in the super-Macdonald to $(0,0)$ or to $(\infty, \infty)$:
\[
s_\La = P_\La(0,0), \qquad \bar{s}_\La = P_\La(\infty, \infty)
\]
Since they are  a special case of the super-Macdonalds, both Schur superpolynomials are of course unitriangular in the monomial basis.    However, since the $(q,t)$-scalar product defining the super-Macdonalds is degenerate whenever $q=t=0$ or $=\infty$, one cannot characterize the super-Schur with a combinatorial definition similar to that of the super-Jacks, or super-Macdonalds (i.e.~from triangularity \emph{and} orthogonality).  But, as it turns out, the two families of Schur superpolynomials are essentially dual to each other.

This is obtained as follows.  Recall property \eqref{rhIIIqtto11} of the ring homomorphism III, and the duality  \eqref{dualiqtMacs} between Macdonald superpolynomials.  We can write 
\beq
(-1)^{\binom{m}{2}} \delta_{\La \Om} = \psc{P_\La(q,t)}{Q_\Om(q,t)}_{q,t} =   \psc{P_\La(q,t)}{  (-1)^{\binom{m}{2}} \widehat{\omega} P_{\Om'}(t^{-1},q^{-1})}.
\eeq
Setting $q=t=0$ in this last expression, and letting 
\beq\label{setwithomegasbar}
s^*_\La = (-1)^{\binom{m}{2}} \widehat{\omega}( \bar{s}_{\La'})
\eeq
for any superpartition, we thus have
\beq\label{eq_ortho_sschurzzz}
\psc{s_\La^{\phantom*}}{s^*_\Om} = (-1)^{\binom{m}{2}} \delta_{\La \Om}.
\eeq
Thus the $s^*_\La$'s, which are obtained from the $\bar s_\Lambda$'s by acting with $ \widehat{\omega}$, are dual to the $s_\Lambda$'s. In the following, when describing the Pieri rules, we shall work with the  $s^*_\La$ basis instead of with the $\bar s_\Lambda$ basis.

In this section, we first present Pieri formula for the two families of super-Schurs. We then show how to obtain the super-Schurs independently of the Macdonald superpolynomials by using creation operators (known as super Bernstein operators).   We finally give tableaux generating series for the Schur functions in superspace $s_\Lambda$ and $\bar s_\Lambda$, and more generally for their skew generalizations (for which we also show how they connect with the Littlewood-Richardson coefficients in superspace).

\medskip

\subsection{Pieri rules} Let $\Ga\in\{(r;\,), (\, ; r), (0;1^r), (\, ; 1^r)\}$ for $r\geq 0$, that is,  $\Ga$ corresponds to a one row or a one column diagram  (either bosonic or fermionic).    For any superpartition $\La$, the Pieri rule refers to  a (combinatorial) formula for the multiplication
\[
s_\La \, s_\Ga = \sum_\Om a_{\La \Om}^\Ga \, s_\Om
\]
which in this case will be such that  $a_{\La \Om}^\Ga= 0,\pm1$ (and equivalently for the $s_\La^*$'s).  
Note that, in  terms of the standard basis, the super-Schurs associated to a one row or one column diagram reduce to  
\[
\begin{tabular}{|c||c|c| c| c |} 
\hline
$\Ga$ &  $(r; \,)$ &   $(\, ; r)$  &  $(0;1^r)$  &  $(\, ; 1^r)$ \\
\hline\hline
$s_\Ga$&$\tilde p_r$&$h_{r}$ &$\tilde{e}_r$ & $e_r$ \\
      $s_\Ga^*$&$\tilde h_r $&$ h_r$&  $\tilde{e}_0 e_r$ & $e_r$    \\  
\hline
\end{tabular}
\]
\medskip

There are thus eight Pieri rules to consider for the two super-Schur families.  We first list here the different Pieri rules, and refer to the subsections that follow for the description of the diagrams upon which the summation is performed (namely RMI, CMI, RMII, CMII).   They read:
\begin{itemize}

\item[\bf{A-}] For the super-Schur $s_\La$ and the one row diagrams, we have
\beq 
s_\La \, s_{(r;\,)} = \sum_\Om (-1)^{\#\ell(\circledast)} s_{\Omega},
\qquad \qquad
s_\La \, s_{(\,;r)} = \sum_\Om s_{\Omega},
\eeq
where $\#\ell(\circledast)$ denotes the number of circles that lie below the one added to the diagram, and where the sum is over diagrams of type RMI.  For example (see  \eqref{Ex1RMI} below), we can write (taking into account the sign)
\[
\begin{split}
s_{(2;4,1)} \, s_{(3;\,)} =& -s_{(7,2 ; 1) } - s_{(6,3; 1)} - s_{( 6,2;2)} - s_{(6,2 ;1,1 )} -s_{( 5,3;1,1 )}
- s_{( 5,2; 2,1)} - s_{( 5,3;2 )} - s_{( 5,2; 3)} 
\\
&
- s_{( 4,3;2,1 )} - s_{( 4,2; 3,1)}- s_{( 3,2; 4,1)} - s_{( 4,2; 4)}.
\end{split}
\]

\medskip

\item[\bf{B-}] For the super-Schur $s_\La$ and the one column diagrams, we have
\beq 
s_\La \, s_{(0;1^r)} = \sum_\Om (-1)^{\#\ell(\circledast)} s_{\Omega},
\qquad \qquad
s_\La \, s_{(\,;1^r)} = \sum_\Om s_{\Omega},
\eeq
where $\#\ell(\circledast)$ denotes the number of circles that lie below the one added to the diagram, and where the sum is over diagrams of type CMI.

\medskip

\item[\bf{C-}] For the super-Schur $s^*_\La$ and the one row diagrams, we have
\beq 
s_\La^* \, s^*_{(r;\,)} = \sum_\Om (-1)^{\#\ell(\circledast)} s^*_{\Omega},
\qquad \qquad
s_\La^* \, s^*_{(\,;r)} = \sum_\Om s_{\Omega},
\eeq
where $\#\ell(\circledast)$ denotes the number of circles that lie below the one added in the diagram, and where the sum is over diagrams of type RMII.  For example (see  \eqref{Ex1RMII} below), we have
\[
s^*_{( 2; 4,1)} \, s^*_{(3 ;\, )} =  s^*_{( 2,0; 7,1 )} + s^*_{( 1,0;6,3  )} + s^*_{(2,0 ;6,2 )} + s^*_{( 2,1; 6,1)} + s^*_{(1,0 ; 5,4)} + s^*_{( 2,0; 5,3)} + s^*_{(2,0 ;4,4 )}- s^*_{( 3,2;4,1 )} .
\]

\medskip

\item[\bf{D-}] For the super-Schur $s^*_\La$ and the one column diagrams, we have
\beq 
s_\La^* \, s^*_{(0;1^r)} = \sum_\Om (-1)^{\#\ell(\circledast)} s^*_{\Omega},
\qquad \qquad
s_\La^* \, s^*_{(\,;1^r )} = \sum_\Om s_{\Omega},
\eeq
where $\#\ell(\circledast)$ denotes the number of circles that lie below the one added in the diagram, and where the sum is over diagrams of type CMII.

\end{itemize}

\medskip

We now give the definitions of the different sets of superpartitions associated to the Pieri rules that we just presented

\medskip

\subsubsection{Row multiplication of type I (RMI)}   Let $\La$ be a superpartition,  and let $\Ga\in \{(r;\,), (\,; r) \}$ with $r\geq0$.
A superpartition $\Om$ belongs to the set RMI (row multiplication of type I)
if it satisfies the following rules.  
\begin{itemize}

\item $\Om^*/\La^*$ is a horizontal $r$-strip;

\item The circles of $\La$ can be moved subject to the restrictions:
\begin{enumerate}
\item[(i)] a circle in the first row can be moved horizontally without 
restrictions;
\item[(ii)] a circle not in the first row can be moved horizontally along the same row  as 
long as there is a square in the row just above it in the original 
diagram $\La$;
\item[(iii)] a circle can be moved vertically in the same column by at most one row.
\end{enumerate}
\item If $\Ga$ is fermionic, then the new circle (the only one which wasn't moved) in  $\Om$ needs to be   in the rightmost position;

\end{itemize}
We illustrate the set RMI with the example (the boxes of 
  $\Om^*/\La^*$ and the new circle are colored) 
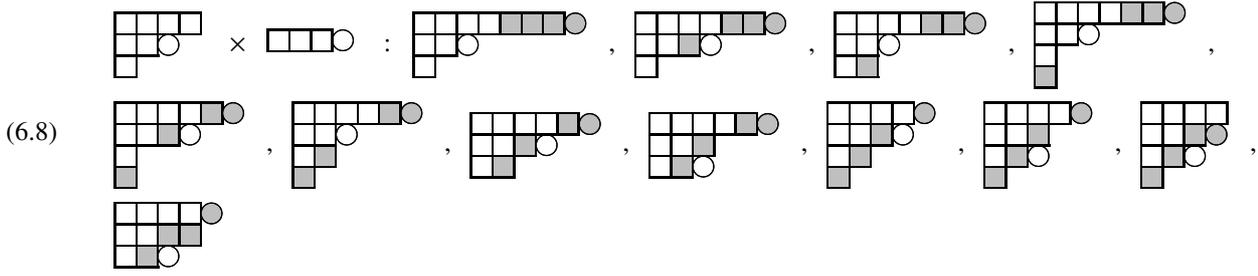
\beq\label{Ex1RMI}\begin{split}
&
\scalebox{1.1}{$\superY{
 \, & \, & \, & \, \\
 \, & \, & \yF{}\\ 
 \,  \\
  }$} 
\times
\scalebox{1.1}{$\superY{
 \, & \, & \,& \yF{}\\ 
  }$} 
  :
\scalebox{1.1}{$\superY{
 \, & \, & \, & \, & *(lightgray) & *(lightgray) & *(lightgray) &\yFgray{}  \\
 \, & \, & \yF{}\\ 
 \,  \\
  }$} 
,
\scalebox{1.1}{$\superY{
 \, & \, & \, & \, & *(lightgray) & *(lightgray) & \yFgray{}  \\
 \, & \, & *(lightgray)& \yF{}\\ 
 \,  \\
  }$} 
,
\scalebox{1.1}{$\superY{
 \, & \, & \, & \, & *(lightgray) & *(lightgray) & \yFgray{}  \\
 \, & \, &  \yF{}\\ 
 \, & *(lightgray) \\
  }$} 
,
\scalebox{1.1}{$\superY{
 \, & \, & \, & \, & *(lightgray) & *(lightgray) & \yFgray{}  \\
 \, & \, &  \yF{}\\ 
 \,  \\
 *(lightgray)\\
  }$} 
,
\\
&
 \scalebox{1.1}{$\superY{
 \, & \, & \, & \, & *(lightgray)  & \yFgray{}  \\
 \, & \, & *(lightgray)& \yF{}\\ 
 \,  \\
 *(lightgray)\\
  }$} 
,
 \scalebox{1.1}{$\superY{
 \, & \, & \, & \, & *(lightgray)  & \yFgray{}  \\
 \, & \, & \yF{}\\ 
 \, &*(lightgray) \\
 *(lightgray)\\
  }$} 
,
 \scalebox{1.1}{$\superY{
 \, & \, & \, & \, & *(lightgray)  & \yFgray{}  \\
 \, & \, & *(lightgray)& \yF{}\\ 
 \, & *(lightgray) \\
  }$} 
,
 \scalebox{1.1}{$\superY{
 \, & \, & \, & \, & *(lightgray)  & \yFgray{}  \\
 \, & \, & *(lightgray)\\ 
 \, & *(lightgray) & \yF{} \\
  }$} 
,
\scalebox{1.1}{$\superY{
 \, & \, & \, & \,   & \yFgray{}  \\
 \, & \, &*(lightgray) & \yF{}\\ 
 \, &*(lightgray) \\
 *(lightgray)\\
  }$} 
,
\scalebox{1.1}{$\superY{
 \, & \, & \, & \,   & \yFgray{}  \\
 \, & \, &*(lightgray) \\ 
 \, &*(lightgray) & \yF{} \\
 *(lightgray)\\
  }$}
,
\scalebox{1.1}{$\superY{
 \, & \, & \, & \,    \\
 \, & \, &*(lightgray) & \yFgray{}  \\ 
 \, &*(lightgray) & \yF{} \\
 *(lightgray)\\
  }$}
,
\\
&
\scalebox{1.1}{$\superY{
 \, & \, & \, & \,   & \yFgray{}  \\
 \, & \, &*(lightgray) & *(lightgray) \\ 
 \, &*(lightgray) & \yF{} \\
  }$}
\end{split}
\eeq

\medskip

\subsubsection{Column multiplication of type I (CMI)}   Let $\La$ be a superpartition,  and let $\Ga\in \{(0;1^r), (\,; 1^r) \}$ with $r\geq0$.  
A superpartition $\Om$ belongs to the set CMI (column multiplication of type I)
if it satisfies the following rules.  
\begin{itemize}

\item $\Om^*/\La^*$ is a vertical $r$-strip;

\item The circles of $\La$ can be moved subject to the restrictions:
\begin{enumerate}
\item[(i)] a circle in the first column can be moved vertically without restrictions;
\item[(ii)] a circle not in the first column can be moved vertically along the same column  as 
long as there is a square in the column immediately to its left in the original diagram $\La$;
\item[(iii)] a circle can be moved horizontally  in the same row by at most one column;
\end{enumerate}
\item If $\Ga$ is fermionic, then the new circle (the only one which wasn't moved) in  $\Om$ needs to be   in the lowermost position.
\end{itemize}

As an example of CMI, one may consider  the transpose of every diagrams appearing in \eqref{Ex1RMI}.

\medskip

\subsubsection{Row multiplication of type II (RMII)}   Let $\La$ be a superpartition,  and let $\Ga\in \{(r;\,), (\,; r) \}$ with $r\geq0$.  
A superpartition $\Om$ belongs to the set RMII (row multiplication of type II) if it satisfies the following rules.

\begin{itemize}
 
\item $\Om^*/\La^*$ is a horizontal $r$-strip;

\item
The $i$-th circle, starting from below, of $\Om$ is either in the same row  as the
$i$-th circle of $\La$ if $\Om^*/\La^*$ does not contain a box in that row or one row below that  of the $i$-th circle of $\La$ if $\Om^*/\La^*$  contains a box in the row  of the $i$-th circle of $\La$;

\item If $\Ga$ is fermionic, then every column to the left of the new circle must have a new box.  
\end{itemize}
We illustrate the set RMII using the same superpartitions as
in \eqref{Ex1RMI}.
\beq\label{Ex1RMII}\begin{split}
&
\scalebox{1.1}{$\superY{
 \, & \, & \, & \, \\
 \, & \, & \yF{}\\ 
 \,  \\
  }$} 
\times
\scalebox{1.1}{$\superY{
 \, & \, & \,& \yF{}\\ 
  }$} 
  :
\scalebox{1.1}{$\superY{
 \, & \, & \, & \, & *(lightgray) & *(lightgray) & *(lightgray)   \\
 \, & \, & \yF{}\\ 
 \,  \\
 \yFgray{} }$} 
,
\scalebox{1.1}{$\superY{
 \, & \, & \, & \, & *(lightgray) & *(lightgray)    \\
 \, & \, &  *(lightgray)\\ 
 \,  & \yF{} \\
 \yFgray{} }$} 
,
\scalebox{1.1}{$\superY{
 \, & \, & \, & \, & *(lightgray) & *(lightgray)    \\
 \, & \, &  \yF{} \\ 
 \,  & *(lightgray) \\
 \yFgray{} }$} 
,
\scalebox{1.1}{$\superY{
 \, & \, & \, & \, & *(lightgray) & *(lightgray)    \\
 \, & \, &  \yF{} \\ 
 \,  & \yFgray{} \\
 *(lightgray) }$}
,
\\
&
\scalebox{1.1}{$\superY{
 \, & \, & \, & \, & *(lightgray)     \\
 \, & \, &  *(lightgray) & *(lightgray) \\ 
 \,  & \yF{} \\
 \yFgray{} }$} 
,
\scalebox{1.1}{$\superY{
 \, & \, & \, & \, & *(lightgray)     \\
 \, & \, &  *(lightgray)  \\ 
 \,  &  *(lightgray) & \yF{} \\
 \yFgray{} }$}
,
\scalebox{1.1}{$\superY{
 \, & \, & \, & \,      \\
 \, & \, &  *(lightgray) & *(lightgray)  \\ 
 \,  &  *(lightgray) & \yF{} \\
 \yFgray{} }$}
,
\scalebox{1.1}{$\superY{
 \, & \, & \, & \,      \\
 \, & \, &  *(lightgray) & \yFgray{} \\ 
 \,  &  *(lightgray) & \yF{} \\
*(lightgray)  }$}
\end{split}
\eeq

\medskip

\subsubsection{Column multiplication of type II (CMII)}   Let $\La$ be a superpartition,  and let $\Ga\in \{(0;1^r), (\,; 1^r) \}$ with $r\geq0$.  
A superpartition $\Om$ belongs to the set CMII (column multiplication of type II) if it satisfies the following rules. 

A superpartition $\Om$ results from the column insertion, or column multiplication, of diagrams $\Ga$ and $\La$, if it can be obtained from $\La$ by applying the following rules.
\begin{itemize}
\item $\Om^*/\La^*$ is a vertical $r$-strip;

\item The 
circles of $\La$ can be moved subject to the following restrictions:
\begin{enumerate}

\item[(i)] a circle cannot overpass another one;

\item[(ii)]  a circle cannot be moved in a row which has an added box;

\item[(iii)] {the addition of  a bosonic box to a fermionic row bumps the circle to the end of the subsequent row (if it is bosonic)  and this bumping can be done repeatedly.}

\end{enumerate}
\item If $\Ga$ is fermionic, then the first column of $\Om$ is fermionic.

\end{itemize}
We consider the (simple) example:
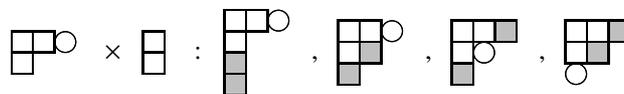
\[
\scalebox{1.1}{$\superY{
\,&\,& \yF{}\\
\, 
}$}
 \times
\scalebox{1.1}{$\superY{
\,\\
\, 
}$}
:
 \scalebox{1.1}{$\superY{
\,&\,& \yF{}\\
\, \\
*(lightgray)\\
*(lightgray)
}$}
,
\scalebox{1.1}{$\superY{
\,&\,& \yF{}\\
\, &*(lightgray) \\
*(lightgray)
}$}
,
\scalebox{1.1}{$\superY{
\,&\,& *(lightgray)\\
\, & \yF{}\\
*(lightgray)
}$}
,
\scalebox{1.1}{$\superY{
\,&\,& *(lightgray)\\
\, & *(lightgray)\\
\yF{}
}$}
\]

\bigskip

\n {\textsc{References:}}  The super-Schurs and the Pieri rules are treated in great details in \cite{BM1, JL17}, and more recently in \cite{ABLM18}.  The Pieri rules (A,B,C) are proved in  \cite{JL17} while the Pieri rule (D) is proved in \cite{ABLM18}.

\medskip

\subsection{The Bernstein superoperators}   
We now turn to the construction of the super-Schurs from vertex-type operators.  In the following, it will be useful to write, for an element $f\in \mathsf A$,
\beq
f[t_1, t_2, \ldots, \tau_0, \tau_1, \ldots]
\eeq
to make  explicit the dependence of $f$ into power-sums, where (here) each $p_i$ (for $i=1,2,\ldots$) is replaced by $t_i$ ($i=1,2,\ldots$) and each $\tilde{p}_j$ (for $j=0,1,\ldots$) is replaced by $\tau_j$ ($j=0,1,\ldots$).  In other words, we will use the notation $[*]$ to indicate that the variables considered are the power-sums (i.e.~not the indeterminates $x,\ta$).  

Let
\beq\label{LesdeuxLzero}
\mathcal L_0^{\phantom\perp} = \sum_{r\geq 0}h_r \partial_{\tilde{p}_r}, \qquad
\text{and}
\qquad
\mathcal L_0^\perp = \sum_{r\geq 0} \tilde{p}_{r} h_r[\hat{\partial}_p]
\eeq
where $\hat{\partial}_p$ denotes the replacement of the (even) power-sum $(p_1, p_2, p_3, \ldots)$ by $(\partial_{p_1}, 2 \partial_{p_2}, 3 \partial_{p_3}, \ldots)$.  The operator $\mathcal L_0^\perp$  in \eqref{LesdeuxLzero} is the adjoint of $\mathcal L_0^{\phantom\perp}$ with respect to the scalar product, i.e.~$\psc{\mathcal L_0^{\phantom \perp}f}{g}= \psc{f}{\mathcal L_0^\perp g}$ for any $f,g\in \mathsf A$.  

Let $t$ and $\tau$ be formal even and odd parameters (respectively), and define the (super)vertex-type operators $B(t,\tau)$ and $C(t, \tau)$, the formal generating series determined by
\beq \label{LessuperBC}\begin{split}
B(t,\tau) &= \mathrm{e}^{\tau \mathcal L_0}  \Bigl(  \exp  \sum_{n>0}\frac{t^n}{n} p_n \Bigr) \Bigl( \sum_{r \geq 0} (-t)^r \tilde{e}_r \Bigr) \Bigl(  \exp  - \sum_{m>0} t^{-m} \partial_{p_m} \Bigr),
\\
C(t,\tau) & =  \mathrm{e}^{\tau \mathcal L_0^\perp}  \Bigl(  \exp  \sum_{n>0}\frac{t^n}{n} p_n \Bigr) \Bigl(  \exp  - \sum_{m>0} t^{-m} \partial_{p_m} \Bigr).
\end{split}
\eeq
Let $B_n^{(k)}$, for $k=\{0,1\}$, $n\in \mathbb Z$, denotes the mode obtained from the operator $B(t,\tau)$ as
\beq
B_n^{(k)}= 
\begin{cases}
 \oint \dd t \int \dd \tau \, t^{-n-1} B(t,\tau)  \qquad \text{if $k=0$} \\
\oint \dd t \int \dd \tau  \, \tau t^{-n-1} B(t,\tau) \qquad \text{if $k=1$} 
\end{cases}
\eeq
and equivalently let $C_n^{(k)}$, for $k=\{0,1\}$, $n\in \mathbb Z$, denotes the mode obtained from the operator $C(t,\tau)$ as
\beq
C_n^{(k)}= 
\begin{cases}
 \oint \dd t \int \dd \tau \, \tau t^{-n-1} C(t,\tau)  \qquad \text{if $k=0$} \\
\oint \dd t \int \dd \tau  \,  t^{-n-1} C(t,\tau) \qquad \text{if $k=1$} 
\end{cases}
\eeq
In the following, we shall be interested in the positive modes ($n\geq0$) only.  Explicitly, we can write these modes as
\beq
B_n^{(0)}= \mathcal L_0 B_n^{(1)}, \qquad 
B_n^{(1)}= \sum_{r\geq 0} (-1)^r \tilde{p}_{n+r} e_r[\hat{\partial}_p], \qquad
C_n^{(0)}= \sum_{r \geq 0} (-1)^r h_{n+r} e_r[\hat{\partial}_p], \qquad
C_n^{(1)}= \mathcal L_0^\perp  C_n^{(0)}.  
\eeq

For a superpartition $\La$, let $\mathsf k_{i}(\La)= \mathsf k_i = \La^\cd_i - \La^*_i=\{0,1\}$ for $i=1,2, \ldots, \ell$.   For any superpartition $\La$ of $\ell$ parts, we can compute the associated super-Schur functions from:
\beq
s_\La = B_{\La^*_1}^{(\mathsf k_1)} \ldots B_{\La_\ell^*}^{(\mathsf k_\ell)}.1; \qquad \quad
s^*_\La = C_{\La^*_1}^{(\mathsf k_1)} \ldots C_{\La_\ell^*}^{(\mathsf k_\ell)}.1,
\eeq
that is, starting from an action on the identity and then successively acting with an ordered sequence of $B$ or $C$ mode operators.

We end this section by  giving  an example to illustrate the construction.  Consider $\La= (2,0;2,1)$, so that $s_\La^*$ is given by
\[
s^*_{(2,0;2,1) } = C_2^{(1)} C_2^{(0)} C_1^{(0)} C_0^{(1)}.1
\]
To use in our computation, we list the first expressions of the homogenous and elementary functions in terms of power-sum:
\[
\begin{tabular}{|c||c|c|} 
\hline
 &  $h_r $ &   $e_r$   \\
\hline\hline
$r=1$ & $p_1$ & $p_1$  \\
 $r=2$ & $\frac12(p_1^2 + p_2) $&$\frac12(p_1^2-p_2)$    \\  
 $r=3 $ & $ \frac16 (p_1^3 + 3 p_2p_1 + 2p_3)$ & $  \frac16 (p_1^3 - 3 p_2p_1 + 2p_3)$ \\
 $r=4$ & $\frac{1}{24} ( p_1^4 +6 p_2p_1^2 +3 p_2^2 +8 p_{3}p_1+6p_4)$ & - \\
 $r=5$ & $\frac{1}{120}(p_1^5+10 p_2 p_1^4+15 p_2^2 p_1+20 p_3p_1^2+20 p_3p_2+30 p_4 p_1+24 p_5)$ & - \\
\hline
\end{tabular}
\]
Then, we proceed by recursion, we have
\[
C_0^{(1)}.1 = \mathcal L_0^\perp .1 = \tilde p_0; \qquad \quad
C_1^{(0)} \, \tilde p_0 = h_1 \tilde p_0 = \tilde p_0 p_1; \qquad \quad
C_2^{(0)} (  \tilde p_0 p_1) = (h_2-h_3 \partial_{p_1})( \tilde p_0 p_1) = \frac13 \tilde p_0 (p_1^3-p_3);
\]
and
\[
\begin{split} 
C_2^{(0)} (  \tilde p_0 (p_1^3-p_3)/3) &= \frac13 \tilde p_0 \Bigl( h_2 - h_3 \partial_{p_1} +\tfrac12 h_4\, (\partial_{p_1}^2 - 2 \partial_{p_2}) - \tfrac16 h_5 \, (\partial_{p_1}^3 - 6\partial_{p_2} \partial_{p_1} + 6 \partial_{p_3}) \Bigr) (p_1^3-p_3) \\
&= \frac{1}{24} \tilde p_0 (p_1^5 - 2 p_2 p_1^3 + 3 p_2^2 p_1-4p_3p_1^2 - 4p_3p_2 + 6 p_4p_1).
\end{split}
 \tag{$\star$}
\]
The (remaining) action of the mode $C_2^{(1)}$ is obtained by acting with $\mathcal L_0^\perp$ on expression $(\star)$, which is
\[\begin{split}
\mathcal L_0^\perp (\star) &=
\frac{1}{24}( \tilde p_1 \tilde p_0  \partial_{p_1} + \tfrac12 \tilde p_2 \tilde p_0 ( \partial_{p_1}^2 + 2 \partial_{p_2}) )  (p_1^5 - 2 p_2 p_1^3 + 3 p_2^2 p_1-4p_3p_1^2 - 4p_3p_2 + 6 p_4p_1)
\\
&= \tilde p_1 \tilde p_0 \bigl( \tfrac{5}{24} p_1^4 - \tfrac{1}{4} p_2 p_1^2 + \tfrac18 p_2^2 - \tfrac13 p_3p_1 + \tfrac14 p_4\bigr) + \tilde p_2 \tilde p_0 \bigl( \tfrac13 p_1^3 - \tfrac13 p_3\bigr)
\\
&=\tfrac{5}{24} p_{(1,0;1^4)} - \tfrac{1}{4}p_{(1,0;2,1,1)} +\tfrac18 p_{(1,0;2,2)}- \tfrac13 p_{(1,0;3,1)} + \tfrac14 p_{(1,0;4)} + \tfrac13 p_{(2,0;1^3)}- \tfrac13 p_{(2,0;3)}
\end{split}\]
which corresponds exactly to the superpolynomial $s^*_{(2,0;2,1)}$, expanded into the power-sum basis.  Note that acting with $(-1) \widehat{\omega}$ on this last expression (which is straightforward because it is already in terms of power-sum), we obtain from  \eqref{setwithomegasbar} the expression for $\bar{s}_{(3,0;2)}$.

\bigskip

\n {\textsc{References:}}  The operators $B(t,\tau)$ and $C(t,\tau)$ are called super-Bernstein operators, as they are the (superspace) analogues of the standard Bernstein operators which construct the Schur symmetric functions \cite{MacSym95}.  The proofs that these super-Bernstein operators build the different families of super-Schurs rely on the Pieri rules presented earlier and
can be found in \cite{ABLM18}.

\bigskip

\subsection{Schur functions in superspace and tableaux}
We now describe how the Schur functions in superspace $s_\Lambda$
and $\bar s_\Lambda$ are generating series of certain types of tableaux.  

We will refer to $\{\bar 0, \bar 1, \bar 2,\bar 3,\dots \}$ as the set of {\it fermionic} nonnegative integers.  In this spirit,
we will also refer to the set of nonnegative
integers  $\{0, 1, 2,3,\dots \}$ as the set of {\it bosonic} nonnegative integers.  For  
$\alpha \in   \{0,\bar 0,1,\bar 1,2,\bar 2, \dots \}$, we will say that ${\rm type}(\alpha)$ is bosonic or fermionic
depending on whether the corresponding integer is fermionic or bosonic.  Finally, define
\begin{equation}
|\alpha|= 
\left\{
\begin{array}{ll}
a & \text{if~} \alpha=\bar a \text{~is fermionic} \\
a & \text{if~} \alpha=  a \text{~is bosonic}
\end{array} \right.
\end{equation}

\subsubsection{$s$-tableaux}  
We say that the sequence $\Omega=\Lambda_{(0)},\Lambda_{(1)},\dots, \Lambda_{(n)}=\Lambda$ is an $s$-tableau 
of shape $\Lambda/\Omega$ and weight $(\alpha_1,\dots,\alpha_n)$, where 
$\alpha_i\in  \{0,\bar 0, 1,\bar 1,2,\bar 2,\dots \}$, if $\Omega=\Lambda_{(i)}$ and $\Lambda=\Lambda_{(i-1)}$ 
obey the conditions of type RMII with $r=\alpha_i$ whenever $\alpha_i$ is bosonic or $r=|\alpha_i|$ whenever $\alpha_i$ is fermionic.  
An $s$-tableau can be represented by a diagram constructed recursively in the following way: 
\begin{enumerate}
\item  the cells of $\Lambda_{(i)}^*/\Lambda_{(i-1)}^*$, which form a horizontal strip,  are filled with the letter $i$.  In 
the fermionic case, the new circle is also filled with a letter $i$. 
\item the circles of $\Lambda_{(i-1)}$ that are moved a row below keep their fillings.
\end{enumerate}
The sign of an $s$-tableau $T$, which corresponds to the product of the signs appearing in the fermionic horizontal strips, 
can be extracted quite efficiently from an $s$-tableau.  Read the fillings of the circles from top to bottom
to obtain a word (without repetition):
the sign of the tableau $T$ is  then equal to $(-1)^{{\rm inv} (T)}$, 
where ${{\rm inv} (T)}$ is the number of inversions of the word.

Given a diagram of an $s$-tableau, we define the path of a given circle (filled let's say with letter $i$) in the following way.  Let $c$ be the leftmost column that does not contain a square (a cell of $\Omega^*$) 
filled with an $i$.  The path starts in the position of the smallest entry 
larger than $i$ (let's say $j$) in column  $c$.  The path then moves to the smallest entry (let's say $k$) larger than $j$ in the row below (if there are many such $k$'s the path goes through the leftmost such $k$).   We continue this way
until
we reach the row above that of the circle filled with an $i$.

It is important to realize that a tableau can be identified with its diagram given that the sequence  $\Omega=\Lambda_{(0)},\Lambda_{(1)},\dots, \Lambda_{(n)}=\Lambda$ can be recovered from the diagram.    We obtain the
diagram corresponding to $\Omega=\Lambda_{(0)},\Lambda_{(1)},\dots, \Lambda_{(n-1)}$ by removing the letters $n$ from the diagram 
(including, possibly, the circled one), and by moving the remaining circle one row above according to the following rule.  
A circle (filled let's say with letter $i$) in a given row $r$ 
is moved to row $r-1$ if there is an $n$ in row $r-1$ that 
belongs to its path.  Otherwise the circle
in row $r$ stays in its position. 

\smallskip

Consider the tableau $\scriptsize {\tableau[scY]{1&2&2&3& 4 \\ 2&4&6  \\4&5&\bl\tcercle{4} \\ 6&6 \\ \bl\tcercle{2} }}$ of weight $(1,\bar 3, 1, \bar 3, 1,3)$ and shape $(2,0;5,3,2)$ ($\Omega=\emptyset$ in the example). 
 The path for the $\tcercle{2}$ is  ${\scriptsize {\tableau[scY]{1&2&2&\bf{3} & 4 \\ 2&\bf{4}&6  \\4&\bf{5}&\bl\tcercle{4} \\ \bf{6} & 6 \\ \bl\tcercle{2} }}}$ which is
seen as follows:  the leftmost column without a non-circled 2 is column 4.  Since there is only one entry, a 3, in that column, the path starts there.   The smallest entry larger than 3 in the row below, the second one, is 4.  Then  the smallest entry larger than 4 in the third row is 5.  Finally, the smallest entry larger than 4 in row 4 is 6, and the path goes through the leftmost 6.  The path then stops since the circled 2 is in row 5.
The
path for $\tcercle{4}$ can similarly be seen to be  ${\scriptsize {\tableau[scY]{1&2&2&3& 4 \\ 2&4&\bf{6}  \\4&5&\bl\tcercle{4} \\ 6&6 \\ \bl\tcercle{2} }}}$.  
It is obvious from the example that the non-circled letters of an $s$-tableau form an ordinary tableau.  
However it is not obvious where the circled letters can be added, and the paths above
could only be constructed because the tableau was valid.

\smallskip

\noindent The sequence of superpartitions associated to that $s$-tableau can then be recovered by stripping successively the tableau of its largest letter and possibly moving circled letters one row above according 
to their paths:

  {\scriptsize
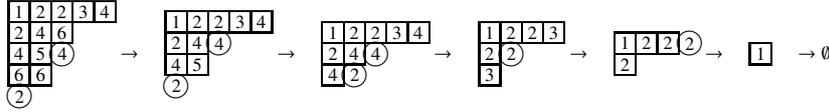
\begin{tabular}{lllllll} 
${\tableau[scY]{1&2&2&3& 4 \\ 2&4&6  \\4&5&\bl\tcercle{4} \\ 6&6 \\ \bl\tcercle{2} }}\rightarrow$&
${\tableau[scY]{1&2&2&3& 4 \\ 2&4 &\bl\tcercle{4}  \\4&5 \\ \bl\tcercle{2}\\  }}\rightarrow$&
${\tableau[scY]{1&2&2&3&4 \\ 2&4 &\bl\tcercle{4}  \\4&\bl\tcercle{2}  \\  }}\rightarrow$&
${\tableau[scY]{1&2&2&3 \\ 2& \bl\tcercle{2}  \\3 }}\rightarrow$&
${\tableau[scY]{1&2&2& \bl\tcercle{2}   \\ 2}}\rightarrow$&
${\tableau[scY]{1}}$ & $\rightarrow \emptyset$ 
\end{tabular}}

\medskip

Now, define the skew Schur function in superspace $s_{\Lambda/\Omega}$ as
\begin{equation}
s_{\Lambda/\Omega} = \sum_{T} (-1)^{{\rm inv}(T)} (x\theta)^T
\end{equation}
where the sum is over all $s$-tableaux of shape $\Lambda/\Omega$, and where
\begin{equation}
(x\theta)^T= \prod_i x_i^{|\alpha_i|} \prod_{j \, : \, {\rm type}(\alpha_j)={\rm fermionic}} \theta_j
\end{equation}
if $T$ is of weight $(\alpha_1,\dots,\alpha_n)$.  We stress that the product over anticommuting variables is ordered
from left to right over increasing indices.

\begin{proposition}\cite{JL17} \label{propsymfun}
 $s_{\Lambda/\Omega}$ is a symmetric function in superspace.  Moreover,
\begin{equation}
s_{\Lambda/\Omega}= \sum_{\Gamma} \bar K_{\Lambda/\Omega, \Gamma} \, m_\Gamma 
\end{equation}
where $\bar K_{\Lambda/\Omega, \Gamma}=\sum_T  (-1)^{{\rm inv}(T)}$, with the sum over 
 all $s$-tableaux $T$ of weight 
$$(\bar \Gamma_1,\dots,\bar \Gamma_m,\Gamma_{m+1},\dots,\Gamma_N)  \qquad \text{\rm for} \quad \Gamma=(\Gamma_1,\dots, \Gamma_m;\Gamma_{m+1},\dots,\Gamma_N)$$ 
\end{proposition}

\begin{corollary} \cite{JL17} \label{coroschur}
We have that $s_{\Lambda/\emptyset}=s_{\Lambda}$.  Hence
\begin{equation}
s_{\Lambda} = \sum_{T} (-1)^{{\rm inv}(T)} (x\theta)^T
\end{equation}
where the sum is over all $s$-tableaux $T$ of shape $\Lambda$.
\end{corollary}

We thus obtain the monomial expansion of  $s_{(3,1;2,1,1)}$ by listing every filling of the shape $(3,1;2,1,1)$ whose weight corresponds to a superpartition:

$$
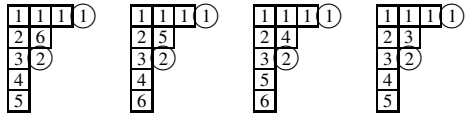
\scriptsize{
{\tableau[scY]{1&1&1&\bl\tcercle{1}\\2&6\\3&\bl\tcercle{2} \\ 4\\5 }}\qquad
{\tableau[scY]{1&1&1&\bl\tcercle{1}\\2&5\\3&\bl\tcercle{2} \\ 4\\6 }}\qquad
{\tableau[scY]{1&1&1&\bl\tcercle{1}\\2&4\\3&\bl\tcercle{2} \\ 5\\6 }}\qquad
{\tableau[scY]{1&1&1&\bl\tcercle{1}\\2&3\\3&\bl\tcercle{2} \\ 4\\5 }}
}
$$
Therefore, $s_{(3,1;2,1,1)}=3\, m_{(3,1;1,1,1,1)}+m_{(3,1;2,1,1)}$ since there are 3 tableaux of weight
$(\bar 3,\bar 1, 1,1,1,1 )$ and one tableau of weight $(\bar 3, \bar 1,2,1,1)$.  We stress that
we don't have an easy criteria in general to determine whether a given filling is a valid tableau.  
However, in the example above, the rules for constructing tableaux immediately imply that we need to have
three non-circled 1's and one non-circled 2 (otherwise the circled 2 could never be in the second
column).  Then there are very few possibilities to fill the rest of the tableau with a weight corresponding to a superpartition. The case of weight
$(\bar 3,\bar 1, 1,1,1,1 )$ where a 3 is above the circled 2 is not allowed since again this would prevent the circled 2 from being in the
second column.

Note that when $\Omega = \emptyset$, the coefficient $\bar K_{\Lambda/\Omega, \Gamma}$ is always a nonnegative integer.

\subsubsection{$\bar s$-tableaux} 
We say that the sequence $\Omega=\Lambda_{(0)},\Lambda_{(1)},\dots, \Lambda_{(n)}=\Lambda$ is an $\bar s$-tableau 
of shape $\Lambda/\Omega$ and weight $(\alpha_1,\dots,\alpha_n)$, where 
$\alpha_i\in  \{0,\bar 0, 1,\bar 1,2,\bar 2,\dots \}$, if $\Omega=\Lambda_{(i)}$ and $\Lambda=\Lambda_{(i-1)}$ 
obey the conditions of type RMI with $\ell=\alpha_i$ whenever $\alpha_i$ is bosonic or with $\ell=|\alpha_i|$ whenever $\alpha_i$ is fermionic.  
An $\bar s$-tableau can be represented by a diagram constructed recursively in the following way: 
\begin{enumerate}
\item  the cells of $\Lambda_{(i)}^*/\Lambda_{(i-1)}^*$ are filled with the letter $i$.  In 
the fermionic case, the new circle is also filled with a letter $i$ 
\item the circles of $\Lambda_{(i-1)}$ that moved along a column or a row keep their fillings.
\end{enumerate}
As is the case for $s$-tableaux, the sign of an $\bar s$-tableau $T$ is equal to $(-1)^{{\rm inv} (T)}$, where ${{\rm inv} (T)}$ is 
the number of inversions of the word obtained by reading the filling of the circles from top to bottom.

It is important to realize that the sequence  $\Omega=\Lambda_{(0)},\Lambda_{(1)},\dots, \Lambda_{(n)}=\Lambda$ can be recovered from the diagram.  We obtain the
diagram corresponding to $\Omega=\Lambda_{(0)},\Lambda_{(1)},\dots, \Lambda_{(n-1)}$ by removing the letters $n$ from the diagram 
(including, possibly, the circled one), and by moving the circled letters one cell above if there is a letter $n$ above them
or to their left if there are letters $n$ to their left and none above them.  For instance, if one considers the $\bar s$-tableau $T$ below of weight $(3,\bar 1, \bar 2,1,\bar 3,5,\bar 0)$, the sequence of superpartitions associated to it can then be recovered by stripping successively the tableaux of their largest letter:

\medskip

{\scriptsize
\begin{tabular}{llllllll} 
$T={\tableau[scY]{1&1&1&5 & 6 & 6 & \bl \tcercle{5}  \\2&3&4 & 6 & \bl \tcercle{7} \\3&5& 6 & \bl \tcercle{3}\\5 & 6 & \bl\tcercle{2}\\}} \! \! \! \rightarrow$&
${\tableau[scY]{1&1&1&5 & 6 & 6 & \bl \tcercle{5}  \\2&3&4 & 6 \\3&5& 6 & \bl \tcercle{3}\\5 & 6 & \bl\tcercle{2}\\}}
\! \! \! \rightarrow$&
${\tableau[scY]{1&1&1&5 &\bl \tcercle{5}  \\2&3&4 & \bl \tcercle{3}\\3&5&\bl\tcercle{2} \\5 \\}}\!\! \rightarrow$&
${\tableau[scY]{1&1&1 & \bl \tcercle{3}  \\2&3 & 4 \\3&\bl\tcercle{2}\\ }} \! \rightarrow$&
${\tableau[scY]{1&1&1 &\bl \tcercle{3}  \\2&3  \\ 3 & \bl\tcercle{2}\\ }}\! \rightarrow$&
${\tableau[scY]{1&1 & 1 \\2 &\bl\tcercle{2} \\  }}\, \, \rightarrow$&
${\tableau[scY]{1 &1 &1 \\  }}$&
\end{tabular}}

We should stress that there is no immediate criteria to determine whether a given filling of a shape is a valid tableau.  It is only after checking that every removal of a letter corresponds to an application of a Pieri rule that we know that the filling is valid.  In the example above, removing the 6 corresponds to
an application of the Pieri rule $h_5$ since the the 6's form a horizontal strip and when moving 
the circled 2 and 3 above and the circled 5 to its left there are no collisions (two circles in the same row or
column).  Similarly, removing the 5's corresponds to the Pieri rule $\tilde h_3$ since the 5's form a horizontal strip with the circled one being the rightmost and when moving the circled 2 to its left and
the circled 3 above there are no collisions.

\smallskip

As was done in the previous subsection, define the skew Schur function in superspace $\bar s_{\Lambda/\Omega}$ as
\begin{equation}
\bar s_{\Lambda/\Omega} = \sum_{T} (-1)^{{\rm sign}(T)} (x\theta)^T
\end{equation}
where the sum is over all $\bar s$-tableaux of shape $\Lambda/\Omega$. 

\begin{proposition} \cite{JL17}  $\bar s_{\Lambda/\Omega}$ is a symmetric function in superspace.  Moreover,
\begin{equation}
\bar s_{\Lambda/\Omega}= \sum_{\Gamma}  K_{\Lambda/\Omega, \Gamma} \, m_\Gamma 
\end{equation}
where $K_{\Lambda/\Omega, \Gamma}=\sum_T  (-1)^{{\rm sign}(T)}$, the sum over 
 all $\bar s$-tableaux $T$ of weight 
$$(\bar \Gamma_1,\dots,\bar \Gamma_m,\Gamma_{m+1},\dots,\Gamma_N)  \qquad \text{\rm for} \quad \Gamma=(\Gamma_1,\dots, \Gamma_m;\Gamma_{m+1},\dots,\Gamma_N)$$ 
\end{proposition}
\begin{corollary} \cite{JL17} \label{coroschurb}
We have that $\bar s_{\Lambda/\emptyset}=\bar s_{\Lambda}$.  Hence
\begin{equation}
\bar s_{\Lambda} = \sum_{T} (-1)^{{\rm sign}(T)} (x\theta)^T
\end{equation}
where the sum is over all $\bar s$-tableaux in superspace of shape $\Lambda$. 
\end{corollary}
The monomial expansion of $\bar s_{(2,0;3)}$ is thus obtained by listing every filling of the shape $(2,0;3)$ whose weight is that of a superpartition
$$
\scriptsize{
{\tableau[scY]{1&4&6\\3&5&\bl\tcercle{1}\\ \bl\tcercle{2}  }}\,\, \, 
{\tableau[scY]{1&4&5\\3&6&\bl\tcercle{1}\\ \bl\tcercle{2}  }}\,\, \, 
{\tableau[scY]{1&3&6\\4&5&\bl\tcercle{1}\\ \bl\tcercle{2}  }}\qquad
{\tableau[scY]{1&3&4\\3&5&\bl\tcercle{1}\\ \bl\tcercle{2}  }} \,\, \, 
{\tableau[scY]{1&3&5\\3&4&\bl\tcercle{1}\\ \bl\tcercle{2}  }}\qquad
{\tableau[scY]{1&3&4\\3&4&\bl\tcercle{1}\\ \bl\tcercle{2}  }}\qquad
{\tableau[scY]{1&3&3\\3&4&\bl\tcercle{1}\\ \bl\tcercle{2}  }}\qquad
{\tableau[scY]{1&1&5\\3&4&\bl\tcercle{1}\\ \bl\tcercle{2}  }}\qquad
{\tableau[scY]{1&1&4\\3&3&\bl\tcercle{1}\\ \bl\tcercle{2}  }}\qquad
{\tableau[scY]{1&1&3\\3&3&\bl\tcercle{1}\\ \bl\tcercle{2}  }}
}
$$
Hence,
$$\bar s_{(2,0;3)}=3\, m_{(1,0;1,1,1,1)}+2\,m_{(1,0;2,1,1)}+m_{(1,0;2,2)}+m_{(1,0;3,1)}+m_{(2,0;1,1,1)}+m_{(2,0;2,1)}+m_{(2,0;3)}\, .$$ 
As was mentioned before, 
we don't have an easy criteria in general to determine whether a given filling is a valid tableau.  For instance, the tableau
$\scriptsize{
{\tableau[scY]{1&3&5\\4&6&\bl\tcercle{1}\\ \bl\tcercle{2}  }}}$ is not valid because the circled 1 and
the circled 2 collide at the moment of removing letter 4:
$$
\scriptsize{
{\tableau[scY]{1&3&5\\4&6&\bl\tcercle{1}\\ \bl\tcercle{2}  }}} \quad  \rightarrow \quad \scriptsize{
{\tableau[scY]{1&3&5\\4&\bl\tcercle{1}\\ \bl\tcercle{2}  }}} \quad \rightarrow \quad \scriptsize{
{\tableau[scY]{1&3\\4&\bl\tcercle{1}\\ \bl\tcercle{2}  }}} 
$$

We note that, as is the case for $\bar K_{\Lambda/\Omega, \Gamma}$,
the coefficient $K_{\Lambda/\Omega, \Gamma}$ is a nonnegative integer when $\Omega = \emptyset$.

\medskip
\n\textsc{References:} Note that a somewhat different combinatorial formula in terms of tableaux for the expansion coefficients of the super-Schurs into the monomial basis is presented as a conjecture in \cite{BM1}.

\subsection{Skew Schur functions in superspace and Littlewood-Richardson coefficients}
We first give the relation between the skew Schur functions in superspace
and Schur functions in superspace through 
the scalar product 
$\langle \! \langle \cdot \, , \cdot \rangle \! \rangle $.  We then connect the generalization to superspace of the
Littlewood-Richardson coefficients to skew Schur functions in superspace.  
These basic properties of skew Schur functions in superspace 
generalize well known properties in the classical case ($m=0$).

\begin{corollary} We have
\begin{equation}
\langle \! \langle s_{\Omega}^* \, f, s_{\Lambda} \rangle \! \rangle =
\langle \! \langle f , s_{\Lambda/\Omega} \rangle \! \rangle   \qquad {\rm and} \qquad
 \langle \! \langle   \bar s_{\Omega}^* \, f, \bar s_{\Lambda} \rangle \! \rangle =
\langle \! \langle f , \bar s_{\Lambda/\Omega} \rangle \! \rangle
\end{equation}
for all symmetric functions in superspace $f$.
\end{corollary}

Define $\bar c^\Lambda_{\Gamma \Omega}$ and $c^\Lambda_{\Gamma \Omega}$ to be respectively such that 
\begin{equation} 
\bar s_{\Gamma}\,  \bar s_{\Omega} = \sum_{\Lambda} \bar c^\Lambda_{\Gamma \Omega}\,  \bar s_{\Lambda}
 \qquad {\rm and} \qquad    
s_{\Gamma}\,  s_{\Omega} = \sum_{\Lambda} c^\Lambda_{\Gamma \Omega}\,  s_{\Lambda}.
\end{equation}
It is immediate from the (anti-)commutation relations between the Schur functions in superspace
 that if $\Gamma$ and $\Omega$ are respectively of fermionic degrees $a$ and $b$, then 
$\bar c^\Lambda_{\Gamma \Omega}=(-1)^{ab} \,  \bar c^\Lambda_{\Omega \Gamma}$ and 
$c^\Lambda_{\Gamma \Omega}=(-1)^{ab}  \, c^\Lambda_{\Omega \Gamma}$.  Even though $\bar c^\Lambda_{\Gamma \Omega}$ and 
$c^\Lambda_{\Gamma \Omega}$ are not always nonnegative from these relations, 
we can consider them as generalizations to superspace
of the Littlewood-Richardson coefficients. 

 We now extend to superspace the well-known connection between 
Littlewood-Richardson coefficients and skew Schur functions.
\begin{proposition} We have
\begin{equation}
s_{\Lambda/\Omega} =  \sum_{\Gamma}  \bar c^{\Lambda'}_{\Gamma' \Omega'} \, s_\Gamma
\qquad {\rm and} \qquad  
\bar s_{\Lambda/\Omega} = \sum_{\Gamma}  c^{\Lambda}_{\Omega \Gamma} \, \bar s_\Gamma.
\end{equation}
Furthermore, 
 $c^\Lambda_{\Omega \Gamma}=c^{\Lambda'}_{\Gamma' \Omega'}$.

\begin{open} An interesting problem would be to obtain Littlewood-Richardson rules for the coefficients $c^\Ga_{\La \Om}, \bar c^\Ga_{\La \Om}$.  
\end{open}
\end{proposition}

Somewhat surprisingly, the coefficient  $\bar c^\Lambda_{\Omega \Gamma}$ does not
have in general any symmetry under conjugation.
 For instance, it can be checked that
if $\Gamma=(1;)$, $\Omega=(0;)$ and $\Lambda=(1,0;)$ then 
$\bar c^{\Lambda}_{\Omega \Gamma}=1$ while
$\bar c^{\Lambda'}_{\Gamma' \Omega'}=0$.

\bigskip

\n {\textsc{References:}}
The relation between generalization of Littlewood-Richardson coefficients and skew Schur superpolynomials was considered in \cite{JL17}.  

\bigskip

%

\bibliography{ref2}

\begin{thebibliography}{10}

\bibitem{MacSym95}
I~G Macdonald.
\newblock {\em Symmetric functions and {Hall} polynomials}.
\newblock Oxford Mathematical Monographs. Clarendon Press, 2nd edition, 1995.

\bibitem{Ruij99}
S~N~M Ruijsenaars.
\newblock Systems of {C}alogero-{M}oser type.
\newblock In {\em Particles and Fields}, pages 251--352. Springer, 1999.

\bibitem{Mimachi95}
K~Mimachi and Y~Yamada.
\newblock Singular vectors of the {V}irasoro algebra in terms of {J}ack
  symmetric polynomials.
\newblock {\em Comm. Math. Phys.}, 174:447--455, 1995.

\bibitem{Awata95}
H~Awata, Y~Matsuo, S~Odake, and J~Shiraishi.
\newblock Excited states of the {C}alogero-{S}utherland model and singular
  vectors of the ${W}_n$ algebra.
\newblock {\em Nucl. Phys.}, B449:347--374, 1995.
\newblock \textsf{arXiv:1706.02243 [\mbox{math-ph}]}.

\bibitem{SSAFR}
R~Sakamoto, J~Shiraishi, D~Arnaudon, L~Frappat, and E~Ragoucy.
\newblock Correspondence between conformal field theory and
  {C}alogero-{S}utherland model.
\newblock {\em Nucl. Phys.}, B704:490--509, 2005.
\newblock \textsf{arXiv:hep-th/0407267}.

\bibitem{AGT}
L~F Alday, D~Gaiotto, and Y~Tachikawa.
\newblock Liouville correlation functions from four-dimensional gauge theories.
\newblock {\em LMP}, 91:167--197, 2010.
\newblock \textsf{arXiv:0906.3219 [\mbox{hep-th}]}.

\bibitem{AY}
H~Awata and Y~Yamada.
\newblock Five-dimensional {A}{G}{T} conjecture and the deformed {V}irasoro
  algebra.
\newblock {\em J. High Energy Phys.}, 125:1--11, 2010.
\newblock \textsf{arXiv:0910.4431 [\mbox{math-ph}]}.

\bibitem{MMS}
A~Mironov, A~Morozov, and S~Shakirov.
\newblock A direct proof of {A}{G}{T} conjecture at $\beta = 1$.
\newblock {\em J. High Energy Phys.}, 02:1--40, 2011.
\newblock \textsf{arXiv:1012.3137 [\mbox{hep-th}]}.

\bibitem{MMSS}
A~Mironov, A~Morozov, S~Shakirov, and A~Smirnov.
\newblock Proving {A}{G}{T} conjecture as {H}{S} duality: extension to five
  dimensions.
\newblock {\em Nucl. Phys.}, B855:128--151, 2012.
\newblock \textsf{arXiv:1105.0948 [\mbox{hep-th}]}.

\bibitem{Yan}
S~Yanagida.
\newblock Whittaker vectors of the {V}irasoro algebra in terms of {J}ack
  symmetric polynomial.
\newblock {\em J. Algebra}, 333:272--294, 2011.
\newblock \textsf{arXiv:1003.1049 [\mbox{math.QA}]}.

\bibitem{DLMnpb}
P~Desrosiers, L~Lapointe, and P~Mathieu.
\newblock Supersymmetric {C}alogero--{M}oser--{S}utherland models and {J}ack
  superpolynomials.
\newblock {\em Nucl. Phys.}, B606:547--582, 2001.
\newblock \textsf{arXiv:hep-th/0210190}.

\bibitem{DLM_sJack}
P~Desrosiers, L~Lapointe, and P~Mathieu.
\newblock Jack polynomials in superspace.
\newblock {\em Comm. Math. Phys.}, 242:331--360, 2003.
\newblock \textsf{arXiv:hep-th/0209074}.

\bibitem{DLM_class}
P~Desrosiers, L~Lapointe, and P~Mathieu.
\newblock Classical symmetric functions in superspace.
\newblock {\em J. Algebr. Comb.}, 24:209--238, 2006.
\newblock \textsf{arXiv:0509408 [\mbox{math.CO}]}.

\bibitem{DLMadv}
P~Desrosiers, L~Lapointe, and P~Mathieu.
\newblock Orthogonality of {J}ack polynomials in superspace.
\newblock {\em Adv. Math.}, 212:361--388, 2007.
\newblock \textsf{arXiv:math-ph/0509039}.

\bibitem{BF2}
O~Blondeau-Fournier, P~Desrosiers, L~Lapointe, and P~Mathieu.
\newblock Macdonald polynomials in superspace as eigenfunctions of commuting
  operators.
\newblock {\em J. of Comb.}, 3:495--561, 2012.
\newblock \textsf{arXiv:1202.3922 [\mbox{math-ph}]}.

\bibitem{ABLM18}
L~Alarie-V{\'e}zina, O~Blondeau-Fournier, L~Lapointe, and P~Mathieu.
\newblock Bernstein operators and super-{S}chur functions: combinatorial
  aspects.
\newblock {\em Lett. Math. Phys.}, pages 1--40, 2018.
\newblock \textsf{arXiv:1802.01705 [\mbox{math-ph}]}.

\bibitem{GJL}
J~Gatica, M~Jones, and L~Lapointe.
\newblock Pieri rules for the {J}ack polynomials in superspace and the 6-vertex
  model.
\newblock 2017.
\newblock \textsf{arXiv:1712.02416 [\mbox{math.CO}]}.

\bibitem{Freedman90}
D~Z Freedman and P~F Mende.
\newblock An exactly solvable ${N}$-particle system in supersymmetric quantum
  mechanics.
\newblock {\em Nucl. Phys.}, B344:317--343, 1990.

\bibitem{Shas93}
B~S Shastry and B~Sutherland.
\newblock Super {L}ax pairs and infinite symmetries in the $1/r^2$ system.
\newblock {\em Phys. Rev. Lett.}, 70:4029, 1993.

\bibitem{DLM_fauxsJack}
P~Desrosiers, L~Lapointe, and P~Mathieu.
\newblock Jack superpolynomials, superpartition ordering and determinantal
  formulas.
\newblock {\em Comm. Math. Phys.}, 263:383--402, 2003.
\newblock \textsf{arXiv:hep-th/0105107}.

\bibitem{DLM_eval}
P~Desrosiers, L~Lapointe, and P~Mathieu.
\newblock Evaluation and normalization of {J}ack superpolynomials.
\newblock {\em Int. Math. Res. Not.}, 23:5267--5327, 2012.
\newblock \textsf{arXiv:1104.3260 [\mbox{math.CO}]}.

\bibitem{BDM15}
O~Blondeau-Fournier, P~Desrosiers, and P~Mathieu.
\newblock Supersymmetric {R}uijsenaars-{S}chneider model.
\newblock {\em Phys. Rev. Lett.}, 114:121602, 2015.
\newblock \textsf{arXiv:1403.4667 [\mbox{hep-th}]}.

\bibitem{BF1}
O~Blondeau-Fournier, P~Desrosiers, L~Lapointe, and P~Mathieu.
\newblock Macdonald polynomials in superspace: conjectural definition and
  positivity conjectures.
\newblock {\em Lett. Math. Phys.}, 101:27--47, 2012.
\newblock \textsf{arXiv:1112.5188 [\mbox{math-ph}]}.

\bibitem{BM1}
O~Blondeau-Fournier and P~Mathieu.
\newblock Schur superpolynomials: combinatorial definition and {P}ieri rule.
\newblock {\em SIGMA}, 11, 2015.
\newblock \textsf{arXiv:1408.2807 [\mbox{math-ph}]}.

\bibitem{JL17}
M~Jones and L~Lapointe.
\newblock Pieri rules for {S}chur functions in superspace.
\newblock {\em J. of Comb. Theor.}, 148:57--115, 2017.
\newblock \textsf{arXiv:1608.08577 [\mbox{math.CO}]}.

\bibitem{SageComp}
Sage{M}ath compendium, a worksheet on symmetric superpolynomials.
\newblock \url{https://github.com/LAV42/SageCompendium}.

\bibitem{corteel2004overpartitions}
S~Corteel and J~Lovejoy.
\newblock Overpartitions.
\newblock {\em Trans. Amer. Math. Soc.}, 356:1623--1635, 2004.

\bibitem{LV}
L~Lapointe and L~Vinet.
\newblock Exact operator solution of the {C}alogero-{S}utherland model.
\newblock {\em Comm. Math. Phys.}, 178:425--452, 1996.
\newblock \textsf{arXiv:q-alg/9509003}.

\bibitem{ICher}
I~Cherednik.
\newblock A unification of {K}nizhnik-{Z}amolodchikov and {D}unkl operators via
  affine {H}ecke algebras.
\newblock {\em Inven. Math.}, 106:411--431, 1991.

\bibitem{LLBN}
L~Lapointe, Y~Le Borgne, and P~Nadeau.
\newblock A normalization formula for the {J}ack polynomials in superspace and
  an identity on partitions.
\newblock {\em Electronic J. Comb}, 16:70, 2009.
\newblock \textsf{arXiv:0803.4182 [\mbox{math.CO}]}.

\bibitem{Knop}
F~Knop and S~Sahi.
\newblock A recursion and a combinatorial formula for {J}ack polynomials.
\newblock {\em Inven. Math.}, 128:9--22, 1997.
\newblock \textsf{arXiv:q-alg/9610016}.

\bibitem{DLM_clust}
P~Desrosiers, L~Lapointe, and P~Mathieu.
\newblock Jack superpolynomials with negative fractional parameter: clustering
  properties and super-{V}irasoro ideals.
\newblock {\em Comm. Math. Phys.}, 316:395--440, 2012.
\newblock \textsf{arXiv:1109.2832 [\mbox{math-ph}]}.

\bibitem{Feigin}
B~Feigin, M~Jimbo, T~Miwa, and E~Mukhin.
\newblock A differential ideal of symmetric polynomials spanned by {J}ack
  polynomials at $\beta=-(r-1)/(k+1)$.
\newblock {\em Int. Math. Res. Not.}, 23:1223--1237, 2002.
\newblock \textsf{arXiv:0112127 [\mbox{math.QA}]}.

\bibitem{melzer1994}
E~Melzer.
\newblock Supersymmetric analogs of the {G}ordon-{A}ndrews identities, and
  related {T}{B}{A} systems.
\newblock 1994.
\newblock \textsf{arXiv:9412154 [\mbox{hep-th}]}.

\bibitem{DLM_SCFT}
P~Desrosiers, L~Lapointe, and P~Mathieu.
\newblock Superconformal field theory and {J}ack superpolynomials.
\newblock {\em J. High Energy Phys.}, 9:1--43, 2012.
\newblock \textsf{arXiv:1205.0784 [\mbox{hep-th}]}.

\bibitem{LM}
L~Lapointe and P~Mathieu.
\newblock From {J}ack to double {J}ack polynomials via the supersymmetric
  bridge.
\newblock {\em SIGMA}, 11:051, 2015.
\newblock \textsf{arXiv:1503.09029 [\mbox{hep-th}]}.

\bibitem{ADM}
L~Alarie-V{\'{e}}zina, P~Desrosiers, and P~Mathieu.
\newblock Ramond singular vectors and {J}ack superpolynomials.
\newblock {\em J. Phys.}, A47:1--17, 2013.
\newblock \textsf{arXiv:1309.7965 [\mbox{hep-th}]}.

\bibitem{BMRW}
O~Blondeau-Fournier, P~Mathieu, D~Ridout, and S~Wood.
\newblock The super-{V}irasoro singular vectors and {J}ack superpolynomials
  relationship revisited.
\newblock {\em Nucl. Phys.}, 913:34--63, 2016.
\newblock \textsf{arXiv:1605.08621 [\mbox{math-ph}]}.

\bibitem{Yana2015}
S~Yanagida.
\newblock Singular vectors of ${N}=1$ super-{V}irasoro algebra via {U}glov
  symmetric functions.
\newblock 2015.
\newblock \textsf{arXiv:1508.06036 [\mbox{math.QA}]}.

\bibitem{Bel2013}
A~Belavin, M~Bershtein, and G~M Tarnopolsky.
\newblock Bases in coset conformal field theory from {A}{G}{T} correspondence
  and {M}acdonald polynomials at the roots of unity.
\newblock {\em J. High Energy Phys.}, 2013:19, 2013.
\newblock \textsf{arXiv:1211.2788 [\mbox{hep-th}]}.

\bibitem{OBFphdth}
O~Blondeau-Fournier.
\newblock {\em Les polyn\^omes de {M}acdonald dans le superespace et le
  mod\`ele {R}uijsenaars-{S}chneider supersym\'etrique}.
\newblock {PhD} thesis, Laval University, 2015.

\bibitem{GL2018}
C~Gonz\'alez and L~Lapointe.
\newblock The norm and the evaluation of the {M}acdonald polynomials in
  superspace.
\newblock 2018.
\newblock \textsf{arXiv:1808.04941 [\mbox{math.CO}]}.

\bibitem{ICher_NS}
I~Cherednik.
\newblock Nonsymmetric {M}acdonald polynomials.
\newblock {\em Int. Math. Res. Not.}, 1995:483--515, 1995.

\bibitem{BLM15}
O~Blondeau-Fournier, L~Lapointe, and P~Mathieu.
\newblock Double {M}acdonald polynomials as the stable limit of {M}acdonald
  superpolynomials.
\newblock {\em J. Algebr. Comb.}, 41:397--459, 2015.
\newblock \textsf{arXiv:1211.3186 [\mbox{math-ph}]}.

\bibitem{Lasc03}
A~Lascoux.
\newblock {\em Symmetric functions and combinatorial operators on polynomials},
  volume~99.
\newblock American Mathematical Soc., 2003.

\bibitem{Hag08}
J~Haglund.
\newblock {\em The $q, t$-{C}atalan numbers and the space of diagonal
  harmonics: with an appendix on the combinatorics of {M}acdonald polynomials},
  volume~41.
\newblock American Mathematical Soc., 2008.

\bibitem{Berg09}
F~Bergeron.
\newblock {\em Algebraic combinatorics and coinvariant spaces}.
\newblock CRC Press, 2009.

\end{thebibliography}
\bibliographystyle{unsrt}
\end{document}